# Acyclicity Notions for Existential Rules and
# Their Application to Query Answering in Ontologies

**Bernardo Cuenca Grau**                    BERNARDO.CUENCA.GRAU@CS.OX.AC.UK
**Ian Horrocks**                                  IAN.HORROCKS@CS.OX.AC.UK
**Markus Krötzsch**                        MARKUS.KROETZSCH@CS.OX.AC.UK
**Clemens Kupke**                              CLEMENS.KUPKE@CS.OX.AC.UK
**Despoina Magka**                            DESPOINA.MAGKA@CS.OX.AC.UK
**Boris Motik**                                    BORIS.MOTIK@CS.OX.AC.UK
**Zhe Wang**                                          ZHE.WANG@CS.OX.AC.UK
*Department of Computer Science, University of Oxford*
*Parks Road, Oxford OX1 3QD, United Kingdom*

## Abstract

Answering conjunctive queries (CQs) over a set of facts extended with existential rules is a prominent problem in knowledge representation and databases. This problem can be solved using the *chase* algorithm, which extends the given set of facts with fresh facts in order to satisfy the rules. If the chase terminates, then CQs can be evaluated directly in the resulting set of facts. The chase, however, does not terminate necessarily, and checking whether the chase terminates on a given set of rules and facts is undecidable. Numerous *acyclicity notions* were proposed as sufficient conditions for chase termination. In this paper, we present two new acyclicity notions called *model-faithful acyclicity* (MFA) and *model-summarising acyclicity* (MSA). Furthermore, we investigate the landscape of the known acyclicity notions and establish a complete taxonomy of all notions known to us. Finally, we show that MFA and MSA generalise most of these notions.

Existential rules are closely related to the Horn fragments of the OWL 2 ontology language; furthermore, several prominent OWL 2 reasoners implement CQ answering by using the chase to *materialise* all relevant facts. In order to avoid termination problems, many of these systems handle only the OWL 2 RL profile of OWL 2; furthermore, some systems go beyond OWL 2 RL, but without any termination guarantees. In this paper we also investigate whether various acyclicity notions can provide a principled and practical solution to these problems. On the theoretical side, we show that query answering for acyclic ontologies is of lower complexity than for general ontologies. On the practical side, we show that many of the commonly used OWL 2 ontologies are MSA, and that the number of facts obtained by materialisation is not too large. Our results thus suggest that principled development of materialisation-based OWL 2 reasoners is practically feasible.

## 1. Introduction

Existential rules are first-order implications between conjunctions of function-free atoms that may contain existentially quantified variables in the implication's consequent (Baget, Leclère, Mugnier, & Salvat, 2011a; Calì, Gottlob, Lukasiewicz, Marnette, & Pieris, 2010a). Such rules are used in a variety of ways in databases, knowledge representation, and logic programming. In database theory, existential rules are known as *tuple-generating dependencies* (Abiteboul, Hull, & Vianu, 1995) and are used to capture a wide range of schema constraints. Furthermore, they are also used as declarative data transformation rules in *data exchange*—the process of transforming a database structured according to a source schema into a database structured according to a target schema (Fagin, Kolaitis, Miller, & Popa, 2005). Existential rules also provide the foundation for several prominent knowledge representation formalisms, such as Datalog$^\pm$ (Calì, Gottlob, & Pieris, 2010b; Calì et al., 2010a), and they are also closely related to logic programs with function symbols in the head. Practical applications of existential rules range from bioinformatics (Mungall, 2009) to modelling complex structures of chemical compounds (Magka, Motik, & Horrocks, 2012; Hastings, Magka, Batchelor, Duan, Stevens, Ennis, & Steinbeck, 2012).

Answering *conjunctive queries* (CQs) over a set of facts extended with existential rules is a fundamental, yet undecidable (Beeri & Vardi, 1981) reasoning problem for existential rules. The problem can be characterised using *chase* (Johnson & Klug, 1984; Maier, Mendelzon, & Sagiv, 1979)—a technique closely related to the hypertableau calculus (Motik, Shearer, & Horrocks, 2009b; Baumgartner, Furbach, & Niemelä, 1996). In a forward-chaining manner, the chase extends the original set of facts with facts that can be derived using the rules. The result of the chase is a *universal* model, in the sense that an arbitrary CQ over the original facts and rules can be answered by evaluating the query in this model.

## 1.1 Chase Termination and Acyclicity Notions

Rules with existentially quantified variables in the head—so-called *generating rules*—require the introduction of fresh individuals. Cyclic applications of generating rules may prevent the chase from terminating, and in fact determining whether chase terminates on a set of rules and facts is undecidable (Deutsch, Nash, & Remmel, 2008). However, several decidable classes of existential rules have been identified, and the existing proposals can be classified into two main groups. In the first group, rules are restricted such that their possibly infinite universal models can be represented using finitary means. This group includes rules with universal models of bounded treewidth (Baget et al., 2011a), guarded rules (Calì et al., 2010a), and 'sticky' rules (Calì, Gottlob, & Pieris, 2011). In the second group, one uses a sufficient (but not necessary) *acyclicity* notion that ensures chase termination.

Roughly speaking, acyclicity notions analyse the information flow between rules to ensure that no cyclic applications of generating rules are possible. *Weak acyclicity* (WA) (Fagin et al., 2005) was one of the first such notions, and it was extended to notions such as *safety* (Meier, Schmidt, & Lausen, 2009), *stratification* (Deutsch et al., 2008), *acyclicity of a graph of rule dependencies* (aGRD) (Baget, Mugnier, & Thomazo, 2011b), *joint acyclicity* (JA) (Krötzsch & Rudolph, 2011), and *super-weak acyclicity* (SWA) (Marnette, 2009). Syntactic acyclicity criteria have also been investigated in the context of logic programs with function symbols in the rule heads, where the goal is to recognise logic programs with finite stable models. Several such notions have been implemented in state of the art logic programming engines, such as omega-restrictedness (Syrjänen, 2001) from the SMODELS system (Syrjänen & Niemelä, 2001), lambda-restrictedness from the ASP grounder GRINGO (Gebser, Schaub, & Thiele, 2007), argument-restrictedness (Lierler & Lifschitz, 2009) from the DLV system (Leone, Pfeifer, Faber, Eiter, Gottlob, Perri, & Scarcello, 2006), and many others (Calimeri, Cozza, Ianni, & Leone, 2008; Greco, Spezzano, & Trubitsyna, 2012; De Schreye & Decorte, 1994).

## 1.2 Applications of Acyclicity Notions

Acyclicity notions are interesting for several reasons. First, unlike guarded rules, acyclic rules can axiomatise structures of arbitrary shapes, as long as these structures are bounded in size. Second, the result of the chase for acyclic rules can be stored and manipulated as if it were a database; this is important, for example, in data exchange, where the goal is to materialise the transformed database.

In this paper, we further argue that acyclicity notions are also relevant to *description logics* (DLs)—knowledge representation formalisms underpinning the OWL 2 ontology language (Cuenca Grau, Horrocks, Motik, Parsia, Patel-Schneider, & Sattler, 2008). CQ answering over DL ontologies is a key reasoning service in many DL applications, and the problem was studied for numerous different DLs (Calvanese, De Giacomo, Lembo, Lenzerini, & Rosati, 2007; Krötzsch, Rudolph, & Hitzler, 2007; Glimm, Horrocks, Lutz, & Sattler, 2008; Ortiz, Calvanese, & Eiter, 2008; Lutz, Toman, & Wolter, 2009; Pérez-Urbina, Motik, & Horrocks, 2009; Rudolph & Glimm, 2010; Kontchakov, Lutz, Toman, Wolter, & Zakharyaschev, 2011). Answering CQs over ontologies, however, is quite technical and often of high computational complexity. Therefore, practical OWL 2 reasoners frequently solve this problem using *materialisation*—a reasoning technique in which the relevant consequences of the ontology are precomputed using chase, allowing queries to be directly evaluated in the materialised set of facts. Examples of materialisation-based systems include Oracle's Semantic Data Store (Wu, Eadon, Das, Chong, Kolovski, Annamalai, & Srinivasan, 2008), Sesame (Broekstra, Kampman, & van Harmelen, 2002), OWLIM (Kiryakov, Ognyanov, & Manov, 2005), Jena (Carroll, Dickinson, Dollin, Reynolds, Seaborne, & Wilkinson, 2004), and DLE-Jena (Meditskos & Bassiliades, 2008). Such reasoning is possible if (i) the ontology is *Horn* (Hustadt, Motik, & Sattler, 2005) and thus does not require disjunctive reasoning, and (ii) the chase is guaranteed to terminate. To satisfy the second assumption, reasoners often consider only axioms in the OWL 2 RL profile (Motik, Cuenca Grau, Horrocks, Wu, Fokoue, & Lutz, 2009a); this systematically excludes generating rules and thus trivially ensures chase termination, but it also makes the approach incomplete. Generating rules are partially supported in systems such as OWLim (Bishop & Bojanov, 2011) and Jena, but such support is typically ad hoc and provides no completeness and/or termination guarantees. Acyclicity notions can be used to address these issues: if an ontology is Horn and acyclic, a complete materialisation can be computed without the risk of non-termination.

## 1.3 Our Contributions

Motivated by the practical importance of chase termination, in this paper we present two new acyclicity notions: *model-faithful acyclicity* (MFA) and *model-summarising acyclicity* (MSA). Roughly speaking, these acyclicity notions use a particular model of the rules to analyse the implications between existential quantifiers, which is why we call them *model based*. In particular, MFA uses the actual 'canonical' model induced by the facts and the rules, which makes the notion very general. We prove that checking whether a set of existential rules is MFA is 2ExpTime-complete, and it becomes ExpTime-complete if the predicates in the rules are of bounded arity. Due to the high complexity, MFA may be unsuitable for practical application. Thus, we introduce MSA, which can be understood as MFA in which the analysis is performed over models that 'summarise' (or overestimate) the actual models. Checking MSA of existential rules can be realised via checking entailment of ground atoms in datalog programs. We use this close connection between MSA and datalog to prove that checking MSA is EXPTIME-complete for general existential rules, and that it becomes coNP-complete if the arity of rule predicates is bounded.

We next conduct a detailed investigation of the landscape of known acyclicity notions, augmented with MFA and MSA. For the class of logic programs that correspond to existential rules with skolemised existential quantifiers, we show that MSA and MFA strictly subsume existing acyclicity notions known from logic programming. We also show that MSA is strictly more general than SWA—one of the most general acyclicity notions known in database theory. Furthermore, we investigate the relationship between the known notions and thus complete the picture with respect to their relative expressiveness.

Both MSA and MFA can be applied to general existential rules *without* equality. Equality can be incorporated via *singularisation*—a technique proposed by Marnette (2009) that transforms the rules to encode the effects of equality. Singularisation is orthogonal to acyclicity: after computing the transformed rules, one can use MFA, MSA, or in fact any notion to check whether the result is acyclic; if so, the chase of the singularised rules terminates, and the chase result can be used in a particular way to answer arbitrary CQs. Unfortunately, singularisation is nondeterministic: some ways of transforming the rules may produce acyclic rule sets, but not all ways are guaranteed to do so. In this paper, we refine singularisation to obtain practically useful upper and lower bounds for acyclicity. We also show that, when used with JA, our lower bound actually coincides with WA.

We next turn our attention to theoretical and practical issues of using acyclicity for materialisation-based CQ answering over ontologies. On the theoretical side, we show that checking MFA and MSA of Horn-$\mathcal{SROIF}$ ontologies is EXPTIME- and PTIME-complete, respectively, and that answering CQs over acyclic Horn-$\mathcal{SROIF}$ ontologies is EXPTIME-complete as well. Furthermore, we show that, for Horn-$\mathcal{SHIF}$ ontologies, the complexity of checking MFA and of answering CQs drops to PSPACE. Answering CQs is EXPTIME-complete for general (i.e., not acyclic) Horn-$\mathcal{SHIF}$ ontologies (Eiter, Gottlob, Ortiz, & Simkus, 2008; Ortiz, Rudolph, & Simkus, 2011), so acyclicity makes this problem easier. Furthermore, Horn ontologies can be extended with arbitrary SWRL rules (Horrocks & Patel-Schneider, 2004) without affecting decidability or worst-case complexity, provided that the union of the ontology and SWRL rules is acyclic; this is in contrast to the general case, where SWRL extensions of DLs easily lead to undecidability.

On the practical side, we explore the limits of reasoning with acyclic OWL 2 ontologies via materialisation. We checked MFA, MSA, and JA for 336 Horn ontologies; furthermore, to estimate the impact of materialisation, we compared the size of the materialisation with the number of facts in the original ontologies. Our experiments revealed that many ontologies are MSA, and that some complex ones are MSA but not JA; furthermore, the universal models obtained via materialisation are typically not too large. Thus, our results suggest that principled, materialisation-based reasoning for ontologies beyond the OWL 2 RL profile may be practically feasible.

This is an extended version of a paper by Cuenca Grau, Horrocks, Krötzsch, Kupke, Magka, Motik, and Wang (2012) published at KR 2012.

## 2. Preliminaries

In this section we introduce definitions and notation used in the rest of our paper.

### 2.1 First-Order Logic

We use the standard notions of constants, function symbols, and predicate symbols, where $\approx$ is the equality predicate, $\top$ is universal truth, and $\bot$ is universal falsehood. Each function or predicate symbol is associated with a nonnegative integer arity. Variables, terms, substitutions, atoms, first-order formulae, sentences, interpretations (i.e., structures), and models are defined as usual. By a slight abuse of notation, we often identify a conjunction with the set of its conjuncts. Furthermore, we often abbreviate a vector of terms $t_1, \ldots, t_n$ as $\vec{t}$; we define $|\vec{t}| = n$; and we often identify $\vec{t}$ with the set of indexed terms $\{t_1, \ldots, t_n\}$. With $\varphi(\vec{x})$ we stress that $\vec{x} = x_1, \ldots, x_n$ are the free variables of a formula $\varphi$, and with $\varphi\sigma$ we denote the result of applying a substitution $\sigma$ to $\varphi$. A term, atom, or formula is *ground* if it does not contain variables; a *fact* is a ground atom. The *depth* $\mathsf{dep}(t)$ of a term $t$ is defined as 0 if $t$ is a constant or a variable, and $\mathsf{dep}(t) = 1 + \max_{i=1}^{n} \mathsf{dep}(t_i)$ if $t = f(t_1, \ldots, t_n)$. A term $t'$ is a *subterm* of a term $t$ if $t' = t$ or $t = f(\vec{s})$ and $t'$ is a subterm of some $s_i \in \vec{s}$; if additionally $t' \neq t$, then $t'$ is a *proper subterm* of $t$. A term $s$ is *contained* in an atom $P(\vec{t})$ if $s \in \vec{t}$, and $s$ *occurs* in $P(\vec{t})$ if $s$ is a subterm of some term $t_i \in \vec{t}$; thus, if $s$ is contained in $P(\vec{t})$, then $s$ also occurs in $P(\vec{t})$, but the converse may not hold. A term $s$ is *contained* (resp. *occurs*) in a set of atoms $I$ if $s$ is *contained* (resp. *occurs*) in some atom in $I$.

In first-order logic, the equality predicate $\approx$ is commonly assumed to have a predefined interpretation—that is, every first-order interpretation is required to interpret $\approx$ as the smallest reflexive relation over the domain. Satisfaction of a sentence $\varphi$ in an interpretation $I$ where $\approx$ is interpreted in this way is written $I \models \varphi$, and entailment of a sentence $\psi$ from a sentence $\varphi$ is written $\varphi \models \psi$. Unless otherwise stated, we use this standard interpretation of equality throughout this paper.

Equality, however, can also be treated as an ordinary predicate with an explicit axiomatisation. Let $\Sigma$ be an arbitrary set of function-free first-order formulae. Then, $\Sigma_{\approx} = \emptyset$ if $\approx$ does not occur in $\Sigma$; otherwise, $\Sigma_{\approx}$ contains formula (1)–(3) and an instance of formula (4) for each $n$-ary predicate $P$ occurring in $\Sigma$ different from $\approx$, and for each $1 \leq i \leq n$. Note that all variables in all of these formulae are (implicitly) universally quantified.

$$\to x \approx x \tag{1}$$

$$x_1 \approx x_2 \to x_2 \approx x_1 \tag{2}$$

$$x_1 \approx x_2 \wedge x_2 \approx x_3 \to x_1 \approx x_3 \tag{3}$$

$$P(x_1, \ldots, x_i, \ldots, x_n) \wedge x_i \approx x_i' \to P(x_1, \ldots, x_i', \ldots, x_n) \tag{4}$$

If $\approx$ is treated as an ordinary predicate, satisfaction of a formula $\varphi$ in a model $I$ is written $I \models_{\approx} \varphi$, and entailment of a formula $\psi$ from formula $\varphi$ is written $\varphi \models_{\approx} \psi$. Please note that, according to our definitions, $I \models_{\approx} \varphi$ can hold even if interpretation $I$ interprets predicate $\approx$ in an arbitrary way; in contrast, $I \models \varphi$ can hold only if interpretation $I$ interprets predicate $\approx$ as the identity relation on the model's domain. The consequences of $\Sigma$ w.r.t. $\models$ and of $\Sigma \cup \Sigma_{\approx}$ w.r.t. $\models_{\approx}$ coincide—that is, for each first-order sentence $\psi$ constructed using the symbols from $\Sigma$, we have $\Sigma \models \psi$ if and only if $\Sigma \cup \Sigma_{\approx} \models_{\approx} \psi$.

## 2.2 Rules and Queries

An *instance* is a finite set of function-free facts. An *existential rule* (or just *rule*) is a function-free sentence of the form

$$\forall \vec{x} \forall \vec{z}. [\varphi(\vec{x}, \vec{z}) \rightarrow \exists \vec{y}. \psi(\vec{x}, \vec{y})] \tag{5}$$

where $\varphi(\vec{x}, \vec{z})$ and $\psi(\vec{x}, \vec{y})$ are conjunctions of atoms, and tuples of variables $\vec{x}$, $\vec{y}$, and $\vec{z}$ are pairwise disjoint. Formula $\varphi$ is the *body* and formula $\psi$ is the *head* of the rule. For brevity, quantifiers $\forall \vec{x} \forall \vec{z}$ are often omitted. For convenience, we sometimes identify a rule body or head with the set of the respective conjuncts. A *datalog rule* is a rule where $\vec{y}$ is empty. A rule is *equality-free* if it does not contain the equality predicate $\approx$. A term $s$ *occurs* in an existential rule if $s$ occurs in a head or body atom of the rule, and these definitions are extended to a set of rules in the obvious way; existential rules do not contain function symbols, so an analogous notion of $s$ being *contained* in a rule coincides with this one. Two variables are *directly connected* in a rule if they occur together in a body atom of the rule; furthermore, *connected* is the transitive closure of *directly connected*; finally, a rule of the form (5) is *connected* if all pairs of variables $w, w' \in \vec{x} \cup \vec{z}$ are connected in the rule.

A *conjunctive query* (CQ) is a formula of the form $Q(\vec{x}) = \exists \vec{y}. \varphi(\vec{x}, \vec{y})$, where $\varphi(\vec{x}, \vec{y})$ is a conjunction of atoms; the query is *Boolean* if $\vec{x}$ is empty. A substitution $\theta$ mapping $\vec{x}$ to constants is an *answer* to $Q(\vec{x})$ w.r.t. a set of rules $\Sigma$ and instance $I$ if $\Sigma \cup I \models Q(\vec{x})\theta$. Answering CQs is the core reasoning problem in many applications of existential rules.

When answering a conjunctive query $Q(\vec{x})$ over a set of rules $\Sigma$ and an instance $I$, in the rest of this paper we implicitly assume that $Q(\vec{x})$ and $I$ contain only the predicates from $\Sigma$. This simplifies the presentation since it allows us to define various transformations of $\Sigma$ without having to take into account possible predicates that occur in $Q(\vec{x})$ or $I$ only. This assumption is w.l.o.g., as we can always extend $\Sigma$ with tautological rules of the form $P(\vec{x}) \rightarrow P(\vec{x})$ for each predicate $P$ occurring in $Q(\vec{x})$ or $I$ but not in $\Sigma$.

Furthermore, we assume that $\approx$ does not occur in the body of any rule in $\Sigma$ or in the query $Q(\vec{x})$. This is w.l.o.g. since we can eliminate each atom of the form $x \approx t$ in a rule body and further replace $x$ with $t$ in the rest of the rule; furthermore, to eliminate body atoms of the form $a \approx b$ with $a$ and $b$ constants, we can introduce a fresh predicate $O_a$, add a new rule $\rightarrow O_a(a)$, replace each body atom $a \approx b$ with conjunction $O_a(x) \wedge x \approx b$ in which $x$ is a fresh variable, and finally eliminate atom $x \approx b$ as before. Similarly, we do not provide an explicit support for the inequality predicate $\not\approx$. Inequality in rule heads can be simulated using an ordinary predicate: each atom of the form $s \not\approx t$ occurring in a rule head can be replaced with $\mathsf{NotEqual}(s, t)$, where $\mathsf{NotEqual}$ is a fresh ordinary predicate that is explicitly axiomatised as irreflexive; note that, if $\approx$ is handled as a regular predicate explicitly axiomatised by rules (1)–(4), then the replacement axioms (4) must be instantiated for $P = \mathsf{NotEqual}$ as well. In contrast, atoms involving the inequality predicate occurring in rule bodies generally require disjunctive reasoning, which is not supported by existential rules.

Finally, we assume that conjunctions $\varphi(\vec{x}, \vec{z})$ and $\psi(\vec{x}, \vec{y})$ in each rule of the form (5) are both not empty. We also assume that $\top$ and $\bot$ are treated as ordinary unary predicates, and that the semantics of $\top$ is captured explicitly in $\Sigma$ by instantiating the following rule for each $n$-ary predicate $P$ occurring in $\Sigma$:

$$P(x_1, \ldots, x_n) \rightarrow \top(x_1) \wedge \ldots \wedge \top(x_n) \tag{6}$$

These assumptions ensure that $I \cup \Sigma$ is always satisfiable, but that $\Sigma \cup I \models \exists y. \bot(y)$ if and only if $I \cup \Sigma$ is unsatisfiable w.r.t. the conventional treatment of $\top$ and $\bot$. By allowing body atoms of the form $\top(x)$, without loss of generality we can require each existential rule to be *safe* (i.e., that each universally quantified variable occurring in a head atom also occurs in a body atom of the rule), which greatly simplifies many of our definitions.

In database theory, satisfaction and entailment are often considered only w.r.t. *finite* interpretations under the *unique name assumption* (UNA); the latter ensures that distinct constants are interpreted as distinct elements. In contrast, such assumptions are not customary in ontology-based KR. In this paper, we do not assume UNA, as UNA can be axiomatised explicitly if needed using the inequality predicate (or a simulation thereof). Furthermore, in this paper we investigate theories that are satisfiable in finite models (i.e., for which the chase is finite); thus, the difference between finite and infinite satisfiability is immaterial to our results.

We frequently use *skolemisation* to interpret rules in *Herbrand* interpretations, which are defined as possibly infinite sets of ground atoms. In particular, for each rule $r$ of the form (5) and each variable $y_i \in \vec{y}$, let $f_r^i$ be a function symbol globally unique for $r$ and $y_i$ of arity $|\vec{x}|$; furthermore, let $\theta_{\mathsf{sk}}$ be the substitution such that $\theta_{\mathsf{sk}}(y_i) = f_r^i(\vec{x})$ for each $y_i \in \vec{y}$. Then, the skolemisation $\mathsf{sk}(r)$ of $r$ is the following rule:

$$\varphi(\vec{x}, \vec{z}) \to \psi(\vec{x}, \vec{y})\theta_{\mathsf{sk}} \tag{7}$$

The skolemisation $\mathsf{sk}(\Sigma)$ of a set of rules $\Sigma$ is obtained by skolemising each rule in $\Sigma$. Skolemisation does not affect the answers to CQs—that is, for each conjunctive query $Q(\vec{x})$ formed from only the predicates in $\Sigma$, each instance $I$, and each substitution $\sigma$, we have $\Sigma \cup I \models Q(\vec{x})\sigma$ if and only if $\mathsf{sk}(\Sigma) \cup \Sigma_{\approx} \cup I \models_{\approx} Q(\vec{x})\sigma$.

## 2.3 The Skolem Chase

Answering CQs can be characterised using *chase*, and in this paper we use the *skolem chase* variant (Marnette, 2009). Let $r = \varphi \to \psi$ be a skolemised rule and let $I$ be a set of ground atoms. A set of ground atoms $S$ is a *consequence* of $r$ on $I$ if substitution $\sigma$ exists mapping the variables in $r$ to the terms occurring in $I$ such that $\varphi\sigma \subseteq I$ and $S \subseteq \psi\sigma$. The result of *applying* $r$ to $I$, written $r(I)$, is the union of all consequences of $r$ on $I$. For $\Omega$ a set of skolemised rules, $\Omega(I) = \bigcup_{r \in \Omega} r(I)$. Let $I$ be a finite set of ground atoms, let $\Sigma$ be a set of rules, let $\Sigma' = \mathsf{sk}(\Sigma) \cup \Sigma_{\approx}$, and let $\Sigma'_f$ and $\Sigma'_n$ be the subsets of $\Sigma'$ containing rules with and without function symbols, respectively. The *chase sequence* for $I$ and $\Sigma$ is a sequence of sets of facts $I_{\Sigma}^0, I_{\Sigma}^1, \ldots$ where $I_{\Sigma}^0 = I$ and, for each $i > 0$, set $I_{\Sigma}^i$ is defined as follows:

- if $\Sigma'_n(I_{\Sigma}^{i-1}) \not\subseteq I_{\Sigma}^{i-1}$, then $I_{\Sigma}^i = I_{\Sigma}^{i-1} \cup \Sigma'_n(I_{\Sigma}^{i-1})$,

- otherwise $I_{\Sigma}^i = I_{\Sigma}^{i-1} \cup \Sigma'_f(I_{\Sigma}^{i-1})$.

The *chase* of $I$ and $\Sigma$ is defined as $I_{\Sigma}^{\infty} = \bigcup_i I_{\Sigma}^i$; note that $I_{\Sigma}^{\infty}$ can be infinite. The chase can be used as a 'database' for answering CQs: a substitution $\sigma$ is an answer to $Q$ over $\Sigma$ and $I$ if and only if $I_{\Sigma}^{\infty} \models_{\approx} Q\sigma$. The chase of $I$ and $\Sigma$ *terminates* if $i \geq 0$ exists such that $I_{\Sigma}^i = I_{\Sigma}^j$ for each $j \geq i$; the chase of $\Sigma$ terminates *universally* if the chase of $I$ and $\Sigma$ terminates for each $I$. If the skolem chase of $I$ and $\Sigma$ terminates, then both the *nonoblivious chase* (Fagin et al., 2005) and the *core chase* (Deutsch et al., 2008) of $I$ and $\Sigma$ terminate as well.

The *critical instance* $I_\Sigma^*$ for a set of rules $\Sigma$ contains all facts that can be constructed using all predicates occurring in $\Sigma$, all constants occurring in the body of a rule in $\Sigma$, and one special fresh constant $*$. The skolem chase for $I_\Sigma^*$ and $\Sigma$ terminates if and only if the skolem chase of $\Sigma$ terminates universally (Marnette, 2009).

### 2.4 Acyclicity Notions

Checking whether the skolem chase terminates on a given instance is undecidable, and checking universal skolem chase termination is conjectured to be undecidable as well. Consequently, various sufficient *acyclicity notions* have been proposed in the literature. Formally, an acyclicity notion $X$ is a class of finite sets of rules; such a definition allows us to talk about (proper) containment between acyclicity notions. We sometimes write '$\Sigma$ is $X$', by which we mean '$\Sigma \in X$'. We next introduce weak and joint acyclicity: the former is one of the first such notions considered in the literature; and as we show in Section 3, the latter notion is relatively powerful, yet still easy to understand. We use these two notions throughout the paper to present examples and state various technical claims. In Section 3 we present the definitions of many other acyclicity notions known in the literature.

In the following, let $\Sigma$ be a set of rules where no variable occurs in more than one rule. A *position* is an expression of the form $P|_i$ where $P$ is an $n$-ary predicate and $i$ is an integer with $1 \leq i \leq n$. Given a rule $r$ of the form (5) and a variable $w$ occurring in $r$, the set $\mathsf{Pos}_B(w)$ of *body positions* of $w$ contains each position $P|_i$ such that $P(t_1, \ldots, t_n) \in \varphi(\vec{x}, \vec{z})$ and $t_i = w$ for some vector $\vec{t}$ of terms. The set $\mathsf{Pos}_H(w)$ of *head positions* is defined analogously, but w.r.t. the head atoms of $r$. Note that, since each variable occurs in at most one rule in $\Sigma$, sets $\mathsf{Pos}_B(w)$ and $\mathsf{Pos}_H(w)$ are (indirectly) associated with the rule that contains $w$. In the rest of this paper, whenever we use notation such as $\mathsf{Pos}_H(w)$ or $\mathsf{Pos}_B(w)$, we silently assume that no variable occurs in more than one rule and so the notation is unambiguous. This is clearly w.l.o.g. as one can always arbitrarily rename variables in different rules.

*Weak acyclicity* (WA) (Fagin et al., 2005) can be applied to existential rules that contain the equality predicate. The *WA dependency graph* $WA(\Sigma)$ for $\Sigma$ contains positions as vertices; furthermore, for each rule $r \in \Sigma$ of the form (5), each variable $x \in \vec{x}$, each position $P|_i \in \mathsf{Pos}_B(x)$, and each variable $y \in \vec{y}$, graph $WA(\Sigma)$ contains

- a *regular edge* from $P|_i$ to each $Q|_j \in \mathsf{Pos}_H(x)$ such that $Q \neq \approx$ and,

- a *special edge* from $P|_i$ to each $Q|_j \in \mathsf{Pos}_H(y)$ such that $Q \neq \approx$.

Set $\Sigma$ is WA if $WA(\Sigma)$ does not contain a cycle that involves a special edge. Equality atoms are effectively ignored by WA.

*Joint acyclicity* (JA) (Krötzsch & Rudolph, 2011) generalises WA, but it is applicable only to equality-free rules. For an existentially quantified variable $y$ in $\Sigma$, let $\mathsf{Move}(y)$ be the smallest set of positions such that

- $\mathsf{Pos}_H(y) \subseteq \mathsf{Move}(y)$, and

- for each existential rule $r \in \Sigma$ and each universally quantified variable $x$ occurring in $r$, if $\mathsf{Pos}_B(x) \subseteq \mathsf{Move}(y)$, then $\mathsf{Pos}_H(x) \subseteq \mathsf{Move}(y)$.

The *JA dependency graph* $JA(\Sigma)$ of $\Sigma$ is defined as follows. The vertices of $JA(\Sigma)$ are the existentially quantified variables occurring in $\Sigma$. Given arbitrary two such variables $y_1$ and $y_2$, the JA dependency graph $JA(\Sigma)$ contains an edge from $y_1$ to $y_2$ whenever the rule that contains $y_2$ also contains a universally quantified variable $x$ such that $\mathsf{Pos}_H(x) \neq \emptyset$ and $\mathsf{Pos}_B(x) \subseteq \mathsf{Move}(y_1)$. Set $\Sigma$ is JA if $JA(\Sigma)$ does not contain a cycle.

## 2.5 Rule Normalisation

Existential rules can often be transformed into other existential rules by replacing parts of the rule head or body with atoms involving fresh predicates. Such a transformation is called *normalisation*, and is often used as a preprocessing step to bring the rules into a suitable form. For example, Horn OWL 2 axioms can be translated into existential rules by using the well known transformations of first-order logic, and the latter can then be normalised to a form we describe in Section 6. In this section we introduce a definition of rule normalisation that captures all similar methods known to us.

Let $r$ be a rule of the form (8), where $\varphi_1$, $\varphi_2$, $\psi_1$, and $\psi_2$ are conjunctions of atoms satisfying $\vec{x}_1 \cup \vec{x}_2 = \vec{x}_3 \cup \vec{x}_4$, $\vec{z}_2 \cap \vec{z}_3 = \emptyset$, and $\vec{y}_2 \cap \vec{y}_3 = \emptyset$.

$$\varphi_1(\vec{x}_1, \vec{z}_1, \vec{z}_2) \wedge \varphi_2(\vec{x}_2, \vec{z}_1, \vec{z}_3) \rightarrow \exists \vec{y}_1, \vec{y}_2, \vec{y}_3.[\psi_1(\vec{x}_3, \vec{y}_1, \vec{y}_2) \wedge \psi_2(\vec{x}_4, \vec{y}_1, \vec{y}_3)] \qquad (8)$$

A *normalisation step* replaces a conjunction in either the head or the body of the rule with an atom involving a fresh predicate. More precisely, a *head normalisation step* replaces $\psi_1(\vec{x}_3, \vec{y}_1, \vec{y}_2)$ with atom $Q(\vec{x}_3, \vec{y}_1)$ where $Q$ is a fresh predicate, thus replacing $r$ with rule (9), and it adds rule (10).

$$\varphi_1(\vec{x}_1, \vec{z}_1, \vec{z}_2) \wedge \varphi_2(\vec{x}_2, \vec{z}_1, \vec{z}_3) \rightarrow \exists \vec{y}_1, \vec{y}_3.[Q(\vec{x}_3, \vec{y}_1) \wedge \psi_2(\vec{x}_4, \vec{y}_1, \vec{y}_3)] \qquad (9)$$

$$Q(\vec{x}_3, \vec{y}_1) \rightarrow \exists \vec{y}_2.\psi_1(\vec{x}_3, \vec{y}_1, \vec{y}_2) \qquad (10)$$

Alternatively, a *body normalisation step* replaces $\varphi_1(\vec{x}_1, \vec{z}_1, \vec{z}_2)$ with atom $Q(\vec{x}_1, \vec{z}_1)$ where $Q$ is a fresh predicate, thus replacing $r$ with rule (11), and it adds rule (12).

$$Q(\vec{x}_1, \vec{z}_1) \wedge \varphi_2(\vec{x}_2, \vec{z}_1, \vec{z}_3) \rightarrow \exists \vec{y}_1, \vec{y}_2, \vec{y}_3.[\psi_1(\vec{x}_3, \vec{y}_1, \vec{y}_2) \wedge \psi_2(\vec{x}_4, \vec{y}_1, \vec{y}_3)] \qquad (11)$$

$$\varphi_1(\vec{x}_1, \vec{z}_1, \vec{z}_2) \rightarrow Q(\vec{x}_1, \vec{z}_1) \qquad (12)$$

Given a set of existential rules $\Sigma$, normalisation steps are often applied to $\Sigma$ iteratively. If the predicate $Q$ introduced in each step is always fresh, we call such normalisation *without structure sharing*. In contrast, normalisation *with structure sharing* allows the predicate $Q$ to be reused across different normalisation steps. For example, once a predicate $Q$ is introduced in a head normalisation step to replace $\varphi_1(\vec{x}_1, \vec{z}_1, \vec{z}_2)$, then a conjunction of the form $\varphi_1(\vec{x}'_1, \vec{z}'_1, \vec{z}'_2)$ where $\vec{x}'_1$, $\vec{z}'_1$, $\vec{z}'_2$ are renamings of $\vec{x}_1$, $\vec{z}_1$, $\vec{z}_2$ can be replaced with $Q(\vec{x}'_3, \vec{y}'_1)$ without introducing the corresponding rule (10). An analogous optimisation can be used in a body normalisation step.

Let $\Sigma'$ be a set of rules obtained via normalisation (with or without structure sharing) from $\Sigma$. It is well known that $\Sigma'$ is a conservative extension of $\Sigma$. Consequently, for each instance $I$ and each BCQ $Q$ that does not use the freshly introduced predicates, we have $\Sigma \cup I \models Q$ if and only if $\Sigma' \cup I \models Q$.

## 3. Novel Acyclicity Notions

Weak acyclicity has considerably influenced the field of data exchange in databases, but it is a rather strict notion and so it may not be sufficient in many applications of existential rules. Joint acyclicity significantly relaxes weak acyclicity and was developed mainly for rule based knowledge representation applications.

In Section 3.1 we show that even joint acyclicity—one of the most general acyclicity notions developed so far—does not capture rules corresponding to axioms commonly found in ontologies for which the chase terminates universally. To address this important limitation, we propose in Section 3.2 *model-faithful acyclicity* (MFA)—a novel, very general, notion that can be used to successfully ensure chase termination for many ontologies used in practice. The computational cost of checking MFA is, however, rather high; hence, in Section 3.3 we introduce *model-summarising acyclicity* (MSA)—a more strict notion that is easier to check and produces the same results as MFA on most existing ontologies.

### 3.1 Limitations of Existing Acyclicity Notions

To motivate our new acyclicity notions, we first present an example that shows how known acyclicity notions, such as JA, are not satisfied by rules that are equivalent to very simple axioms that abound in OWL ontologies.

**Example 1.** *Let $\Sigma$ be the set of rules* (13)–(17).

$$r_1 = \qquad\qquad\qquad A(x_1) \rightarrow \exists y_1.R(x_1, y_1) \wedge B(y_1) \qquad\qquad (13)$$

$$r_2 = \qquad\qquad R(x_2, z_1) \wedge B(z_1) \rightarrow A(x_2) \qquad\qquad\qquad (14)$$

$$r_3 = \qquad\qquad\qquad B(x_3) \rightarrow \exists y_2.R(x_3, y_2) \wedge C(y_2) \qquad\qquad (15)$$

$$r_4 = \qquad\qquad\qquad C(x_4) \rightarrow D(x_4) \qquad\qquad\qquad\qquad (16)$$

$$r_5 = \qquad\qquad R(x_5, z_2) \wedge D(z_2) \rightarrow B(x_5) \qquad\qquad\qquad (17)$$

*Rules $r_1$ and $r_2$ correspond to the description logic axiom $A \equiv \exists R.B$, rule $r_3$ corresponds to axiom $B \sqsubseteq \exists R.C$, rule $r_4$ corresponds to axiom $C \sqsubseteq D$, and rule $r_5$ corresponds to axiom $\exists R.D \sqsubseteq B$. Such axioms are very common in OWL ontologies.*

*By the definition of JA from Section 2, we have $\mathsf{Move}(y_1) = \{R|_2, B|_1, R|_1, A|_1\}$. Thus, the JA dependency graph contains an edge from $y_1$ to itself, so the set of axioms $\Sigma$ is not JA. In contrast, the following table shows the chase sequence for $I^*_\Sigma$ and $\Sigma$.*

| | | | |
|---|---|---|---|
| $A(*)$ | $R(*, f(*))$ | $R(f(*), g(f(*)))$ | $D(g(f(*)))$ |
| $B(*)$ | $B(f(*))$ | $C(g(f(*)))$ | |
| $C(*)$ | $R(*, g(*))$ | $D(g(*))$ | |
| $D(*)$ | $C(g(*))$ | | |
| $R(*, *)$ | | | |

*Rule $r_2$ is not applicable to $R(f(*), g(f(*)))$ since $I_3$ does not contain the fact $B(g(f(*)))$ necessary to match the atom $B(z_1)$ from the rule. Thus, the chase terminates.* $\diamond$

All existing acyclicity notions essentially try to estimate whether an application of a rule can produce facts that can (possibly by applying chase to other rules) repeatedly trigger the same rule in an infinite manner. The key difference between various notions is how rule applicability is determined. In particular, JA considers each variable in a rule in isolation and does not check satisfaction of all body atoms at once; for example, rule (14) is not applicable to the facts generated by rule (15), but this can be determined only by considering variables $x_2$ and $z_1$ in rule (14) simultaneously. These notions thus overestimate rule applicability and, as a result, they can fail to detect chase termination.

### 3.2 Model-Faithful Acyclicity (MFA)

Our main intuition for addressing this problem is that more precise chase termination guarantees can be obtained by tracking rule applicability more 'faithfully'. A simple solution is to be completely precise about rule applicability: one can run the skolem chase and then use sufficient checks to identify cyclic computations. Since no sufficient, necessary, and computable test can be given for the latter, we must adopt a practical approach. For example, we can 'raise the alarm' and stop the process if the chase derives a 'cyclic' term $f(\vec{t})$, where $f$ occurs in $\vec{t}$. This idea can be further refined; for example, one could stop only if $f$ occurs nested in a term some fixed number of times. The choice of the appropriate test thus depends on an application; however, as our experiments show, checking only for one level of nesting suffices in many cases. In particular, no term $f(\vec{t})$ with $f$ occurring in $\vec{t}$ is generated in the chase of the set of rules $\Sigma$ from Example 1.

**Definition 2.** *A term $t$ is* cyclic *if a function symbol $f$ exists such that some term $f(\vec{s})$ is a subterm of $t$, and some term $f(\vec{u})$ is a proper subterm of $f(\vec{s})$.*

Our notion of acyclicity is declarative: the given set of rules $\Sigma$ is transformed into a new set of rules $\Sigma'$ that tracks rule dependencies using fresh predicates; then, $\Sigma$ is identified as being acyclic if $\Sigma'$ does not entail a special nullary predicate $\mathsf{C}$. Since acyclicity is defined via entailment, it can be decided using any theorem proving procedure for existential rules that is sound and complete. Acyclicity guarantees termination of the skolem chase, which also guarantees termination of nonoblivious chase and core chase. We call our notion *model-faithful acyclicity* because it estimates rule application precisely, by examining the actual structure of the universal model of $\Sigma$.

**Definition 3.** *For each rule $r = \varphi(\vec{x}, \vec{z}) \to \exists \vec{y}.\psi(\vec{x}, \vec{y})$ and each variable $y_i \in \vec{y}$, let $\mathsf{F}_r^i$ be a fresh unary predicate unique for $r$ and $y_i$; furthermore, let $\mathsf{S}$ and $\mathsf{D}$ be fresh binary predicates, and let $\mathsf{C}$ be a fresh nullary predicate. Then, $\mathsf{MFA}(r)$ is the following rule:*

$$\varphi(\vec{x}, \vec{z}) \to \exists \vec{y}. \left[ \psi(\vec{x}, \vec{y}) \wedge \bigwedge_{y_i \in \vec{y}} \left[ \mathsf{F}_r^i(y_i) \wedge \bigwedge_{x_j \in \vec{x}} \mathsf{S}(x_j, y_i) \right] \right]$$

*For a set $\Sigma$ of rules, $\mathsf{MFA}(\Sigma)$ is the smallest set that contains $\mathsf{MFA}(r)$ for each rule $r \in \Sigma$, rules (18)–(19), and rule (20) instantiated for each $\mathsf{F}_r^i$ corresponding to some $r \in \Sigma$:*

$$\mathsf{S}(x_1, x_2) \to \mathsf{D}(x_1, x_2) \tag{18}$$

$$\mathsf{D}(x_1, x_2) \wedge \mathsf{S}(x_2, x_3) \to \mathsf{D}(x_1, x_3) \tag{19}$$

$$\mathsf{F}_r^i(x_1) \wedge \mathsf{D}(x_1, x_2) \wedge \mathsf{F}_r^i(x_2) \to \mathsf{C} \tag{20}$$

The set $\Sigma$ is model-faithful acyclic *(MFA) w.r.t. an instance $I$ if* $I \cup \mathsf{MFA}(\Sigma) \not\models \mathsf{C}$; *furthermore,* $\Sigma$ *is* universally MFA[1] *if* $\Sigma$ *is MFA w.r.t.* $I_\Sigma^*$.

**Example 4.** *Let $\Sigma$ be the set of rules from Example 1. Then, $\mathsf{MFA}(r_1)$ and $\mathsf{MFA}(r_3)$ are given by (21) and (22), respectively; since $r_1$ and $r_3$ contain a single existentially quantified variable each, we omit the superscripts in $\mathsf{F}_{r_1}$ and $\mathsf{F}_{r_3}$ for the sake of clarity. Thus, $\mathsf{MFA}(\Sigma)$ consists of rules (14), (16), and (17), rules (21)–(22), rules (18)–(19), and rule (20) instantiated for $\mathsf{F}_{r_1}$ and $\mathsf{F}_{r_3}$.*

$$A(x_1) \to \exists y_1.R(x_1, y_1) \wedge B(y_1) \wedge \mathsf{F}_{r_1}(y_1) \wedge \mathsf{S}(x_1, y_1) \tag{21}$$

$$B(x_3) \to \exists y_2.R(x_3, y_2) \wedge C(y_2) \wedge \mathsf{F}_{r_3}(y_2) \wedge \mathsf{S}(x_3, y_2) \tag{22}$$

*It is straightforward to see that the chase of $I_\Sigma^*$ and $\mathsf{MFA}(\Sigma)$ consists of the facts presented in Example 1, augmented with the following facts:*

| | | |
|---|---|---|
| $\mathsf{S}(*, f(*))$ | $\mathsf{D}(*, f(*))$ | $\mathsf{D}(f(*), g(f(*)))$ |
| $\mathsf{S}(*, g(*))$ | $\mathsf{D}(*, g(*))$ | $\mathsf{D}(*, g(f(*)))$ |
| $\mathsf{F}_{r_1}(f(*))$ | $\mathsf{S}(f(*), g(f(*)))$ | |
| $\mathsf{F}_{r_3}(g(*))$ | $\mathsf{F}_{r_3}(g(f(*)))$ | |

*The chase of $I_\Sigma^*$ and $\mathsf{MFA}(\Sigma)$ does not contain $\mathsf{C}$, which implies that $I \cup \mathsf{MFA}(\Sigma) \not\models \mathsf{C}$. As a result, $\Sigma$ is universally MFA.* ◇

MFA is formulated as a semantic, rather than a syntactic notion, and is thus mainly independent from algorithmic details: entailment $I \cup \mathsf{MFA}(\Sigma) \not\models \mathsf{C}$ can be checked using an arbitrary sound and complete first-order calculus. In Section 4 we discuss the relationship between MFA and existing notions, and we show that MFA generalises most of them.

The following proposition shows that MFA characterises the derivations of the skolem chase in which no cyclic terms occur.

**Proposition 5.** *A set $\Sigma$ of rules is not MFA w.r.t. an instance $I$ if and only if $I_{\mathsf{MFA}(\Sigma)}^\infty$ contains a cyclic term.*

*Proof.* Let $\Sigma' = \mathsf{MFA}(\Sigma)$, and let $I_{\Sigma'}^0, I_{\Sigma'}^1, \dots$ be the chase sequence for $I$ and $\Sigma'$. Moreover, let $f_r^i$ be the function symbol used to skolemise the $i$-th existentially quantified variable in rule $r$, as defined in Section 2.2. We next prove that the following claims hold for all terms $t$ and $t'$ occurring in $I_{\Sigma'}^k$, each rule $r$, each integer $i$, and each integer $k$, as well as $k = \infty$.

1. Term $t$ is of the form $f_r^i(\vec{u})$ if and only if $\mathsf{F}_r^i(t) \in I_{\Sigma'}^k$.

2. Term $t$ is of the form $f_r^i(\vec{u})$ and $t' \in \vec{u}$ if and only if $\mathsf{S}(t', t) \in I_{\Sigma'}^k$.

3. If $t'$ is a proper subterm of $t$, then $\mathsf{D}(t', t) \in I_{\Sigma'}^{k+2}$; furthermore, $\mathsf{D}(t', t) \in I_{\Sigma'}^\infty$ if and only if $t'$ is a proper subterm of $t$.

---

1. In the rest of this paper we often omit 'universally'; furthermore, when used as an acyclicity notion, MFA means 'universally MFA'.

(Claims 1 and 2, direction $\Rightarrow$) The proof is by induction on $k$. Set $I_{\Sigma'}^0$ does not contain functional terms, and so it clearly satisfies both claims. For the induction step, assume that both claims hold for $I_{\Sigma'}^{k-1}$ and consider $I_{\Sigma'}^k$. Since $I_{\Sigma'}^{k-1} \subseteq I_{\Sigma'}^k$, both claims clearly hold for each term $t$ that occurs in $I_{\Sigma'}^{k-1}$. Consider an arbitrary term $t$ of the form $f_r^i(\vec{u})$ that does not occur in $I_{\Sigma'}^{k-1}$, and an arbitrary term $t' \in \vec{u}$. Clearly, $t$ is introduced into $I_{\Sigma'}^k$ by an application of the skolemisation of $\mathsf{MFA}(r)$ for some rule $r \in \Sigma$. Since the head of $\mathsf{MFA}(r)$ contains atoms $\mathsf{F}_r^i(y_i)$ and $\mathsf{S}(x_j, y_i)$ for each $x_j \in \vec{x}$, we have $\mathsf{F}_r^i(t) \in I_{\Sigma'}^k$ and $\mathsf{S}(t', t) \in I_{\Sigma'}^k$ for each $t' \in \vec{u}$, and so we have $\mathsf{F}_r^i(t) \in I_{\Sigma'}^\infty$ and $\mathsf{S}(t', t) \in I_{\Sigma'}^\infty$ for each $t' \in \vec{u}$ as well. Finally, since $I_{\Sigma'}^\infty = \bigcup_k I_{\Sigma'}^k$, these claims clearly hold for $k = \infty$.

(Claims 1 and 2, direction $\Leftarrow$) Predicate $\mathsf{S}$ and each predicate $\mathsf{F}_r^i$ occur in $\Sigma'$ only in head atoms of the form $\mathsf{F}_r^i(y_i)$ and $\mathsf{S}(x_j, y_i)$; hence, the skolemised rules contain these predicates only in head atoms of the form $\mathsf{F}_r^i(f_r^i(\vec{x}))$ and $\mathsf{S}(x_j, f_r^i(\vec{x}))$, which clearly implies our claim.

(Claim 3, the first part for $k \neq \infty$) The proof is by induction on $k$. The base case holds vacuously since $I_{\Sigma'}^0$ does not contain functional terms. Assume now that the claim holds for some $k - 1$, and consider an arbitrary term $t = f_r^i(\vec{u})$ occurring in $I_{\Sigma'}^k$ such that $t'$ is a subterm of some $t_i \in \vec{u}$. By Claim 2, we have $\mathsf{S}(t_i, t) \in I_{\Sigma'}^k$; furthermore, $t_i$ occurs in $I_{\Sigma'}^{k-1}$, so by the induction assumption we have $\mathsf{D}(t', t_i) \in I_{\Sigma'}^{k+1}$. Finally, the rules without functional terms are applied before the rules with functional terms; hence, by rule (19) we have $\mathsf{D}(t', t) \in I_{\Sigma'}^{k+2}$, as required.

(Claim 3, the second part) The 'proper subterm' relation is transitive, and rules (18) and (19) effectively define $\mathsf{D}$ as the transitive closure of $\mathsf{S}$, which clearly implies this claim.

Assume now that $I_{\Sigma'}^\infty$ contains a cyclic term $t$. Then, some term $t_1 = f_r^i(\vec{s})$ is a subterm of $t$ and some term $t_2 = f_r^i(\vec{u})$ is a proper subterm of $t_1$. By Claims 1 and 3, then we have $\{\mathsf{F}_r^i(t_2), \mathsf{D}(t_2, t_1), \mathsf{F}_r^i(t_1)\} \subseteq I_{\Sigma'}^\infty$. But then, since $\Sigma'$ contains rule (20), we have $\mathsf{C} \in I_{\Sigma'}^\infty$, so $\Sigma$ is not MFA. For the converse claim, assume that $\Sigma$ is not MFA w.r.t. an instance $I$. Then, by Definition 3 we have that $I \cup \mathsf{MFA}(\Sigma) \models \mathsf{C}$. Since the special nullary predicate $\mathsf{C}$ occurs only on the right-hand side of rule (20), there exist terms $t_1$ and $t_2$, a rule $r \in \Sigma$, and a predicate $\mathsf{F}_r^i$ such that $\{\mathsf{F}_r^i(t_1), \mathsf{D}(t_1, t_2), \mathsf{F}_r^i(t_2)\} \subseteq I_{\Sigma'}^\infty$. Since $\mathsf{F}_r^i(t_1)$ and $\mathsf{F}_r^i(t_2)$ are contained in $I_{\Sigma'}^\infty$, Claim 1 implies that $t_1$ and $t_2$ are of the form $t_1 = f_r^i(\vec{u_1})$ and $t_2 = f_r^i(\vec{u_2})$, respectively. Finally, $\mathsf{D}(t_1, t_2) \in I_{\Sigma'}^\infty$ and Claim 3 imply that $t_1$ is a proper subterm of $t_2$, so $I_{\Sigma'}^\infty$ contains a cyclic term. $\square$

This characterisation implies termination of skolem chase of MFA rules $\Sigma$ in 2ExpTime. In particular, a term $t$ derived by the skolem chase of $\Sigma' = \mathsf{MFA}(\Sigma)$ cannot be cyclic by Proposition 5; such $t$ can then be seen as a tree with branching factor bounded by the maximum arity of a function symbol in $\mathsf{sk}(\Sigma')$ and with depth bounded by the number of function symbols in $\mathsf{sk}(\Sigma')$. The chase can thus generate at most a doubly exponential number of different terms and atoms. The 2ExpTime bound already holds if the rules are WA (Calì et al., 2010b), so CQ answering for MFA rules is not harder than for WA rules.

**Proposition 6.** *If a set of rules $\Sigma$ is MFA w.r.t. an instance $I$, then the skolem chase for $I$ and $\Sigma$ terminates in double exponential time.*

*Proof.* Let $\Sigma' = \mathsf{MFA}(\Sigma)$, let $c$, $f$, and $p$ be the number of constants, function symbols, and predicate symbols, respectively, occurring in $\mathsf{sk}(\Sigma')$, let $\ell$ be the maximum arity of a function symbol, and let $a$ be the maximum arity of a predicate symbol in $\mathsf{sk}(\Sigma')$. Consider now an arbitrary term $t$ occurring in $I_{\Sigma'}^\infty$; clearly, $t$ can be seen as a tree with branching factor $\ell$ containing constants in the leaf nodes and function symbols in the internal nodes; furthermore, since $t$ is not cyclic, $\mathsf{dep}(t) \le f$, the number of leaves is bounded by $\ell^f$, and the total number of nodes is bounded by $f \cdot \ell^f$. Each node is assigned a constant or a function symbol, so the number of different terms occurring in $I_{\Sigma'}^\infty$ is bounded by $\wp = (c + f)^{f \cdot \ell^f}$, and the number of different atoms in $I_{\Sigma'}^\infty$ is bounded by $p \cdot \wp^a$, which is clearly doubly exponential in $\Sigma$ and $I$. Consequently, the size of $I_{\Sigma'}^\infty$ is at most doubly exponential in $\Sigma$ and $I$. Furthermore, for an arbitrary set of facts $I'$ and rule $r$, the set $r(I')$ can be computed by examining all mappings of the variables in $r$ to the terms occurring in $I'$, which requires exponential time in the size of $r$ and polynomial time in the size of $I'$. Consequently, $I_{\Sigma'}^\infty$ can be computed in time that is double exponential in $I$ and $\Sigma$. Finally, it is straightforward to see that $I_\Sigma^\infty \subseteq I_{\Sigma'}^\infty$, so $I_\Sigma^\infty$ can be computed in double exponential time as well. $\qquad\square$

By Proposition 6, answering a BCQ over MFA rules is in 2ExpTime. We next prove that checking MFA w.r.t. a specific instance $I$ is also in 2ExpTime, and that checking universal MFA is 2ExpTime-hard. This provides tight complexity bounds for both problems. Towards this goal, we first establish in Lemma 7 a relationship between answering certain kinds of queries over certain kinds of rules and checking whether a related set of rules is universally MFA; we use this relationship in several hardness proofs in the rest of this paper. Then, in Theorem 8 we present our main complexity result.

**Lemma 7.** *Let $\Sigma$ be a set of weakly acyclic, constant-free, equality-free, and connected rules with predicates of nonzero arity, let $A$ and $B$ be unary predicates, let $R$ be a fresh binary predicate, let $a$ be a constant, and let $\Omega$ be $\Sigma$ extended with rule (23).*

$$R(z, x) \wedge B(x) \to \exists y.[R(x, y) \wedge A(y)] \tag{23}$$

*Then, we have $\{A(a)\} \cup \Sigma \not\models B(a)$ if and only if $\Omega$ is universally MFA.*

*Proof.* Let $I = \{A(a)\}$, and let $I_\Sigma^0, I_\Sigma^1, \dots$ be the chase sequence for $I$ and $\Sigma$. Furthermore, let $\Omega' = \mathsf{MFA}(\Omega)$, let $J = I_\Omega^*$, let $J_{\Omega'}^0, J_{\Omega'}^1, \dots$ be the chase sequence for $J$ and $\Omega'$, and let $f$ be the function symbol used to skolemise the existential quantifier in rule (23). Set $\Sigma$ is constant-free, so $a$ is the only constant occurring in each set $I_\Sigma^i$.

We next show that the facts in $J_{\Omega'}^j$ are of a certain form. To this end, for each $\ell \ge 0$, let $t_\ell = f(\dots f(*) \dots)$ where the function symbol $f$ is repeated $\ell$ times (by this definition, we have $t_0 = *$); also, each term or fact obtained from $t_\ell$ by zero or more applications of predicates or function symbols not in $\{f, \mathsf{D}, \mathsf{S}, \mathsf{C}, R\}$ is of *level* $\ell$. By induction on the chase sequence for $J$ and $\Omega'$, we next prove that the sequence satisfies the following property ($\blacklozenge$):

for each fact $F \in J_{\Omega'}^j$, some integer $\ell$ exists such that $F$ is of the form $R(*, *)$ or $R(t_\ell, t_{\ell+1})$, or the predicate of $F$ is contained in $\{\mathsf{D}, \mathsf{S}, \mathsf{C}\}$, or $F$ is an $\ell$-level fact and the predicate of $F$ is not contained in $\{\mathsf{D}, \mathsf{S}, \mathsf{C}, R\}$.

Set $J_{\Omega'}^0 = J$ clearly satisfies property ($\blacklozenge$) since each fact in it is clearly of level 0. Now assume that $J_{\Omega'}^j$ satisfies property ($\blacklozenge$) for some $j$, and consider an application of a rule $r \in \Omega'$. If $r$ corresponds to rule (18), (19), (20), or (23), then the result of the rule application clearly satisfies property ($\blacklozenge$). Otherwise, $r$ is safe and no body atom contains a predicate in $\{\mathsf{D}, \mathsf{S}, \mathsf{C}, R\}$; by induction assumption, then some atom is matched to a fact of some level $\ell$; the body atoms of $r$ are connected, so all body atoms are matched to facts of the same level; finally, the head atoms of $r$ contain function symbols different from $f$, but no constants or predicates of zero arity, so each fact derived by an atom in the head of $r$ either contains predicate $\mathsf{S}$ or is of level $\ell$.

We next show that the chase sequences for $I$ and $\Sigma$, and for $J$ and $\Omega'$ are related by the following property ($\Diamond$):

for each fact $F'$ of level 1 and the fact $F$ obtained by replacing each $t_1$ in $F'$ with $a$, we have $F \in I^i_\Sigma$ for some $i$ if and only if $F' \in J^j_{\Omega'} \setminus J$ for some $j$.

The proof of ($\Diamond$) is straightforward: $J$ contains $R(*, *)$ and $B(*)$, so $J^1_{\Omega'}$ contains $R(*, f(*))$ and $A(f(*))$; moreover, due to ($\blacklozenge$), term $t_1$ plays in the chase sequence for $J$ and $\Omega'$ the 'same role' as constant $a$ in the chase sequence for $I$ and $\Sigma$, so the rule applications to facts of level 1 in the former chase sequence correspond one-to-one with rule applications in the latter chase sequence. We omit the formal details for the sake brevity.

Now assume that $\{A(a)\} \cup \Sigma \models B(a)$. Then, $B(a) \in I^i_\Sigma$ holds for some $i$. By property ($\Diamond$), then integer $j$ exists such that $B(f(*)) \in J^j_{\Omega'}$. But then, due to rule (23), some $\ell \geq j$ exists such that $A(f(f(*))) \in J^\ell_{\Omega'}$. By Proposition 5, then $\Omega$ is not universally MFA.

Conversely, assume that $\{A(a)\} \cup \Sigma \not\models B(a)$. Since $\Sigma$ is weakly acyclic and equality-free, $\Sigma$ is super-weakly acyclic (Marnette, 2009); as we will show in Section 4 (see Theorem 19), $\Sigma$ is then MFA as well. Now consider an arbitrary integer $j$ and fact $F \in J^j_{\Omega'}$. If $F$ is of level 0 or 1, since $\Sigma$ is MFA, fact $F$ does not contain a cyclic term. Furthermore, $B(a) \notin I^\infty_\Sigma$ so, by property ($\Diamond$), fact $F$ is not of the form $B(f(*))$; thus, rule (23) does not 'fire' to introduce facts of level greater than 1. Consequently, $F$ does not contain a cyclic term, and so, by Proposition 5, the set $\Omega$ is universally MFA. □

**Theorem 8.** *Given a set of rules $\Sigma$, deciding whether $\Sigma$ is MFA w.r.t. an instance $I$ is in* 2ExpTime*, and deciding whether $\Sigma$ is universally MFA is* 2ExpTime*-hard. Both results hold even if the arity of predicates in $\Sigma$ is bounded.*

*Proof.* (Membership) Let $\Sigma' = \mathsf{MFA}(\Sigma)$, let $I^0_{\Sigma'}, I^1_{\Sigma'}, \ldots$ be the chase sequence for $I$ and $\Sigma'$, and let $\wp$, $p$, and $a$ be as stated in the proof of Proposition 6. The number of different atoms that can be constructed from $\wp$ terms is bounded by $k = p \cdot \wp^a$; note that this is double exponential even if $a$ is bounded. Let $k' = k + 4$; we next show that whether $\Sigma$ is MFA w.r.t. $I$ can be decided by constructing $I^{k'}_{\Sigma'}$ and then checking whether $\mathsf{C} \in I^{k'}_{\Sigma'}$. As in the proof of Proposition 6, the latter can be done in double exponential time.

If $I^k_{\Sigma'} = I^{k'}_{\Sigma'}$, then $I^\infty_{\Sigma'} = I^{k'}_{\Sigma'}$, so $\Sigma$ is not MFA if and only if $\mathsf{C} \in I^{k'}_{\Sigma'}$. Otherwise, we have $I^k_{\Sigma'} \subsetneq I^{k'}_{\Sigma'}$; but then, $I^{k+1}_{\Sigma'}$ clearly contains at least one cyclic term $t = f^i_r(\vec{t})$ such that $t' = f^i_r(\vec{s})$ is a subterm of some $t_i \in \vec{t}$. Since $I^{k+1}_{\Sigma'}$ satisfies Claims 1–3 from the proof of Proposition 5, we have $\mathsf{D}(t_i, t) \in I^{k+3}_{\Sigma'}$; by rule (20) and the fact that rules without functional terms are applied before rules with functional terms, we have $\mathsf{C} \in I^{k'}_{\Sigma'}$; thus, $\mathsf{C} \in I^\infty_{\Sigma'}$, so $\Sigma$ is not MFA by Proposition 5.

(Hardness) We prove the claim by a reduction from the problem of checking $I \cup \Sigma \models Q$, where $\Sigma$ is a *weakly acyclic* set of equality-free rules and with predicates of bounded arity, $I$ is an instance, and $Q = \exists \vec{y}.\xi(\vec{y})$ is a Boolean conjunctive query. Calì et al. (2010b) show that, for such $I$, $\Sigma$, and $Q$, deciding $I \cup \Sigma \models Q$ is 2EXPTIME-complete. We next transform $I$, $\Sigma$, and $Q$ so that we can apply Lemma 7, which proves our claim.

Let $\Sigma_1 = \Sigma \cup \{\xi(\vec{y}) \to B\}$ where $B$ is a fresh predicate of zero arity; clearly, $I \cup \Sigma \models Q$ holds if and only if $I \cup \Sigma_1 \models B$ holds.

Let $\Sigma_2$ and $I_2$ be obtained by eliminating constants from the rules in $\Sigma_1$; that is, we initially set $I_2 = I$ and then, for each rule $r \in \Sigma$ and each constant $c$ occurring in $r$, we replace all occurrences in $c$ with a fresh variable $w_c$, add an atom $O_c(w_c)$ to the body of $r$ where $O_c$ is a fresh predicate uniquely associated with $c$, and add a fact $O_c(c)$ to $I_2$. It is straightforward to see that $I \cup \Sigma_1 \models B$ if and only if $I_2 \cup \Sigma_2 \models B$.

Finally, to transform $\Sigma_2$ and $I_2$ into $\Sigma_3$, we define some notation. Let $\hat{P}$ be a fresh $n+1$-ary predicate unique for each $n$-ary predicate $P$, and let $w$ be a fresh variable not occurring in $\Sigma_2$. For a conjunction of atoms $\varphi$, let $\hat{\varphi} = \bigwedge_{P(\vec{t}) \in \varphi} \hat{P}(\vec{t}, w)$. Rule (24) is obtained from $I_2$ as specified below, where $A$ is a fresh unary predicate, each constant $c$ occurring in $I_2$ is associated with a distinct, fresh variable $v_c$, and $\vec{v}_c$ is the vector of all such variables:

$$A(w) \to \exists \vec{v}_c. \bigwedge_{P(c_1,\ldots,c_k) \in I_2} \hat{P}(v_{c_1}, \ldots, v_{c_k}, w) \tag{24}$$

Finally, the set $\Sigma_3$ contains rule (24) and a rule

$$\hat{\varphi}(\vec{x}, \vec{z}, w) \to \exists \vec{y}.\hat{\psi}(\vec{x}, \vec{y}, w) \qquad \text{for each rule} \qquad \varphi(\vec{x}, \vec{z}) \to \exists \vec{y}.\psi(\vec{x}, \vec{y}) \text{ in } \Sigma_2. \tag{25}$$

Clearly, all predicates in $\Sigma_3$ are of nonzero arity; all rules in $\Sigma_3$ are constant-free and connected; and $A$ occurs only in the body of rule (24) and $\Sigma_2$ is WA, so $\Sigma_3$ is WA as well. Finally, let $I_3 = \{A(a)\}$ where $a$ is a fresh constant; by induction on the chase sequences for $\Sigma_2$ and $I_2$, and $\Sigma_3$ and $I_3$, it is straightforward to show that, for each integer $i$ and each fact $P(c_1, \ldots, c_k)$, we have $P(c_1, \ldots, c_k) \in (I_2)^i_{\Sigma_2}$ if and only if $\hat{P}(f_{c_1}(a), \ldots, f_{c_k}(a), a) \in (I_3)^{i+1}_{\Sigma_3}$, where $f_{c_1}, \ldots, f_{c_k}$ are the skolem functions used to skolemise $v_{c_1}, \ldots, v_{c_k}$ in rule (24). Thus, $I_2 \cup \Sigma_2 \models B$ if and only if $I_3 \cup \Sigma_3 \models \hat{B}(a)$, which by Lemma 7 implies our claim. □

The results of Theorem 8 are somewhat discouraging: known acyclicity notions can typically be checked in PTIME or in NP. We consider MFA to be an 'upper bound' of practically useful acyclicity notions. We see two possibilities for improving these results. In Section 3.3 we introduce an approximation of MFA that is easier to check; our experiments (see Section 7) show that this notion often coincides with MFA in practice. Furthermore, we show next that the complexity is lower for rules of the following shape.

**Definition 9.** *A rule $r$ of the form* (5) *is an $\exists$-1 rule if $\vec{y}$ is empty or $\vec{x}$ contains at most one variable.*

As we discuss in the following sections, $\exists$-1 rules capture (extensions of) Horn DLs. We next show that BCQ answering and MFA checking for $\exists$-1 rules is easier than for general rules. Intuitively, if $\Sigma$ is MFA and contains only $\exists$-1 rules, then all functional terms in $\mathsf{sk}(\mathsf{MFA}(\Sigma))$ are unary and hence the number of different terms and atoms derivable by chase becomes exponentially bounded, as shown by the following theorem.

**Theorem 10.** *Let $\Sigma$ be a set of $\exists$-1 rules, and let $I$ be an instance. Checking whether $\Sigma$ is MFA w.r.t. $I$ is in EXPTIME, and checking whether $\Sigma$ is universally MFA is EXPTIME-hard. Moreover, if $\Sigma$ is MFA w.r.t. $I$, then answering a BCQ over $\Sigma$ and $I$ is EXPTIME-complete.*

*Proof.* We defer the proof of both hardness claims to Section 6, which deals with an even smaller class of rules that correspond to Horn description logic ontologies. In particular, we prove hardness of BCQ answering in Lemma 59, and hardness of checking whether $\Sigma$ is MFA w.r.t. $I$ in Lemma 60. In the rest of this proof, we show both membership results.

Let $\Sigma' = \mathsf{MFA}(\Sigma)$; let $c$ be the number of constants in an instance; and let $f$ be the number of function symbols in the rules. Since $\Sigma$ contains only $\exists$-1 rules, $\Sigma'$ also contains only $\exists$-1 rules; consequently, all functional terms in $\mathsf{sk}(\Sigma')$ are of arity 1. Hence, each noncyclic term can be understood as a sequence of at most $f$ function symbols, so the total number of different noncyclic terms is bounded by $\wp = c \cdot (f + 1)^f$. The total number of atoms is bounded by $p \cdot \wp^a$, where $p$ is the number of predicates and $a$ is the maximum arity of a predicate in $\Sigma'$. Note that this is exponential even if $a$ is fixed. As in the proof of Proposition 6, we can now show that either the chase for $\Sigma'$ and $I$ terminates or a cyclic term is derived in exponential time, which proves that the complexity of checking whether $\Sigma$ is MFA w.r.t. $I$ is in EXPTIME.

Finally, since $I_\Sigma^\infty \subseteq I_{\Sigma'}^\infty$, if $\Sigma$ is MFA, then $I_\Sigma^\infty$ can be computed in exponential time, so a BCQ over $\Sigma$ and $I$ can be answered in EXPTIME. $\square$

### 3.3 Model-Summarising Acyclicity (MSA)

The high cost of checking MFA of $\Sigma$ arises because the arity of function symbols in $\mathsf{sk}(\Sigma)$ is unbounded and the depth of cyclic terms can be linear in $\Sigma$. To obtain an acyclicity notion that is easier to check, we must coarsen the structure used for cycle analysis. We thus next introduce *model-summarising acyclicity*, which 'summarises' the models of $\Sigma$ by reusing the same constant to satisfy an existential quantifier, instead of introducing 'deep' terms.

**Definition 11.** *Let $\mathsf{S}$, $\mathsf{D}$, and $\mathsf{F}_r^i$ be as specified in Definition 3; furthermore, for each rule $r = \varphi(\vec{x}, \vec{z}) \to \exists \vec{y}.\psi(\vec{x}, \vec{y})$ and each variable $y_i \in \vec{y}$, let $\mathsf{c}_r^i$ be a fresh constant unique for $r$ and $y_i$. Then, $\mathsf{MSA}(r)$ is the following rule, where $\theta_{\mathsf{MSA}}$ is the substitution that maps each variable $y_i \in \vec{y}$ to $\mathsf{c}_r^i$:*

$$\varphi(\vec{x}, \vec{z}) \to \psi(\vec{x}, \vec{y})\theta_{\mathsf{MSA}} \wedge \bigwedge_{y_i \in \vec{y}} \left[ \mathsf{F}_r^i(y_i)\theta_{\mathsf{MSA}} \wedge \bigwedge_{x_j \in \vec{x}} \mathsf{S}(x_j, y_i)\theta_{\mathsf{MSA}} \right]$$

*For a set $\Sigma$ of rules, $\mathsf{MSA}(\Sigma)$ is the smallest set that contains $\mathsf{MSA}(r)$ for each rule $r \in \Sigma$, rules (18)–(19), and rule (20) instantiated for each predicate $\mathsf{F}_r^i$. Set $\Sigma$ is model-summarising acyclic (MSA) w.r.t. an instance $I$ if $I \cup \mathsf{MSA}(\Sigma) \not\models \mathsf{C}$; furthermore, $\Sigma$ is universally MSA if $\Sigma$ is MSA w.r.t. $I_\Sigma^*$.*

**Example 12.** *Consider again the set of rules $\Sigma$ from Example 1. $\mathsf{MSA}(r_1)$ and $\mathsf{MSA}(r_3)$ are given by rules (26) and (27), respectively; since $r_1$ and $r_3$ contain a single existentially quantified variable each, we omit the superscripts in $\mathsf{F}_{r_1}$, $\mathsf{F}_{r_3}$, $\mathsf{c}_{r_1}$, and $\mathsf{c}_{r_3}$ for the sake of* *clarity. Thus, $\mathsf{MSA}(\Sigma)$ consists of rules (14), (16), and (17), rules (26)–(27), rules (18)–(19), and rule (20) instantiated for $\mathsf{F}_{r_1}$ and $\mathsf{F}_{r_3}$.*

$$A(x_1) \rightarrow R(x_1, \mathsf{c}_{r_1}) \wedge B(\mathsf{c}_{r_1}) \wedge \mathsf{F}_{r_1}(\mathsf{c}_{r_1}) \wedge \mathsf{S}(x_1, \mathsf{c}_{r_1}) \tag{26}$$

$$B(x_3) \rightarrow R(x_3, \mathsf{c}_{r_3}) \wedge C(\mathsf{c}_{r_3}) \wedge \mathsf{F}_{r_3}(\mathsf{c}_{r_3}) \wedge \mathsf{S}(x_3, \mathsf{c}_{r_3}) \tag{27}$$

*The following table shows the chase sequence for $I_\Sigma^*$ and $\mathsf{MSA}(\Sigma)$.*

| | | | |
|---|---|---|---|
| $A(*)$ | $R(*, \mathsf{c}_{r_1})$ | $R(\mathsf{c}_{r_1}, \mathsf{c}_{r_3})$ | $\mathsf{D}(\mathsf{c}_{r_1}, \mathsf{c}_{r_3})$ |
| $B(*)$ | $R(*, \mathsf{c}_{r_3})$ | $D(\mathsf{c}_{r_3})$ | |
| $C(*)$ | $B(\mathsf{c}_{r_1})$ | $\mathsf{S}(\mathsf{c}_{r_1}, \mathsf{c}_{r_3})$ | |
| $D(*)$ | $C(\mathsf{c}_{r_3})$ | $\mathsf{D}(*, \mathsf{c}_{r_3})$ | |
| $R(*, *)$ | $\mathsf{S}(*, \mathsf{c}_{r_1})$ | $\mathsf{D}(*, \mathsf{c}_{r_1})$ | |
| | $\mathsf{S}(*, \mathsf{c}_{r_3})$ | | |
| | $\mathsf{F}_{r_1}(\mathsf{c}_{r_1})$ | | |
| | $\mathsf{F}_{r_3}(\mathsf{c}_{r_3})$ | | |

*The result of the chase does not contain $\mathsf{C}$, and so $\Sigma$ is universally MSA.* ◇

Note that $\mathsf{MSA}(\Sigma)$ is equivalent to a set of datalog rules: the only minor difference is that the rules in $\mathsf{MSA}(\Sigma)$ can contain several head atoms, but such rules can clearly be transformed into equivalent datalog rules. Thus, MSA can be checked using a datalog reasoner. This connection with datalog and the complexity results by Dantsin, Eiter, Gottlob, and Voronkov (2001) for checking entailment of a ground atom in a datalog program provide us with the upper complexity bound for checking MSA in Theorem 13. The complexity of datalog reasoning is $O(r \cdot n^v)$ where $r$ is the number of rules, $v$ is the maximum number of variables in a rule, and $n$ is the size of the set of facts that the rules are applied to; thus, checking MSA should be feasible if the rules in $\Sigma$ are 'short' and so $v$ is 'small'.

**Theorem 13.** *For $\Sigma$ a set of rules, deciding whether $\Sigma$ is MSA w.r.t. an instance $I$ is in* EXPTIME, *and deciding whether $\Sigma$ is universally MSA is* EXPTIME-*hard. The two problems are in co*NP *and co*NP-*hard, respectively, if the arity of the predicates in $\Sigma$ is bounded.*

*Proof.* (Membership) Let $\Sigma' = \mathsf{MSA}(\Sigma)$, and note that $\Sigma$ is MSA w.r.t. $I$ if and only if $I \cup \Sigma' \not\models \mathsf{C}$ if and only if $\mathsf{C} \notin I_{\Sigma'}^\infty$. The total number of atoms occurring in $I_{\Sigma'}^\infty$ is $p \cdot c^a$, where $p$ is the number of predicates, $c$ is the number of constants, and $a$ is the maximum arity of the predicates in $\Sigma'$; this number is clearly exponential if $a$ is not bounded. The rest of the proof is the same as in Theorem 8.

Assume now that $a$ is bounded; then the number of ground atoms in $I_{\Sigma'}^\infty$ becomes polynomial. Furthermore, by the definition of the chase, $\mathsf{C} \in I_{\Sigma'}^\infty$ if and only if there exist a sequence of rules $r_1, \ldots, r_n$ of the form $r_i = \varphi_i \rightarrow \psi_i$ and a sequence of substitutions $\sigma_1, \ldots, \sigma_n$ such that $\varphi \sigma_i \subseteq I \cup \{\psi_j \sigma_j \mid j < i\} \subseteq I_{\Sigma'}^\infty$ for each $1 \le i \le n$ and $\psi_n \sigma_n = \mathsf{C}$. Clearly, we can assume that $n \le p \cdot c^a$, which is polynomial. Thus, we can guess the two sequences in nondeterministic polynomial time, and we can check the required property in polynomial time. Thus, $I \cup \Sigma' \models \mathsf{C}$ can be checked in nondeterministic polynomial time, so checking whether $\Sigma$ is MSA w.r.t. $I$ is in coNP.

(Hardness) Let $\Sigma$ be a set of datalog rules, let $I$ be an instance, and let $Q$ be ground atom. Checking whether $I \cup \Sigma \models Q$ is ExpTime-complete in general (Dantsin et al., 2001). Furthermore, the problem is NP-hard if the arity of predicates is bounded: a rule in $\Sigma$ can encode an arbitrary Boolean conjunctive query with atoms of bounded arity but arbitrarily many variables, for which answering is well known to be NP-hard.

Let $\Sigma_4$ and $I_4$ be obtained from $\Sigma$ as in the proof of Theorem 8; then, $I \cup \Sigma \models Q$ if and only if $I_4 \cup \Sigma_4 \models B(a)$, and the set of rules $\Omega$ obtained from $\Sigma_4$ as specified in Lemma 7 is universally MFA if and only if $I_4 \cup \Sigma_4 \not\models B(a)$. Finally, the only existential variable in $\Omega$ occurs in a rule of the form (23), so it is straightforward to see that $\Omega$ is universally MFA if and only if $\mathcal{T}'$ is universally MSA. $\qquad\square$

Before concluding this section, we present Theorem 14 and Example 15, which together show that MFA is strictly more general than MSA.

**Theorem 14.** *If a set of rules $\Sigma$ is MSA (w.r.t. an instance $I$), then $\Sigma$ is MFA (w.r.t. $I$) as well.*

*Proof.* Let $\Sigma_1 = \mathsf{MFA}(\Sigma)$ and let $\Sigma_2 = \mathsf{MSA}(\Sigma)$. Furthermore, let $h$ be the mapping of ground terms to constants defined such that $h(t) = \mathsf{c}_r^i$ if $t$ is of the form $f_r^i(\ldots)$, and $h(t) = t$ if $t$ is a constant; for an atom $A = P(t_1, \ldots, t_n)$, let $h(A) = P(h(t_1), \ldots, h(t_n))$; and for an instance $I$, let $h(I) = \{h(A) \mid A \in I\}$. Finally, let $I_{\Sigma_1}^0, I_{\Sigma_1}^1, \ldots$ be the chase sequence for $I$ and $\Sigma_1$, and let $I_{\Sigma_2}^0, I_{\Sigma_2}^1, \ldots$ be the chase sequence for $I$ and $\Sigma_2$. Note that $\mathsf{sk}(\Sigma_2) = \Sigma_2$ differs from $\mathsf{sk}(\Sigma_1)$ only in that the former contains the constant $\mathsf{c}_r^i$ in place of each functional term $f_r^i(\vec{x})$. Please note that, although our definition of the chase applies rules with function symbols after rules without function symbols, one can clearly construct the chase of the function-free set of rules $\Sigma_2$ using any order of rule applications, including the one corresponding to the order of rule applications in the chase of $\Sigma_1$. Assuming this slight modification, one can show by a straightforward induction on $i$ that $h(I_{\Sigma_1}^i) \subseteq I_{\Sigma_2}^i$ for each $i$; this implies $h(I_{\Sigma_1}^\infty) \subseteq I_{\Sigma_2}^\infty$. Consequently, $\mathsf{C} \notin I_{\Sigma_2}^\infty$ clearly implies $\mathsf{C} \notin I_{\Sigma_1}^\infty$; hence, if $\Sigma$ is MSA, then $\Sigma$ is MFA as well, as required. $\qquad\square$

**Example 15.** *Let $\Sigma$ be the set of rules (28)–(31).*

$$r_1 = \qquad\qquad A(x) \to \exists y.R(x,y) \wedge B(y) \qquad\qquad (28)$$

$$r_2 = \qquad\qquad B(x) \to \exists y.S(x,y) \wedge T(y,x) \qquad\qquad (29)$$

$$r_3 = \qquad A(z) \wedge S(z,x) \to C(x) \qquad\qquad (30)$$

$$r_4 = \qquad C(z) \wedge T(z,x) \to A(x) \qquad\qquad (31)$$

*It is straightforward to check that $\Sigma$ is universally MFA, but not universally MSA.* $\qquad\Diamond$

## 3.4 Acyclicity Notions and Normalisation

As mentioned in Section 2.5, existential rules are often normalised into a particular form; however, this cannot destroy acyclicity: if a set of rules $\Sigma$ is MFA, then each set of rules obtained from $\Sigma$ by normalisation is MFA as well. This claim involves certain technical assumptions about the treatment of equality, which is why we postpone a formal proof of this statement until Section 5. Next, however, we show that normalisation can have a positive effect on acyclicity.

**Example 16.** *Let $\Sigma$ be the set containing only the following rule:*

$$A(x) \rightarrow \exists y.[B(x) \land A(y)] \tag{32}$$

*As specified in Section 2.2, this rule is skolemised as follows, which causes that the skolem chase of $\Sigma$ and instance $I = \{A(a)\}$ does not terminate.*

$$A(x) \rightarrow B(x) \land A(f(x)) \tag{33}$$

*Note, however, that atoms $B(x)$ and $A(y)$ in the head of the rule do not share variables, so we can normalise this rule as follows, where $Q$ is a fresh predicate of zero arity:*

$$A(x) \rightarrow B(x) \land Q \tag{34}$$
$$Q \rightarrow \exists y.A(y) \tag{35}$$

*It is straightforward to check that this normalised set of rules is MFA; in fact, the normalised set of rules is even JA. Intuitively, normalisation, as defined in Section 2.5, ensures that each functional symbol introduced during normalisation depends on as few variables in the rule as possible.* ◇

Normalisation, however, can have a negative effect on universal termination, as shown by the following example.

**Example 17.** *Let $\Sigma$ be the set containing only the following rule:*

$$C(z) \land R(z, x) \land B(x) \rightarrow \exists y_1 \exists y_2.[R(x, y_1) \land R(y_1, y_2) \land B(y_2)] \tag{36}$$

*One can readily check that $\Sigma$ is universally MFA. Now let $\Sigma'$ be the following set of rules, which is obtained by replacing conjunction $R(y_1, y_2) \land B(y_2)$ in the rule head with $Q(y_1)$:*

$$C(z) \land R(z, x) \land B(x) \rightarrow \exists y_1.[R(x, y_1) \land Q(y_1)] \tag{37}$$
$$Q(y_1) \rightarrow \exists y_2.[R(y_1, y_2) \land B(y_2)] \tag{38}$$

*Let $f_1$ and $f_2$ be function symbols used to skolemise the existential quantifier in rule (37) and (38), respectively. Since $Q(*) \in I^*_{\Sigma'}$, the chase of $\Sigma'$ and $I^*_{\Sigma'}$ derives $R(*, f_2(*))$ and $B(f_2(*))$; but then, these facts, $C(*)$, and rule (37) derive $Q(f_1(f_2(*)))$, after which rule (38) derives $R(f_1(f_2(*)), f_2(f_1(f_2(*))))$ and $B(f_2(f_1(f_2(*))))$. The chase of $\Sigma'$ and $I^*_{\Sigma'}$ thus contains a cyclic term, so $\Sigma'$ is not universally MFA.*

*Intuitively, this problem occurs because the critical instance $I^*_{\Sigma'}$ for $\Sigma'$ also instantiates the predicate $Q$ introduced during normalisation. Such predicates, however, cannot occur in arbitrary input instances, so we can use the critical instance for $\Sigma$. Since $Q(*) \notin I^*_{\Sigma}$, the skolem chase of $\Sigma'$ and $I^*_{\Sigma}$ does not derive a cyclic term, from which we can conclude that the skolem chase of $\Sigma'$ terminates on each instance $I$ that contains facts constructed using only the predicates occurring in $\Sigma$.* ◇

## 4. Relationship with Known Acyclicity Notions

Many acyclicity notions have been proposed in the literature, but the relationships between them have been only partially investigated. We next investigate the relationship between MFA, MSA, and the acyclicity notions known to us, and we produce a detailed picture of their relative expressiveness. We show that MFA and MSA generalise most of these notions.

### 4.1 Acyclicity in Databases

Acyclicity notions have been considered in databases in data integration and data exchange scenarios. Weak acyclicity (Fagin et al., 2005) was one of the first such notions, and it has spurred on the research into more sophisticated notions for ensuring chase termination.

#### 4.1.1 Super-Weak Acyclicity

Marnette (2009) proposed *super-weak acyclicity* (SWA), which generalises weak acyclicity provided that the rules are equality-free. We next recapitulate the definition of SWA, and then we show that MSA and MFA are strictly more general than SWA.

**Definition 18.** *Let $\Sigma$ be a set of existential rules in which no variable occurs in more than one rule, and let $\theta_{\mathsf{sk}}$ be the substitution used to skolemise the rules in $\Sigma$.[2] A place is a pair $\langle A, i \rangle$ where $A$ is an $n$-ary atom occurring in a rule in $\Sigma$ and $1 \leq i \leq n$. A set of places $P'$ covers a set of places $P$ if, for each place $\langle A, i \rangle \in P$, a place $\langle A', i' \rangle \in P'$ and substitutions $\sigma$ and $\sigma'$ exist such that $A\sigma = A'\sigma'$ and $i = i'$. Given a variable $w$ occurring in a rule $r = \varphi \rightarrow \exists \vec{y}.\psi$, sets of places $\mathsf{In}(w)$, $\mathsf{Out}(w)$, and $\mathsf{Move}(w)$ are defined as follows:*

- *set $\mathsf{In}(w)$ contains each place $\langle R(\vec{t}), i \rangle$ such that $R(\vec{t}) \in \varphi$ and $t_i = w$;*

- *set $\mathsf{Out}(w)$ contains each place $\langle R(\vec{t})\theta_{\mathsf{sk}}, i \rangle$ such that $R(\vec{t}) \in \psi$ and $t_i = w$; and*

- *set $\mathsf{Move}(w)$ is the smallest set of places such that*

  - *$\mathsf{Out}(w) \subseteq \mathsf{Move}(w)$ and*
  - *$\mathsf{Out}(w') \subseteq \mathsf{Move}(w)$ for each variable $w'$ that is universally quantified in some rule in $\Sigma$ such that $\mathsf{Move}(w)$ covers $\mathsf{In}(w')$.*

*The SWA dependency graph $SWA(\Sigma)$ of $\Sigma$ contains a vertex for each rule of $\Sigma$, and an edge from a rule $r \in \Sigma$ to a rule $r' \in \Sigma$ if a variable $x'$ occurring in both the body and the head of $r'$ and an existentially quantified variable $y$ occurring in the head of $r$ exists such that $\mathsf{Move}(y)$ covers $\mathsf{In}(x')$. Set $\Sigma$ is super-weakly acyclic (SWA) if $SWA(\Sigma)$ is acyclic.*

Marnette (2009) uses a slightly different definition: the notation for places is the same as our notation for positions; a variable may occur in more than one rule so sets $\mathsf{In}(w)$, $\mathsf{Out}(w)$, and $\mathsf{Move}(w)$ are defined w.r.t. a rule and a variable; and a rule trigger relation is used instead of the SWA dependency graph. For simplicity, Definition 18 introduces SWA in the same style as JA; however, both definitions capture the same class of rules.

The following theorem shows that MSA is more general than SWA. Furthermore, in Example 12 we argued that the set of rules $\Sigma$ from Example 1 is MSA, and one can readily check that $\Sigma$ is not SWA. Consequently, MSA is strictly more general than SWA.

**Theorem 19.** *If a set of rules $\Sigma$ is SWA, then $\Sigma$ is universally MSA.*

---

2. Substitution $\theta_{\mathsf{sk}}$ is unique for each rule in Section 2.2; however, since each variable in $\Sigma$ occurs in at most one rule, w.l.o.g. we can take $\theta_{\mathsf{sk}}$ as the substitution used to skolemise all the rules in $\Sigma$.

*Proof.* Let $\Sigma' = \mathsf{MSA}(\Sigma)$, let $I^0, I^1, \ldots$ be the chase sequence for $I^*_\Sigma$ and $\Sigma'$, and let $I^\infty$ be the chase of $I^*_\Sigma$ and $\Sigma'$. Furthermore, let $\rho$ be the function that maps constants to themselves and that maps ground functional terms as $\rho(f^i_r(\vec{t})) = \mathsf{c}^i_r$, where $f^i_r$ and $\mathsf{c}^i_r$ were introduced in Section 2.2 and Definition 11, respectively. Finally, let $\rho(P(t_1, \ldots, t_n)) = P(\rho(t_1), \ldots, \rho(t_n))$.

We next prove the following property ($\blacklozenge$): for each rule $r \in \Sigma$, each existentially quantified variable $y_i$ occurring in $r$, each $P(\vec{t}) \in I^\infty$ where $P \notin \{\mathsf{S}, \mathsf{D}, \mathsf{C}\}$, and each $t_j \in \vec{t}$ such that $t_j = \mathsf{c}^i_r$, a substitution $\tau$ and a place $\langle A, j \rangle \in \mathsf{Move}(y_i)$ exist such that $P(\vec{t}) = \rho(A\tau)$. The proof is by induction on the length of the chase. Since $I^0 = I^*_\Sigma$ does not contain a constant of the form $\mathsf{c}^i_r$, property ($\blacklozenge$) holds vacuously for $I^0$. Assume now that property ($\blacklozenge$) holds for some $I^{k-1}$, and consider an arbitrary rule $r \in \Sigma$, an existentially quantified variable $y_i$ in $r$, a fact $P(\vec{t}) \in I^k \setminus I^{k-1}$ with $P \notin \{\mathsf{S}, \mathsf{D}, \mathsf{C}\}$, and a term $t_j \in \vec{t}$ such that $t_j = \mathsf{c}^i_r$. Fact $P(\vec{t})$ is derived in $I^k$ from the head atom $H$ of some rule $r^1 \in \mathsf{MSA}(\Sigma)$. Let $\sigma$ be the substitution used in the rule application; clearly, we have $H\sigma = P(\vec{t})$. Furthermore, let $r^2 \in \Sigma$ be the rule such that $r^1 = \mathsf{MSA}(r^2)$, let $r^3 = \mathsf{sk}(r^2)$, and let $H^3$ be the head atom of $r^3$ that corresponds to $H$; clearly, we have $\rho(H^3\sigma) = P(\vec{t})$. Now if $H$ has $\mathsf{c}^i_r$ in position $j$, then $r = r^1$ since $r^1$ is the only rule that contains $\mathsf{c}^i_r$; thus, $\langle H^3, j \rangle \in \mathsf{Out}(y_i) \subseteq \mathsf{Move}(y_i)$, so property ($\blacklozenge$) holds. Otherwise, $H$ contains at position $j$ a universally quantified variable $x$ such that $\sigma(x) = \mathsf{c}^i_r$. Let $B_1, \ldots, B_n$ be the body atoms of $r^1$ that contain variable $x$; clearly, $\{B_1\sigma, \ldots, B_n\sigma\} \subseteq I^{k-1}$. All these atoms satisfy the induction assumption, so for each $B_m \in \{B_1, \ldots, B_n\}$ and each $\ell$ such that $B_m$ contains variable $x$ at position $\ell$, a place $\langle B'_m, \ell \rangle \in \mathsf{Move}(y_i)$ and substitution $\tau^m$ exist such that $B_m\sigma = \rho(B'_m\tau^m)$. Let $\sigma'$ be the substitution obtained from $\sigma$ by setting $\sigma'(w) = \tau^m(w)$ for each variable $w$ for which $\tau^m(w)$ is a functional term; clearly, $B_m\sigma' = B'_m\tau^m$. But then, $\mathsf{Move}(y_i)$ covers $\mathsf{In}(x)$; hence, by the definition of $\mathsf{Move}$, we have that $\langle H^3, j \rangle \in \mathsf{Move}(y_i)$, so property ($\blacklozenge$) holds.

We additionally prove the following property ($\lozenge$): if $\mathsf{S}(\mathsf{c}^i_r, \mathsf{c}^{i'}_{r'}) \in I^\infty$ for some $i$ and $i'$, then $SWA(\Sigma)$ contains an edge from $r$ to $r'$. Consider an arbitrary such fact, let $y_i$ be the existentially quantified variable of $r$ corresponding to $\mathsf{c}^i_r$, and let $k$ be the smallest integer such that $\mathsf{S}(\mathsf{c}^i_r, \mathsf{c}^{i'}_{r'}) \in I^k$. Clearly, $\mathsf{S}(\mathsf{c}^i_r, \mathsf{c}^{i'}_{r'})$ is derived in $I^k$ from the head atom $\mathsf{S}(x, \mathsf{c}^{i'}_{r'})$ of rule $r'$. Let $\sigma$ be the substitution used in the rule application; thus, $\sigma(x) = \mathsf{c}^i_r$. Let $B_1, \ldots, B_n$ be the body atoms of $r$ that contain $x$; clearly, we have $\{B_1, \ldots, B_n\sigma\} \subseteq I^{k-1}$. All these atoms satisfy property ($\blacklozenge$), so for each $B_m \in \{B_1, \ldots, B_n\}$ and each $\ell$ such that $B_m$ contains variable $x$ at position $\ell$, a place $\langle B'_m, \ell \rangle \in \mathsf{Move}(y_i)$ and substitution $\tau^m$ exist such that $B_m\sigma = \rho(B'_m\tau^m)$. But then, as in the previous paragraph we have that $\mathsf{Move}(y_i)$ covers $\mathsf{In}(x)$, so $SWA(\Sigma)$ contains an edge from $r$ to $r'$.

Assume now that $\Sigma$ is not MSA, so $\mathsf{C} \in I^\infty$; then $\{\mathsf{F}^i_r(t), \mathsf{D}(t, t'), \mathsf{F}^i_r(t')\} \subseteq I^\infty$ holds for some $\mathsf{F}^i_r$ due to rules (20). But then, since predicate $\mathsf{F}^i_r$ occurs in $\Sigma'$ only in an atom $\mathsf{F}^i_r(\mathsf{c}^i_r)$, we have $t = t' = \mathsf{c}^i_r$. Finally, since $\mathsf{D}$ is axiomatised in $\Sigma'$ as the transitive closure of $\mathsf{S}$, clearly $SWA(\Sigma)$ contains a path from $r$ to itself, and so $\Sigma$ is not SWA. $\qquad\square$

The rule set in Example 1 is MSA but not SWA. Furthermore, it is known that SWA is more general than JA, and the two notions differ only if at least one rule contains a body atom in which at least one variable occurs more than once (Krötzsch & Rudolph, 2013). The following example shows that SWA is strictly more general than JA.

**Example 20.** *Let $\Sigma$ be the set of the following rules:*

$$r_1 = \qquad\qquad A(x_1) \rightarrow \exists y.R(x_1, y) \wedge R(y, x_1) \wedge R(x_1, x_1) \qquad\qquad (39)$$

$$r_2 = \qquad\qquad R(x_2, x_2) \rightarrow B(x_2) \tag{40}$$

$$r_3 = \qquad\qquad B(x_3) \rightarrow A(x_3) \tag{41}$$

*One can readily verify that $\Sigma$ is SWA, but not JA.* $\diamond$

Theorem 19 holds even if $\Sigma$ contains the equality predicate, but provided that the axiomatisation of equality (cf. Section 2) is taken as part of the input. On such rule sets, however, SWA, JA, MSA, and MFA are not strictly more general than WA. We discuss the underlying problems, as well as possible solutions, in Section 5.

### 4.1.2 Acyclicity by Rewriting

Spezzano and Greco (2010) proposed an acyclicity notion called *Adn-WA*. Roughly speaking, one first rewrites a set of rules $\Sigma$ into another set of rules $\Sigma'$ by adorning the positions in the predicates that can contain infinitely many terms during the chase; then, one checks whether $\Sigma'$ is WA. The rewriting algorithm is rather involved, so we do not recapitulate its definition; instead, we discuss it by means of an example. Spezzano and Greco used this example to show that Adn-WA is not subsumed by SWA, but the same example also shows that Adn-WA is not subsumed by MFA either.

**Example 21.** *Let $\Sigma$ be the set containing the following rules:*

$$A(x) \rightarrow \exists y.R(x, y) \tag{42}$$

$$B(z) \land R(z, x) \rightarrow A(x) \tag{43}$$

*The transformation by Spezzano and Greco (2010) produces a set $\Sigma'$ that consists of three groups of rules. The first group contains rules (44)–(47).*

$$A^b(x) \rightarrow \exists y.R^{bf}(x, y) \tag{44}$$

$$B^b(z) \land R^{bb}(z, x) \rightarrow A^b(x) \tag{45}$$

$$B^b(z) \land R^{bf}(z, x) \rightarrow A^f(x) \tag{46}$$

$$A^f(x) \rightarrow \exists y.R^{ff}(x, y) \tag{47}$$

*For each $n$-ary predicate $P$, the transformation introduces predicates of the form $P^m$, where $m$ is an* adornment—*a string of length $n$ of letters $b$ or $f$. Intuitively, if $m$ contains letter $b$ at position $i$, then during the chase construction the $i$-th position of $P^m$ can contain only constants occurring in an instance. Rules (44)–(47) were derived as follows. Rule (44) is obtained from rule (42) by marking all positions of variable $x$ with $b$, which effectively creates a variant of the rule whose body is applicable only to constants. Variable $y$ in the head of rule (44) occurs under an existential quantifier, so the corresponding position is marked with $f$. Rule (45) is obtained from rule (43) in an analogous way. But then, since facts introduced by rule (44) can trigger an application of rule (43), the latter rule is marked as rule (46); predicate $A^f$ in the head of rule (46) reflects the fact that variable $x$ in the rule body is instantiated by atom $R^{bf}(z, x)$. Finally, facts derived by rule (46) can trigger an application of rule (42), so the latter rule is instantiated as (47). At this point the algorithm terminates: since no rule was instantiated with a marking $B^f$ in the head, it is not possible to use predicate $R^{ff}$ to mark the body of rule (43) in a consistent way.*

*The second group consists of rules (48)–(50), which 'populate' the adorned predicates with the contents of an instance.*

$$R(x_1, x_2) \to R^{bb}(x_1, x_2) \tag{48}$$

$$A(x) \to A^b(x) \tag{49}$$

$$B(x) \to B^b(x) \tag{50}$$

*The third group consists of rules (51)–(56), which 'gather' the content of each adorned predicate $P^m$ into a fresh output predicates $\hat{P}$.*

$$R^{bb}(x_1, x_2) \to \hat{R}(x_1, x_2) \tag{51}$$

$$R^{bf}(x_1, x_2) \to \hat{R}(x_1, x_2) \tag{52}$$

$$R^{ff}(x_1, x_2) \to \hat{R}(x_1, x_2) \tag{53}$$

$$A^b(x) \to \hat{A}(x) \tag{54}$$

$$A^f(x) \to \hat{A}(x) \tag{55}$$

$$B^b(x) \to \hat{B}(x) \tag{56}$$

*It is straightforward to check that $\Sigma$ is not MFA. In contrast, $\Sigma'$ is WA; furthermore, Spezzano and Greco (2010) show that, for each instance $I$ and each vector of ground terms $\vec{t}$, we have $\hat{P}(\vec{t}) \in I_{\Sigma'}^{\infty}$ if and only if $P(\vec{t}) \in I_{\Sigma}^{\infty}$. Since $\Sigma'$ is WA, $I_{\Sigma'}^{\infty}$ is finite, and, by the previously mentioned property, $I_{\Sigma}^{\infty}$ is finite as well.* $\diamond$

The following example shows that MFA is not subsumed by Adn-WA, which indicates that MFA and Adn-WA are incomparable.

**Example 22.** *Let $\Sigma$ be the set containing the following rules:*

$$r_1 = \qquad A(x) \to \exists y. R(x, y) \wedge B(y) \tag{57}$$

$$r_2 = \qquad S(z, x) \wedge B(x) \to \exists y. S(x, y) \tag{58}$$

*The rules in the first group of the set $\Sigma'$ obtained by the transformation are shown below; we do not show the rules in the second and the third group for the sake of brevity.*

$$A^b(x) \to \exists y. R^{bf}(x, y) \wedge B^f(y) \tag{59}$$

$$S^{bb}(z, x) \wedge B^b(x) \to \exists y. S^{bf}(x, y) \tag{60}$$

$$S^{bf}(z, x) \wedge B^f(x) \to \exists y. S^{ff}(x, y) \tag{61}$$

$$S^{ff}(z, x) \wedge B^f(x) \to \exists y. S^{ff}(x, y) \tag{62}$$

*The last rule ensures that the WA dependency graph for $\Sigma'$ contains a special edge from position $S^{ff}|_2$ to itself; thus, $\Sigma'$ is not WA, and therefore $\Sigma$ is not Adn-WA. In contrast, one can readily verify that $\Sigma$ is MFA.* $\diamond$

Spezzano and Greco (2010) also proposed several optimisations of this transformation, the discussion of which is out of scope of this paper. All of them can be seen as 'unfolding' the rules in $\Sigma$ up to a certain number of chase steps. This seems close to an idea by Baget et al. (2011b), who propose to run the chase for some fixed number of steps before checking for potential cycles. A similar effect could be obtained by extending the notion of MFA to check for terms that contain a function symbol nested some fixed number of times.

Finally, note that the transformation by Spezzano and Greco (2010) is independent from the notion used to check the acyclicity of the transformed rule set; hence, given an arbitrary acyclicity notion $X$, one can define Adn-$X$ in the obvious way. Given arbitrary notions $X$ and $Y$ such that $X \subseteq Y$, it is obvious that Adn-$X \subseteq$ Adn-$Y$; consequently, we have Adn-$X \not\subseteq$ MFA for each $X$ such that WA $\subseteq X$. In contrast, however, it not obvious whether the inclusion between Adn-$X$ and Adn-$Y$ is strict whenever the inclusion between $X$ and $Y$ is strict, or whether MFA is contained in Adn-$X$ for some $X$ with WA $\subseteq X$. Finally, we conjecture that $X \subseteq$ Adn-$X$ holds for an arbitrary notion $X$, but we do not have a formal proof of this conjecture. Due to the complex nature of the rewriting, we refrain from further analysis of these relationships.

### 4.1.3 Monitor Graph

Meier et al. (2009) propose an idea that is similar in spirit to MFA. The idea is to track each chase step in an additional data structure called the *monitor graph*. If the chase is infinite, then the monitor graph contains cycles of arbitrary length; conversely, if one can show that the monitor graph does not contain a cycle of some fixed length, then the chase is guaranteed to terminate. While this idea is closely related to MFA, note that the definition of MFA is semantic; hence, one can use an arbitrary theorem proving technique to check whether $\mathsf{MFA}(\Sigma) \models \mathsf{C}$. In contrast, the notion of a monitor graph is specifically tied to the nonoblivious chase. It is well known that the result of the nonoblivious chase depends on the order in which the rules applied; consequently, a set of rules can be identified as cyclic or acyclic depending on the selected rule application strategy. Because of this dependence, it is difficult to compare the monitor graph approach with other acyclicity notions.

## 4.2 Acyclicity in Knowledge Representation

Existential rules can capture knowledge representation formalisms such as Horn fragments of description logics (see Section 6), conceptual graphs (Baget, 2004; Baget et al., 2011a), and datalog$^\pm$ rules (Calì et al., 2010a), and so acyclicity notions allow for materialisation-based query answering over knowledge bases. In this context, Baget (2004) and Baget et al. (2011a) proposed the notion of *acyclic graph rule dependencies* (aGRD). Intuitively, aGRD introduces a rule dependency relation $\prec$ for which $r_1 \prec r_2$ means that an application of rule $r_1$ on an instance $I$ can subsequently trigger an application of rule $r_2$. If the relation $\prec$ is acyclic, then no rule can trigger itself so the skolem chase terminates on an arbitrary instance. This can be formalised as follows.

**Definition 23.** *The* rule dependency relation $\prec \subseteq \Sigma \times \Sigma$ *on a set of rules* $\Sigma$ *is defined as follows. Let* $r_1 = \varphi_1 \to \exists \vec{y}_1.\psi_1$ *and* $r_2 = \varphi_2 \to \exists \vec{y}_2.\psi_2$ *be arbitrary rules in* $\Sigma$, *and let* $\mathsf{sk}(r_1) = \varphi_1 \to \psi_1'$ *and* $\mathsf{sk}(r_2) = \varphi_2 \to \psi_2'$. *Then,* $r_1 \prec r_2$ *if and only if there exist an instance* $I$, *a substitution* $\sigma_1$ *for all variables in* $\mathsf{sk}(r_1)$, *and a substitution* $\sigma_2$ *for all variables in* $\mathsf{sk}(r_2)$ *such that* $\varphi_1 \sigma_1 \subseteq I$, $\varphi_2 \sigma_2 \not\subseteq I$, $\varphi_2 \sigma_2 \subseteq I \cup \psi_1' \sigma_1$, *and* $\psi_2' \sigma_2 \not\subseteq I \cup \psi_1' \sigma_1$. *Set* $\Sigma$ *has an* acyclic graph of rules dependencies *(aGRD) if the relation* $\prec$ *on* $\Sigma$ *is acyclic.*

Definition 23 differs from the original definition by Baget (2004) in several ways. First, Baget uses fresh nulls to capture the effect of existential quantifiers, whereas Definition 23 uses skolem functions; however, this does not change the resulting relation $\prec$ in any way. Second, Baget does not require $\psi_2'\sigma_2 \not\subseteq I \cup \psi_1'\sigma_1$. This condition intuitively ensures that an application of $r_1$ to $I$ enables $r_2$ to derive something new; analogous optimisations were proposed by Deutsch et al. (2008) and Greco et al. (2012). It should be clear that Definition 23 is stronger than the one by Baget. To unify the notions used in various parts of this paper, we included this optimisation into Definition 23; however, we nevertheless call the resulting stronger notion aGRD.

The following example shows that aGRD, even in its weaker form as originally proposed by Baget (2004), is not contained in SWA.

**Example 24.** *Let $\Sigma$ be the set consisting of the following rule:*

$$r = \qquad A(z_1, x, z_2) \wedge B(z_2) \to \exists y_1 \exists y_2. A(x, y_1, y_2) \qquad (63)$$

*To see that $r \prec r$ does not hold, consider the skolemisation $r'$ of $r$:*

$$\mathsf{sk}(r) = \qquad A(z_1, x, z_2) \wedge B(z_2) \to A(x, f_1(x), f_2(x)) \qquad (64)$$

*Now let $I$ be an arbitrary instance, and let $\sigma_1$ and $\sigma_2$ be arbitrary substitutions such that $\{A(z_1, x, z_2)\sigma_1, B(z_2)\sigma_1\} \subseteq I$ and $\{A(z_1, x, z_2)\sigma_2, B(z_2)\sigma_2\} \not\subseteq I$. Since instance $I$ contains only constants, atom $A(x, f_1(x), f_2(x))\sigma_1$ is of the form $A(a, f_1(a), f_2(a))$; but then, for $\{A(z_1, x, z_2)\sigma_2, B(z_2)\sigma_2\} \subseteq I \cup \{A(a, f_1(a), f_2(a))\}$ to hold, it must be that $\sigma_2(z_2) = f_2(a)$; thus, $B(z_2)\sigma_2 = B(f_2(a))$ should be contained in $I$, which is impossible since $I$ is an instance and thus does not contain functional terms. Note that the additional condition by Greco et al. (2012) plays no role here. Thus, we have $r \not\prec r$, so $\Sigma$ is aGRD even in the weaker form by Baget (2004). However, one can easily check that $\Sigma$ is not SWA.* ◇

However, aGRD seems to be a rather weak notion: as the following example shows, even a set of rules without existential quantifiers can be cyclic according to this criterion.

**Example 25.** *Let $\Sigma$ be the set consisting of the following rules:*

$$r_1 = \qquad A(x) \to B(x) \qquad (65)$$

$$r_2 = \qquad B(x) \to C(x) \qquad (66)$$

$$r_3 = \qquad C(x) \to A(x) \qquad (67)$$

*To see that $r_1 \prec r_2$, let $I = \{A(a)\}$, let $\sigma = \{x \mapsto a\}$, and note that $A(x)\sigma \in I$, $B(x)\sigma \notin I$, $B(x)\sigma \in I \cup \{B(x)\sigma\}$, and $C(x)\sigma \notin I \cup \{B(x)\sigma\}$. Analogously, by taking $I = \{B(a)\}$ we get $r_2 \prec r_3$, and by taking $I = \{C(a)\}$ we get $r_3 \prec r_1$. Consequently, $\Sigma$ is not aGRD. However, $\Sigma$ is obviously WA since it does not contain existentially quantified variables.* ◇

Baget et al. (2011a) suggested that rule dependencies become more powerful if they are combined with an arbitrary acyclicity notion $X$. Intuitively, the main idea is to use $\prec$ to partition a set of rules into strongly connected components, and then check whether each component is $X$; we call this notion $X^\prec$. This idea can be formalised as follows.

**Definition 26.** *Let $\Sigma$ be a set of existential rules, and let $\prec$ be the rule dependency relation on $\Sigma$. Relation $\prec$ is extended to arbitrary sets $C \subseteq \Sigma$ and $C' \subseteq \Sigma$ such that $C \prec C'$ if and only if rules $r \in C$ and $r' \in C'$ exist such that $r \prec r'$. A dependency partition of $\Sigma$ is a sequence of sets $\Sigma_1, \ldots, \Sigma_n$ such that $\Sigma = \bigcup_{i=1}^n \Sigma_i$, each $\Sigma_i$ is a strongly connected component of $\prec$, and $\Sigma_j \not\prec \Sigma_i$ for all $i$ and $j$ such that $1 \leq i < j \leq n$.*

*Let $X$ be an arbitrary acyclicity notion. Then, $\Sigma \in X^\prec$ if a dependency partition $\Sigma_1, \ldots, \Sigma_n$ of $\Sigma$ exists such that, for each $1 \leq i \leq n$, we have $\Sigma_i \in X$, or $\Sigma_i$ consists of a single rule $r_i$ such that $r_i \not\prec r_i$.*

If $\Sigma$ is aGRD, then each strongly connected component $\Sigma_i$ contains a single rule $r_i$ such that $r_i \not\prec r_i$. Now if Definition 26 did not consider the special case where $\Sigma_i$ consists of a single rule that does not depend on itself, then SWA$^\prec$ would not extend aGRD; for example, the rule in Example 24 would not be in SWA$^\prec$. The extra condition in Definition 26 thus ensures that aGRD is contained in $X^\prec$ regardless of the choice of $X$, and that aGRD can be understood as $\emptyset^\prec$—the acyclicity notion obtained by extending the empty notion (i.e., the notion under which no rule set is acyclic) with rule dependencies.

We next present two simple results. Proposition 27 precludes inclusions between certain acyclicity notions and will thus help us establish proper inclusions between many acyclicity notions. Furthermore, Proposition 28 shows that combining an acyclicity notion contained in SWA with rule dependencies creates a strictly stronger acyclicity notion; note that this holds even for the weaker form of rule dependencies originally proposed by Baget (2004).

**Proposition 27.** *Let $X$ and $Y$ be acyclicity notions such that $X \subseteq Y$. Then, $X^\prec \subseteq Y^\prec$. Furthermore, if there exists a set $\Sigma \in Y \setminus X$ whose rule dependency relation has a cycle containing all the rules from $\Sigma$, then $Y \nsubseteq X^\prec$, $Y^\prec \nsubseteq X^\prec$, and $X^\prec \subsetneq Y^\prec$.*

*Proof.* Relationship $X^\prec \subseteq Y^\prec$ is immediate from Definition 26. Assume now that there exists a set of rules $\Sigma \in Y \setminus X$ whose rule dependency relation has a cycle containing all the rules from $\Sigma$. By Definition 26, $\Sigma \notin X$ implies $\Sigma \notin X^\prec$, and $\Sigma \in Y$ implies $\Sigma \in Y^\prec$. But then, we clearly have $Y \nsubseteq X^\prec$ and $Y^\prec \nsubseteq X^\prec$, and the latter clearly implies $X^\prec \subsetneq Y^\prec$. $\square$

**Proposition 28.** *For each acyclicity notion $X$ such that $X \subseteq SWA$, we have $X \subsetneq X^\prec$ and $aGRD \nsubseteq X$.*

*Proof.* Set $\Sigma$ from Example 24 is in aGRD and thus in $X^\prec$; however, $\Sigma$ is not in SWA and hence not in $X$ either. $\square$

MSA also does not contain aGRD; however, unlike for SWA, our claim depends on the optimisation in Definition 23. An analysis of the relationship between MSA and the version of rule dependencies originally proposed by Baget (2004) is out of scope of this paper.

**Example 29.** *Let $\Sigma$ be the set consisting of the following rules:*

$$r_1 = \qquad\qquad R(x_1, x_1) \wedge U(x_1, z) \wedge U(x_2, z) \rightarrow R(x_1, x_2) \qquad (68)$$

$$r_2 = \qquad\qquad\qquad\qquad\qquad R(z, x) \rightarrow \exists y.T(x, y) \qquad (69)$$

$$r_3 = \qquad\qquad\qquad\qquad\qquad T(z, x) \rightarrow \exists y.U(x, y) \qquad (70)$$

It is obvious that $r_1 \prec r_2$, $r_1 \not\prec r_3$, $r_2 \not\prec r_1$, $r_2 \not\prec r_2$, $r_2 \prec r_3$, $r_3 \not\prec r_2$, and $r_3 \not\prec r_3$. We next argue that $r_1 \not\prec r_1$ and $r_3 \not\prec r_1$, which implies that $\Sigma$ is aGRD.

To see that $r_1 \not\prec r_1$, assume that an application of $r_1$ to an instance $I$ produces an atom of the form $R(a, b)$; due to atom $R(x_1, x_1)$ in the body of $r_1$, we have $R(a, a) \in I$. Now let $I' = I \cup \{R(a, b)\}$; since $R(a, a) \in I$, the rule application derives 'something new' only if $a \neq b$. Now assume that a substitution $\sigma_2$ exists that makes $r_1$ applicable to $I'$ but not to $I$; this rule application must 'use' the fact $R(a, b)$, which implies that $R(x_1, x_1)\sigma_2 = R(a, b)$; however, this is impossible since $a \neq b$. Consequently, we have $r_1 \not\prec r_1$, and this holds even for the version of rule dependencies by Baget (2004).

Furthermore, to see that $r_3 \not\prec r_1$, assume that $r_3$ is applicable to an instance $I$, and that the rule application derives a fact of the form $U(a, f(a))$. Now let $I' = I \cup \{U(a, f(a))\}$, and assume that a substitution $\sigma_2$ exists that makes $r_1$ applicable to $I'$ but not to $I$; this rule application must 'use' the fact $U(a, f(a))$, which implies that $\sigma_2(x_1) = \sigma_2(x_2) = a$ and $\sigma_2(z) = f(a)$. Furthermore, rule $r_1$ is applicable only if $R(a, a) \in I$; but then, the rule application does not derive 'something new' since $R(x_1, x_2)\sigma_2 = R(a, a)$. Consequently, we have $r_3 \not\prec r_1$; however, unlike in the previous paragraph, this claim depends on the optimisation in Definition 23.

Consider now the chase of $I_\Sigma^*$ and $\mathsf{MSA}(\Sigma)$ as shown below (facts involving the predicates $\mathsf{D}$, $\mathsf{F}_{r_2}$, and $\mathsf{F}_{r_3}$ are omitted for clarity). The chase result contains $\mathsf{C}$, so $\Sigma$ is not in MSA, and thus aGRD $\not\subseteq$ MSA; as a corollary, we also get $MSA \subsetneq MSA^\prec$.

| | | | | | |
|---|---|---|---|---|---|
| $R(*, *)$ | $T(*, \mathsf{c}_{r_2})$ | $U(\mathsf{c}_{r_2}, \mathsf{c}_{r_3})$ | $R(*, \mathsf{c}_{r_2})$ | $T(\mathsf{c}_{r_2}, \mathsf{c}_{r_2})$ | $\mathsf{C}$ |
| $T(*, *)$ | $U(*, \mathsf{c}_{r_3})$ | $S(\mathsf{c}_{r_2}, \mathsf{c}_{r_3})$ | | $S(\mathsf{c}_{r_2}, \mathsf{c}_{r_2})$ | |
| $U(*, *)$ | $S(*, \mathsf{c}_{r_2})$ | | | | |
| | $S(*, \mathsf{c}_{r_3})$ | | | | |

Note that $R(*, \mathsf{c}_{r_2})$ is derived from $R(*, *)$, $U(*, \mathsf{c}_{r_2})$, and $U(\mathsf{c}_{r_2}, \mathsf{c}_{r_3})$, where the latter two facts are obtained from distinct instantiations of $\mathsf{MSA}(r_3)$. Rule dependencies, however, analyse rule applicability w.r.t. $\mathsf{sk}(r_3)$, which is closer to the actual skolem chase. ◇

In contrast to this result, in Theorem 32 we will show that extending MFA with rule dependencies does not create a stronger notion: MFA$^\prec$ coincides with MFA, which implies that $X^\prec \subseteq$ MFA for each notion $X$ such that $X \subseteq$ MFA. Towards this goal, we show in Lemma 30 that independent rule sets can be evaluated independently, and in Lemma 31 that a single rule that does not depend on itself can be applied only once.

**Lemma 30.** *Let $\Sigma_1$ and $\Sigma_2$ be sets of existential rules such that $\Sigma_2 \not\prec \Sigma_1$, and let $F$ be a set of ground facts not containing a function symbol in $\mathsf{sk}(\Sigma_2)$. Then, $F_{\Sigma_1 \cup \Sigma_2}^\infty = (F_{\Sigma_1}^\infty)_{\Sigma_2}^\infty$.*

*Proof.* Let $F_0 = F_{\Sigma_1}^\infty$; let $F_0, F_1, \ldots$ be the chase sequence for $F_0$ and $\Sigma_2$ where, for convenience, we assume each $F_i$ to be obtained from $F_{i-1}$ by a single rule application (this assumption is clearly w.l.o.g.); and let $F' = (F_0)_{\Sigma_2}^\infty$. By the definition of the skolem chase, we clearly have $F' \subseteq F_{\Sigma_1 \cup \Sigma_2}^\infty$. Furthermore, assume that $F_{\Sigma_1 \cup \Sigma_2}^\infty \not\subseteq F'$; then, a skolemised rule $r_1 \in \mathsf{sk}(\Sigma_1)$ of the form $r_1 = \varphi_1(\vec{x}_1) \to \psi_1(\vec{x}_1)$ exists such that $F' \subsetneq r_1(F')$. Fix the smallest $i$ such that $F_i \subsetneq r_1(F_i)$ (we clearly have $i > 0$), and let $\sigma_1$ be the substitution used in the application of $r_1$. Furthermore, let $r_2 \in \mathsf{sk}(\Sigma_2)$ be the skolemised rule of the form $r_2 = \varphi_2(\vec{x}_2) \to \psi_2(\vec{x}_2)$ that is used to derive $F_i$ from $F_{i-1}$, and let $\sigma_2$ be the substitution used in the application of $r_2$. Now consider an arbitrary term $f(\vec{x}_2)$ in the head of $r_2$ and assume that $f(\vec{x}_2)\sigma_2$ occurs in $F_{i-1}$; since the function symbol $f$ is 'private' to $r_2$, the head of $r_2$ must have been already instantiated for $\sigma_2$; but then, $\psi_2\sigma_2 \subseteq F_{i-1}$, which contradicts our assumption that $\psi_2\sigma_2 \in F_i \setminus F_{i-1}$. Thus, we have the following property ($\star$):

for each term $f(\vec{x}_2)$ occurring in the head of $r_2$, ground term $f(\vec{x}_2)\sigma_2$ does not occur in $F_{i-1}$.

Finally, let $\delta$ be a function that maps each ground term in $F_{i-1}$ to a fresh distinct constant; let $I = \delta(F_{i-1})$; let $\sigma'_2$ be the substitution defined by $\sigma'_2(w) = \delta(\sigma_2(w))$ for each variable $w$ in $r_2$; and let $\sigma'_1$ be the substitution defined as follows for each variable $w$ in $r_1$:

- $\sigma'_1(w) = f(\delta(\vec{t}))$ if $\sigma_1(w) = f(\vec{t})$ for $f$ a function symbol 'private' to $r_2$; and

- $\sigma'_1(w) = \delta(\sigma_1(w))$ otherwise.

We clearly have $\varphi_2\sigma'_2 \subseteq I$ and $\psi_2\sigma'_2 \not\subseteq I$; furthermore, by ($\star$), we also have $\varphi_1\sigma'_1 \subseteq I \cup \psi_2\sigma'_2$ and $\psi_1\sigma'_1 \not\subseteq I \cup \psi_2\sigma'_2$. Moreover, $\varphi_1\sigma'_1 \not\subseteq I$ follows from our assumption that $i$ is the smallest integer such that $F_i \subsetneq r_1(F_i)$. But then, by Definition 23, we have $r_2 \prec r_1$ and, consequently, $\Sigma_2 \prec \Sigma_1$ as well, which is a contradiction. $\qquad\square$

**Lemma 31.** *Let $\Sigma = \{r\}$ be a singleton rule set such that $r \not\prec r$, and let $F$ be a set of facts not containing a function symbol in $\mathsf{sk}(\Sigma)$. Then, $F^\infty_\Sigma = \Sigma(F)$.*

*Proof.* Let $F_0 = F$, and let $F_0, F_1, \dots$ be sets of facts such that each $F_{i+1}$ is the union of $F_i$ with the result of a distinct single application of $r$ to $F_0$; clearly, $\bigcup_i F_i = \Sigma(F_0)$. Now assume that $\bigcup_i F_i \subsetneq \Sigma(\bigcup_i F_i)$; then analogously to the proof of Lemma 30, one can show that $r \prec r$, which is a contradiction; we omit the details for the sake of brevity. $\qquad\square$

**Theorem 32.** *Let $\Sigma$ be an arbitrary set of rules and let $I$ be an arbitrary instance. If $\Sigma$ is $\mathrm{MFA}^\prec$ w.r.t. $I$, then $\Sigma$ is also MFA w.r.t. $I$.*

*Proof.* Assume that $\Sigma$ is in $\mathrm{MFA}^\prec$; let $I$ be an arbitrary instance; let $\Sigma_1, \dots, \Sigma_n$ be a dependency partition of $\Sigma$; let $\Upsilon_0 = \emptyset$ and $I_0 = I$; and, for each $1 \le i \le n$, let $\Upsilon_i = \bigcup^i_{\ell=1} \Sigma_\ell$ and $I_i = (I_{i-1})^\infty_{\Sigma_i}$. By the definition of dependency partitions, we have that $\Sigma_i \not\prec \Upsilon_{i-1}$ holds for each $1 \le i \le n$. We next show that, for each $0 \le i \le n$, the following two properties hold:

(a) $I_i = (I_0)^\infty_{\Upsilon_i}$, and

(b) $I_i$ does not contain a cyclic term.

Set $I_0$ does not contain functional terms and hence it trivially satisfies (a) and (b). Now consider arbitrary $0 < i < n$ such that $I_{i-1}$ satisfies (a) and (b). By the induction assumption, Lemma 30, $\Sigma_i \not\prec \Upsilon_{i-1}$, and $\Upsilon_i = \Sigma_i \cup \Upsilon_{i-1}$, we have that $(I_0)^\infty_{\Upsilon_i} = ((I_0)^\infty_{\Upsilon_{i-1}})^\infty_{\Sigma_i}$; thus, $I_i$ satisfies (a). To see that $I_i$ satisfies (b), note that no function symbol used to skolemise the rules in $\Sigma_i$ is used to skolemise the rules in $\Upsilon_{i-1}$; we call this property ($\star$). Now there are two ways to compute $I_i$.

- Assume that $\Sigma_i = \{r_i\}$ such that $r_i \not\prec r_i$. By Lemma 31, we have $I_i = r_i(I_{i-1})$; but then, $I_i$ does not contain a cyclic term due to ($\star$).

- If $\Sigma_i$ is MFA, then $I_i$ does not contain a cyclic term due to ($\star$) and Proposition 5.

From the above claim we have that $I_n = I_{\Upsilon_n}^\infty = I_\Sigma^\infty$ and that $I_n$ does not contain a cyclic term; but then, $\Sigma$ is MFA w.r.t. $I$ by Proposition 5. $\qquad\square$

Combinations of rule dependencies with acyclicity notions have also been considered in databases: Deutsch et al. (2008) proposed a notion of *stratification*, and Meier et al. (2009) further developed this idea and proposed a notion of *c-stratification*. Roughly speaking, each such notion checks whether all strongly connected components of a certain rule dependency graph are WA. The rule dependency notions, however, were developed for the nonoblivious chase and are thus different from Definition 23, as illustrated by the following rule:

$$r = \qquad\qquad R(z, x) \to \exists y. R(x, y) \wedge R(y, y) \qquad\qquad (71)$$

The skolem chase on the critical instance for $r$ is infinite, and $r \prec r$ by Definition 23. In contrast, rule $r$ does not pose problems for the nonoblivious chase. In particular, assume that the rule is matched to an atom $R(t_1, t_2)$, and that it derives $R(t_2, t_3)$ and $R(t_3, t_3)$. Then, rule $r$ is not applicable to $R(t_2, t_3)$ or $R(t_3, t_3)$ since in either case the head atom is satisfied; hence, the rule dependency graphs by Deutsch et al. and Meier et al. are both empty. These results can be summarised as follows: if a rule set $\Sigma$ satisfies the notion by Deutsch et al., then for each instance $I$ there exists a finite nonoblivious chase sequence; furthermore, if $\Sigma$ satisfies the notion by Meier et al., then for each instance $I$ *all* chase sequences (regardless of the rule application strategy) are finite. Meier (2010) discusses in detail the subtle differences between these notions. Since these notions consider a different chase variant, we do not discuss them any further in this paper.

### 4.3 Acyclicity and Logic Programming

Acyclicity notions have also been considered in the context of disjunctive logic programs with function symbols under the answer set semantics, with the goal of ensuring that a given program has finitely many answer sets, all of which are finite. All of these notions must deal with disjunction and nonmonotonic negation, which is one of the main differences to the notions considered thus far. All notions from logic programming, however, are applicable to rules without disjunction and nonmonotonic negation, in which case they ensure termination of the skolem chase. Therefore, in this section we compare such specialisations of the acyclicity notions from logic programming with aGRD, WA, JA, SWA, MSA, and MFA. We simplify all definitions so that they apply only to skolemised existential rules—that is, we do not present parts of definitions that handle disjunctions in the head and nonmonotonic negation and function symbols in the body.

#### 4.3.1 FINITE DOMAIN NOTION

Calimeri et al. (2008) proposed a finite domain (FD) notion. We next recapitulate this definition, but we do so in the style of Greco et al. (2012), which will come useful in Section 4.3.3 when we introduce $\Gamma$-acyclicity. Both approaches use an *argument graph* to determine possible ways for propagating ground terms between positions during chase. The definition of the argument graph is the same as that of the WA dependency graph (see Section 2.4), but without the distinction between regular and special edges. To simplify the presentation, we consistently use the WA dependency graph instead of the argument graph.

**Definition 33.** *Let $\Sigma$ be a set of rules. A position $P|_i$ is $\Sigma$-recursive with a position $Q|_j$ if the WA dependency graph $WA(\Sigma)$ contains a cycle (consisting of regular and/or special edges) going through $P|_i$ and $Q|_j$. The set $\mathsf{Pos}_{FD}(\Sigma)$ of finite domain positions of $\Sigma$ is the largest set of positions in $\Sigma$ such that, for each position $P|_i \in \mathsf{Pos}_{FD}(\Sigma)$, each rule $r \in \Sigma$ of the form $r = \varphi(\vec{x}, \vec{z}) \to \exists \vec{y}.\psi(\vec{x}, \vec{y})$, and each head atom of $r$ of the form $P(\vec{t})$, the following conditions are satisfied:*

- *if the $i$-th component of $\vec{t}$ is a variable $x \in \vec{x}$, then $\mathsf{Pos}_B(x) \cap \mathsf{Pos}_{FD}(\Sigma) \neq \emptyset$; and*

- *if the $i$-th component of $\vec{t}$ is a variable $y \in \vec{y}$, then, for each variable $x \in \vec{x}$, some position $Q|_j \in \mathsf{Pos}_B(x) \cap \mathsf{Pos}_{FD}(\Sigma)$ exists that is not $\Sigma$-recursive with $P|_i$.*

*Set $\Sigma$ is FD if $\mathsf{Pos}_{FD}(\Sigma)$ coincides with the set of all positions in $\Sigma$.*

Note that the notion of $\Sigma$-recursive positions introduced above is symmetric: if $P|_i$ is $\Sigma$-recursive with $Q|_j$, then $Q|_j$ is also $\Sigma$-recursive with $P|_i$. Furthermore, note that Calimeri et al. (2008) defined FD as follows:

A set of rules $\Sigma$ is FD if, for each rule $r = \varphi(\vec{x}, \vec{z}) \to \exists \vec{y}.\psi(\vec{x}, \vec{y})$ in $\Sigma$, each atom $Q(\vec{t})$ in the head of $r$, each $j$-th term of $\vec{t}$ that is an existential variable $y$, and each variable $x \in \vec{x}$, there exists a position $P|_i \in \mathsf{Pos}_B(x)$ such that $Q|_j$ is not $\Sigma$-recursive with $P|_i$.

Conditions in the above definition clearly correspond to the conditions in Definition 33; but then, since $\mathsf{Pos}_{FD}(\Sigma)$ was defined as the maximal set satisfying these conditions, the two definitions of FD coincide.

We next show that WA is strictly contained in FD. To this end, we first prove that WA is contained in FD, and then we present an example showing that the inclusion is strict.

**Proposition 34.** *If a set of rules $\Sigma$ is WA, then $\Sigma$ is FD.*

*Proof.* Let $\Sigma$ be a set of rules that is not FD. Then, there exist a rule $r \in \Sigma$, an atom $Q(\vec{t})$ in the head of $r$, a $j$-th term of $\vec{t}$ equal to an existential variable $y$, and a variable $x \in \vec{x}$ such that each position $P|_i \in \mathsf{Pos}_B(x)$ is $\Sigma$-recursive with $Q|_j$. The set $\mathsf{Pos}_B(x)$ is not empty ($\vec{x}$ contains precisely those variables occurring both in the body and the head of the rule), so choose an arbitrary position $P|_i \in \mathsf{Pos}_B(x)$. The WA dependency graph $WA(\Sigma)$ then contains a special edge from $P|_i$ to $Q|_j$. Furthermore, since $Q|_j$ is $\Sigma$-recursive with $P|_i$, graph $WA(\Sigma)$ contains a cycle going through $P|_i$ and $Q|_j$. Thus, $WA(\Sigma)$ clearly contains a cycle containing a special edge, so $\Sigma$ is not WA. $\square$

**Example 35.** *Let $\Sigma$ be the set containing rules (72) and (73).*

$$r_1 = \qquad\qquad R(z,x) \wedge A(x) \to \exists y.S(x,y) \qquad\qquad (72)$$

$$r_2 = \qquad\qquad S(x_1,x_2) \to R(x_1,x_2) \qquad\qquad (73)$$

*Set $\Sigma$ is not WA since the WA dependency graph contains a special edge from $R|_2$ to $S|_2$ and a regular edge from $S|_2$ to $R|_2$. However, $\Sigma$ is FD because position $S|_2$ is not $\Sigma$-recursive with $A|_1 \in \mathsf{Pos}_B(x)$. Together with Proposition 34, we can conclude that $WA \subsetneq FD$.*

*In addition, we have $r_1 \prec r_2$ and $r_2 \prec r_1$. In Section 4.3.2 we will prove that $FD \subseteq JA$; hence, $FD \subsetneq FD^\prec$, $WA^\prec \subsetneq FD^\prec$, and $FD \not\subseteq WA^\prec$ from Propositions 27, 28, and 34.* ◇

### 4.3.2 ARGUMENT-RESTRICTED RULE SETS

Lierler and Lifschitz (2009) proposed the notion of argument-restricted rule sets, whose definition we summarise next.

**Definition 36.** *An argument ranking for a set of rules $\Sigma$ is a function $\alpha$ that assigns a nonnegative integer to each position in $\Sigma$ such that the following conditions are satisfied for each rule $r \in \Sigma$, each universally quantified variable $x$ in $r$, and each existentially quantified variable $y$ in $r$:*

1. *for each $P|_i \in \mathsf{Pos}_H(x)$, some $Q|_j \in \mathsf{Pos}_B(x)$ exists such that $\alpha(P|_i) \geq \alpha(Q|_j)$; and*

2. *for each $P|_i \in \mathsf{Pos}_H(y)$, some $Q|_j \in \mathsf{Pos}_B(x)$ exists such that $\alpha(P|_i) > \alpha(Q|_j)$.*

*Set $\Sigma$ is argument restricted (AR) if an argument ranking for $\Sigma$ exists.*

An argument-restricted set of rules has a finite skolem chase on an arbitrary instance: by a straightforward induction on the chase sequence, one can show that $\mathsf{dep}(t_i) \leq \alpha(P|_i)$ for each ground fact $P(t_1, \ldots, t_n)$ derived by the chase and each $1 \leq i \leq n$.

We next show that JA is strictly more general than AR. Towards this goal, we first prove an auxiliary lemma that establishes a relationship between the set $\mathsf{Move}$ from the definition of JA and an argument ranking; next, we use this lemma to prove that $AR \subseteq JA$; and finally we present an example that shows this inclusion to be proper.

**Lemma 37.** *Let $\Sigma$ be a set of rules, let $\alpha$ be an argument ranking for $\Sigma$, let $y$ be an existentially quantified variable in $\Sigma$, and let $\mathsf{Move}(y)$ be the set of positions used in the definition of JA. For each position $P|_i \in \mathsf{Move}(y)$, some position $Q|_j \in \mathsf{Pos}_H(y)$ exists such that $\alpha(P|_i) \geq \alpha(Q|_j)$ holds.*

*Proof.* Let $y$ be an existentially quantified variable occurring in some rule $r \in \Sigma$, and consider an arbitrary position $P|_i \in \mathsf{Move}(y)$. We prove the claim by induction on the definition of $\mathsf{Move}(y)$. The base case when $P|_i \in \mathsf{Pos}_H(y)$ is trivial. Assume now that $P|_i \in \mathsf{Pos}_H(x)$ for some variable $x$ occurring in a rule $r' \in \Sigma$, and that $\mathsf{Pos}_B(x) \subseteq \mathsf{Move}(y)$, so $P|_i$ needs to be added to $\mathsf{Move}(y)$. By the definition of an argument ranking and since $P|_i \in \mathsf{Pos}_H(x)$, position $P'|_\ell \in \mathsf{Pos}_B(x)$ exists such that $\alpha(P|_i) \geq \alpha(P'|_\ell)$. But then, since $P'|_\ell \in \mathsf{Pos}_B(x) \subseteq \mathsf{Move}(y)$, by the induction hypothesis we have that position $Q|_j \in \mathsf{Pos}_H(y)$ exists such that $\alpha(P'|_\ell) \geq \alpha(Q|_j)$ holds. Thus, $\alpha(P|_i) \geq \alpha(Q|_j)$ holds, as required. $\square$

**Theorem 38.** *If a set of rules $\Sigma$ is AR, then $\Sigma$ is JA.*

*Proof.* Assume that $\Sigma$ is AR, let $\alpha$ be an argument ranking for $\Sigma$, and let $JA(\Sigma)$ be the JA dependency graph for $\Sigma$. We next prove the following claim: for each edge in $JA(\Sigma)$ from a variable $y_1$ to a variable $y_2$, and for each position $Q|_j \in \mathsf{Pos}_H(y_2)$, there exists a position $P|_i \in \mathsf{Pos}_H(y_1)$ such that $\alpha(P|_i) < \alpha(Q|_j)$. Consider an arbitrary edge from $y_1$ to $y_2$ in $JA(\Sigma)$ and an arbitrary position $Q|_j \in \mathsf{Pos}_H(y_2)$. By the definition of the JA dependency graph, then the rule $r$ that contains $y_2$ also contains a universally quantified variable $x$ such that $x$ occurs in the head of $r$ and $\mathsf{Pos}_B(x) \subseteq \mathsf{Move}(y_1)$. Since $\alpha$ is an argument ranking for $\Sigma$, some position $P'|_\ell \in \mathsf{Pos}_B(x)$ exists such that $\alpha(P'|_\ell) < \alpha(Q|_j)$. Since $P'|_\ell \in \mathsf{Move}(y_1)$, by Lemma 37 position $P|_i \in \mathsf{Pos}_H(y_1)$ exists such that $\alpha(P|_i) \leq \alpha(P'|_\ell)$. Thus, we have $\alpha(P|_i) < \alpha(Q|_j)$, and so our claim holds. But then, this claim clearly implies that the JA dependency graph $JA(\Sigma)$ is acyclic, and therefore $\Sigma$ is JA. $\qquad\square$

**Example 39.** *Let $\Sigma$ be the set consisting of the following rules:*

$$r_1 = \qquad\qquad R(z_1, x_1) \to \exists y_1.S(x_1, y_1) \qquad\qquad (74)$$

$$r_2 = \qquad\qquad R(z_2, x_2) \to \exists y_2.S(y_2, x_2) \qquad\qquad (75)$$

$$r_3 = \qquad\qquad S(x_3, x_4) \to T(x_3, x_4) \qquad\qquad (76)$$

$$r_4 = \qquad T(x_5, x_6) \wedge T(x_6, x_5) \to R(x_5, x_6) \qquad\qquad (77)$$

*Let $\alpha$ be an argument ranking for $\Sigma$. Then, $\alpha(R|_2) < \alpha(S|_2)$ due to (74); $\alpha(R|_2) < \alpha(S|_1)$ due to (75); $\alpha(S|_1) \leq \alpha(T|_1)$ and $\alpha(S|_2) \leq \alpha(T|_2)$ due to (76); and $\alpha(T|_2) \leq \alpha(R|_2)$ or $\alpha(T|_1) \leq \alpha(R|_2)$ due to (77). Together, these observations are contradictory, so such $\alpha$ cannot exist and $\Sigma$ is not AR. In contrast, $\mathsf{Move}(y_1) = \{S|_2, T|_2\}$ and $\mathsf{Move}(y_2) = \{S|_1, T|_1\}$, and so $\Sigma$ is JA.*

*In addition, we have $r_1 \prec r_3$, $r_2 \prec r_3$, $r_3 \prec r_4$, $r_4 \prec r_1$, and $r_4 \prec r_2$; hence, we have $AR \subsetneq AR^\prec$, $AR^\prec \subsetneq JA^\prec$, and $JA \not\subseteq AR^\prec$ from Theorem 38 and Propositions 27 and 28.* $\Diamond$

Lierler and Lifschitz (2009, Thm. 4) proved that AR is strictly more general than FD. We next present an example that shows $FD \subsetneq AR$, but that also settles the relationships between $FD^\prec$ and $AR^\prec$.

**Example 40.** *Let $\Sigma$ be the set consisting of the following rules:*

$$r_1 = \qquad\qquad A(x) \to \exists y.R(x, y) \qquad\qquad (78)$$

$$r_2 = \qquad\qquad R(x_1, x_2) \to S(x_1, x_2) \qquad\qquad (79)$$

$$r_3 = \qquad S(z, x) \wedge B(x) \to A(x) \qquad\qquad (80)$$

*The WA dependency graph for $\Sigma$ contains a special edge from $A|_1$ to $R|_2$, as well as regular edges from $R|_2$ to $S|_2$ and from $S|_2$ to $A|_1$; thus, $R|_2$ is $\Sigma$-recursive with $A|_1$. Consequently, rule (78) cannot satisfy the conditions in Definition 33, so we have $R|_2 \notin \mathsf{Pos}_{FD}(\Sigma)$, and thus $\Sigma$ is not FD. In contrast, $\Sigma$ is AR, as evidenced by the following argument ranking:*

$$\alpha = \{A|_1 \mapsto 0,\ B|_1 \mapsto 0,\ R|_1 \mapsto 0,\ R|_2 \mapsto 1,\ S|_1 \mapsto 0,\ S|_2 \mapsto 1\}$$

*In addition, we have $r_1 \prec r_2$, $r_2 \prec r_3$, and $r_3 \prec r_1$; hence, $FD \subsetneq FD^\prec$, $FD^\prec \subsetneq AR^\prec$, and $AR \not\subseteq FD^\prec$ from Propositions 27 and 28.* $\Diamond$

Finally, we note that $\lambda$-restricted programs by Gebser et al. (2007) and $\omega$-restricted programs by Syrjänen (2001) are both included in FD and AR; thus, when restricted to skolemised existential rules, these notions are also included in JA.

### 4.3.3 Γ-ACYCLICITY

Greco et al. (2012) recently proposed the notion of Γ-acylicity for logic programs with function symbols. The original definition of Γ-acyclicity is rather complex, so we next present a simplified version of Γ-acyclicity that is applicable to existential rules. To unify the naming style for the notions in this paper, we often write Γ-acyclicity as ΓA.

Greco et al. (2012) introduce a notion of an *activation graph*, which tracks whether a rule can trigger another rule. This notion is closely related to the notion of rule dependencies from Definition 23, but with the requirement that $I$ is an arbitrary finite set of ground facts (possibly containing functional terms). To understand why the latter is needed in logic programming, consider the following logic program:

$$r_1 = \qquad\qquad A(x) \wedge B(x) \rightarrow A(f(x)) \qquad\qquad (81)$$

$$r_2 = \qquad\qquad A(x) \wedge B(x) \rightarrow B(f(x)) \qquad\qquad (82)$$

If we restrict the set $I$ in Definition 23 to be an instance, then $r_1 \not\prec r_2$ and $r_2 \not\prec r_1$; however, the skolem chase of $r_1$, $r_2$, and facts $A(a)$ and $B(a)$ is infinite. Intuitively, $r_1$ and $r_2$ contain the same function symbol $f$, so to determine whether an application of $r_1$ can trigger an application of $r_2$, we must allow the set $I$ in Definition 23 to contain facts such as $B(f(a))$. In our setting, however, function symbols are introduced by skolemisation and are thus 'private' to each rule, which allows us to restrict the set $I$ in Definition 23 to facts without functional terms. Thus, in the rest of this section, we simply reuse the rule dependency relation $\prec$ from Definition 23, which gives us a slightly stronger version of ΓA for existential rules than the one proposed by Greco et al. (2012).

Furthermore, Greco et al. (2012) handle logic programming rules with functional terms in the body. Such rules, however, are not considered in this paper, which allows us to omit the definition of a *labelled argument graph* and simplify the notion of a *propagation graph* to a subset of the WA dependency graph.

We are now ready to present a simplified version of Γ-acyclicity that is applicable to existential rules.

**Definition 41.** *Let $\Sigma$ be a set of rules. The rule dependency relation $\prec$ is taken from Definition 23, and the set of finite domain positions $\mathsf{Pos}_{FD}(\Sigma)$ is taken from Definition 33.*

*The set of* safe *positions of $\Sigma$, written $\mathsf{Pos}_S(\Sigma)$, is the least set of the positions of $\Sigma$ such that $\mathsf{Pos}_{FD}(\Sigma) \subseteq \mathsf{Pos}_S(\Sigma)$, and $P|_i \in \mathsf{Pos}_S(\Sigma)$ if and only if, for each rule $r \in \Sigma$, at least one of the following conditions is satisfied:*

- *if $P$ occurs in the head of $r$, then $\prec$ does not contain a cycle going through $r$, or*

- *for each atom $P(\vec{t})$ in the head of $\mathsf{sk}(r)$ and each variable $x$ that occurs in $i$-th component of $\vec{t}$, we have $\mathsf{Pos}_B(x) \cap \mathsf{Pos}_S(\Sigma) \neq \emptyset$.*

*A position is* affected *if it is not safe. The* propagation graph *$PG(\Sigma)$ for $\Sigma$ has the affected positions of $\Sigma$ as vertices, and the edges of $PG(\Sigma)$ are defined as in weak acyclicity, but restricted to affected positions. The set $\Sigma$ is Γ-acyclic (ΓA) if $PG(\Sigma)$ does not contain a cycle that involves a special edge.*

In order to relate $\Gamma$A to the notions considered thus far, we first establish some containment relationships. It is obvious from Definition 41 that FD $\subseteq$ $\Gamma$A: if all positions in $\Sigma$ are finite domain, then they are also safe and so the propagation graph is empty. Furthermore, the set of rules in Example 40 is actually $\Gamma$A (all positions are safe), but not FD; hence, by Proposition 27, we have that FD $\subsetneq$ $\Gamma$A, FD$^{\prec}$ $\subsetneq$ $\Gamma$A$^{\prec}$, and $\Gamma$A $\not\subseteq$ FD$^{\prec}$. Next, Proposition 42 observes that aGRD is contained in $\Gamma$A, and Theorem 43 shows that, perhaps somewhat surprisingly, $\Gamma$A$^{\prec}$ is contained in AR$^{\prec}$.

**Proposition 42.** *If a set of rules $\Sigma$ is aGRD, then $\Sigma$ is $\Gamma$A.*

*Proof.* If the rule dependency relation $\prec$ on $\Sigma$ is acyclic, then by the first safety condition in Definition 41 all positions in $\Sigma$ are safe; but then, $PG(\Sigma)$ is empty, and so $\Sigma$ is $\Gamma$A.  $\square$

**Theorem 43.** *If a set of rules $\Sigma$ is $\Gamma$A$^{\prec}$, then $\Sigma$ is AR$^{\prec}$.*

*Proof.* The claim clearly follows from the following property: if the rule dependency relation $\prec$ for $\Sigma$ has just one strongly connected component and $\Sigma$ is $\Gamma$A, then $\Sigma$ is AR. Thus, assume that each rule $r \in \Sigma$ occurs on a cycle of $\prec$. We next construct a mapping $\alpha$ that assigns a nonnegative integer to each position in $\Sigma$, and then we show that $\alpha$ is an argument ranking for $\Sigma$. In the rest of this proof, we write $p_1 \leadsto p_2$ if $WA(\Sigma)$ (see Section 2.4) contains a path (consisting of regular and/or special edges) from position $p_1$ to position $p_2$.

Due to our assumption on $\Sigma$, the first item in Definition 41 never applies. Furthermore, let $\Psi$ be the function that maps a set $S$ of positions into another set of positions as follows:

$$\Psi(S) = S \cup \{P|_i \mid \mathsf{Pos}_B(x) \cap S \neq \emptyset \text{ for each } r \in \Sigma, \text{ each atom } P(\vec{t}) \text{ in the head of } \mathsf{sk}(r),$$
$$\text{and each variable } x \text{ occurring in the } i\text{-th component of } \vec{t}\}$$

Let $\Psi^0(S) = S$, $\Psi^k(S) = \Psi(\Psi^{k-1}(S))$ for each $k > 0$, and $\Psi^\infty(S) = \bigcup \Psi^k(S)$. From Definition 41 it is obvious that $\mathsf{Pos}_S(\Sigma) = \Psi^\infty(\mathsf{Pos}_{FD}(\Sigma))$.

We next define the mapping $\alpha$. In the rest of this proof, let $Y$ be the set containing each position $p \in \mathsf{Pos}_{FD}(\Sigma)$ for which an existentially quantified variable $y$ in $\Sigma$ exists such that $p \in \mathsf{Pos}_H(y)$. Furthermore, we use a convention that $\max \emptyset = 0$.

- For each position $p \in \mathsf{Pos}_{FD}(\Sigma)$, we define $\alpha(p)$ as follows:

$$\alpha(p) = \begin{cases} |\{p' \in Y \mid p' \leadsto p \text{ and } p' \neq p\}| + 1 & \text{if } p \in Y \\ |\{p' \in Y \mid p' \leadsto p\}| & \text{if } p \notin Y \end{cases}$$

- For each position $p \in \mathsf{Pos}_S(\Sigma) \setminus \mathsf{Pos}_{FD}(\Sigma)$, we define $\alpha(p)$ as follows:

$$\alpha(p) = \Big[\min\{k \mid p \in \Psi^k(\mathsf{Pos}_{FD}(\Sigma))\}\Big] + [\max\{\alpha(q) \mid q \in \mathsf{Pos}_{FD}(\Sigma)\}]$$

- For each position $p$ in $\Sigma$ with $p \notin \mathsf{Pos}_S(\Sigma)$, we define $\alpha(p)$ as follows, where $m(p)$ is the maximum number of special edges occurring in $PG(\Sigma)$ on a path ending at $p$:

$$\alpha(p) = m(p) + 1 + [\max\{\alpha(q) \mid q \in \mathsf{Pos}_S(\Sigma)\}]$$

Since $\Sigma$ is $\Gamma$A$^{\prec}$, $PG(\Sigma)$ does not contain a cycle involving a special edge, so $m(p)$ is always a nonnegative integer and $\alpha(p)$ is correctly defined.

We next show that $\alpha$ is an argument ranking—that is, that it satisfies all conditions of Definition 36. To this end, consider an arbitrary rule $r \in \Sigma$, an arbitrary existentially quantified variable $y$ in $r$, an arbitrary universally quantified variable $x$ in $r$, and an arbitrary position $P|_i \in \mathsf{Pos}_H(y)$; we have the following cases.

- $P|_i \in \mathsf{Pos}_{FD}(\Sigma)$. By Definition 33, position $Q|_j \in \mathsf{Pos}_B(x) \cap \mathsf{Pos}_{FD}(\Sigma)$ exists that is not $\Sigma$-recursive with $P|_i$. Thus, we have $P|_i \not\rightsquigarrow Q|_j$; furthermore, $Q|_j \rightsquigarrow P|_i$ by the definition of $WA(\Sigma)$. Together, the latter two properties imply the following:

$$\{p' \in Y \mid p' \rightsquigarrow Q|_j \text{ and } p' \neq Q|_j\} \subseteq \{p' \in Y \mid p' \rightsquigarrow P|_i \text{ and } p' \neq P|_i\}$$

  If $Q|_j \in Y$, this inclusion is strict since $Q|_j$ is contained in the set on the right-hand side, but not in the set on the left-hand side; thus, $\alpha(Q|_j) < \alpha(P|_i)$ holds, as required. If $Q|_j \notin Y$, then $\alpha(Q|_j) < \alpha(P|_i)$ holds since the definition of $\alpha$ ensures that $\alpha(P|_i) - \alpha(Q|_j)$ is at least 1.

- $P|_i \in \mathsf{Pos}_S(\Sigma) \setminus \mathsf{Pos}_{FD}(\Sigma)$. Let $k$ be the smallest number with $P|_i \in \Psi^k(\mathsf{Pos}_{FD}(\Sigma))$. By the definition of $\Psi$, there exists a position $Q|_j \in \mathsf{Pos}_B(x) \cap \Psi^{k-1}(\mathsf{Pos}_{FD}(\Sigma))$, so $k - 1 \geq \alpha(Q|_j)$ by the definition of $\alpha$. Thus, $\alpha(P|_i) > \alpha(Q|_j)$ holds, as required.

- $P|_i \notin \mathsf{Pos}_S(\Sigma)$. The first possibility is that some position $Q|_j \in \mathsf{Pos}_B(x) \cap \mathsf{Pos}_S(\Sigma)$ exists; but then, by the definition of $\alpha$, we have $\alpha(Q|_j) < \alpha(P|_i)$, as required. The second possibility is that there exists some affected position $Q|_j \in \mathsf{Pos}_B(x)$; but then, $Q|_j$ has at least one less incoming special edge in $PG(\Sigma)$ than $P|_i$; thus, we also have $\alpha(Q|_j) < \alpha(P|_i)$, as required.

To complete the proof, we must also consider an arbitrary position $P|_i \in \mathsf{Pos}_H(x)$; however, the cases are analogous as above, so we omit them for the sake of brevity. □

To place $\Gamma A$ precisely in the landscape of acyclicity notions, we present three examples that disprove relevant containment relationships. Greco et al. (2012) stated that AR is strictly contained in $\Gamma A$, but we were unable to find a formal proof of that statement; in fact, Example 44 shows that this is not the case, and that actually $\Gamma A^\prec \subsetneq AR^\prec$ holds. Moreover, Example 45 shows that $\Gamma A \not\subseteq MSA$. Finally, Example 46 shows that $WA^\prec \not\subseteq \Gamma A$.

**Example 44.** *Let $\Sigma$ be the set consisting of the following rules:*

$$r_1 = \qquad\qquad A(x) \rightarrow \exists y.R(x,y) \qquad\qquad (83)$$
$$r_2 = \qquad\qquad R(x_1,x_2) \rightarrow S(x_1,x_2) \qquad\qquad (84)$$
$$r_3 = \qquad\qquad S(z,x) \wedge B(x) \rightarrow A(x) \qquad\qquad (85)$$
$$r_4 = \qquad\qquad R(z,x) \rightarrow T(x,x) \qquad\qquad (86)$$
$$r_5 = \qquad\qquad T(x,z) \rightarrow R(x,x) \qquad\qquad (87)$$
$$r_6 = \qquad T(z_1,x) \wedge R(z_2,x) \rightarrow \exists y.T(x,y) \qquad\qquad (88)$$

*One can readily verify that the following mapping of positions to nonnegative integers is an argument ranking for $\Sigma$:*

$$\alpha = \{A|_1 \mapsto 0,\; B|_1 \mapsto 0,\; R|_1 \mapsto 1,\; R|_2 \mapsto 1,\; S|_1 \mapsto 1,\; S|_2 \mapsto 1,\; T|_1 \mapsto 1,\; T|_2 \mapsto 2\}$$

*We next argue that $\Sigma$ is not $\Gamma A$. First, the rule dependency relation in $\Sigma$ holds (at least) between the pairs of rules shown below. Thus, each rule in $\Sigma$ occurs in $\prec$ on a cycle, and so $\Sigma$ is the only strongly connected component of $\prec$.*

$$r_1 \prec r_2 \qquad r_2 \prec r_3 \qquad r_3 \prec r_1 \qquad r_1 \prec r_4 \qquad r_4 \prec r_5 \qquad r_5 \prec r_2 \qquad r_5 \prec r_6 \qquad r_6 \prec r_5$$

*Second, the WA dependency graph for $\Sigma$ contains a special edge from $A|_1$ to $R|_2$ due to rule $r_1$, a regular edge from $R|_2$ to $S|_2$ due to rule $r_2$, and a regular edge from $S|_2$ to $A|_1$ due to rule $r_3$; consequently, $R|_2$ is $\Sigma$-recursive with $A|_1$; but then, rule $r_1$ does not satisfy the conditions in Definition 33, and so $R|_2 \notin \mathsf{Pos}_{FD}(\Sigma)$. Furthermore, due to rule $r_4$, we have $T|_1 \notin \mathsf{Pos}_{FD}(\Sigma)$ and $T|_2 \notin \mathsf{Pos}_{FD}(\Sigma)$ as well. Finally, $R|_1 \notin \mathsf{Pos}_{FD}(\Sigma)$ due to rule $r_5$. Consequently, the set of finite domain positions is given by $\mathsf{Pos}_{FD}(\Sigma) = \{A|_1, B|_1, S|_1\}$.*

*Third, we argue that $\mathsf{Pos}_S(\Sigma) = \mathsf{Pos}_{FD}(\Sigma)$. In particular, there is no need to extend $\mathsf{Pos}_S(\Sigma)$ with $R|_2$: position $R|_2$ occurs in the head of rule $r_5$, but since $T|_2$ is not a finite domain position and $r_5$ occurs on a cycle of $\prec$, neither condition from Definition 41 holds. Analogously, positions $T|_1$ and $T|_2$ do not need to be added to $\mathsf{Pos}_S(\Sigma)$ either.*

*Fourth, since positions $R|_2$, $T|_1$, and $T|_2$ are all affected, the propagation graph $PG(\Sigma)$ contains a special edge from $T|_2$ to itself due to rule $r_6$. Consequently, $\Sigma$ is not $\Gamma A$.*

*Finally, since $\Sigma$ is the only strongly connected component of $\prec$, this example also shows that $AR \not\subseteq \Gamma A^\prec$ and $AR^\prec \not\subseteq \Gamma A^\prec$; but then, by Theorem 43, we have $\Gamma A^\prec \subsetneq AR^\prec$.* $\diamond$

**Example 45.** *Let $\Sigma$ be the set of rules from Example 29. As explained in the example, $\Sigma$ is aGRD, but not MSA and thus also not JA, AR, or FD. By Proposition 42, $\Sigma$ is $\Gamma A$, which implies $\Gamma A \not\subseteq MSA$, and thus $\Gamma A \not\subseteq SWA$, $\Gamma A \not\subseteq JA$, $\Gamma A \not\subseteq AR$, and $\Gamma A \not\subseteq FD$.* $\diamond$

**Example 46.** *Let $\Sigma$ be the set consisting of the following rules:*

$$r_1 = \qquad R(x_1, x_1) \rightarrow \exists y_1 \exists y_2.[A(x_1) \wedge S(y_1, x_1) \wedge S(x_1, y_2)] \qquad (89)$$

$$r_2 = \qquad A(x_2) \rightarrow B(x_2) \qquad (90)$$

$$r_3 = \qquad B(x_3) \rightarrow R(x_3, x_3) \qquad (91)$$

$$r_4 = \qquad S(x_4, x_4) \rightarrow \exists y_3 \exists y_4.[C(x_4) \wedge R(y_3, x_4) \wedge R(x_4, y_4)] \qquad (92)$$

$$r_5 = \qquad C(x_5) \rightarrow D(x_5) \qquad (93)$$

$$r_6 = \qquad D(x_6) \rightarrow S(x_6, x_6) \qquad (94)$$

*Note that $r_1 \not\prec r_4$ and $r_4 \not\prec r_1$, so the rule dependency relation $\prec$ in $\Sigma$ has two strongly connected components: the first one consists of $r_1$, $r_2$, and $r_3$, and the second one consists of $r_4$, $r_5$, and $r_6$. Moreover, each strongly connected component is WA, so $\Sigma$ is $WA^\prec$.*

*In contrast, each position in $\Sigma$ is $\Sigma$-recursive with itself, so $\mathsf{Pos}_{FD}(\Sigma) = \emptyset$. Moreover, each position in $\Sigma$ occurs in the head of a rule that (i) appears in $\prec$ in a cycle and (ii) does not satisfy the second safety condition in Definition 41; hence, $\mathsf{Pos}_S(\Sigma) = \emptyset$, and all positions are affected. But then, $PG(\Sigma) = WA(\Sigma)$, and so $\Sigma$ is not $\Gamma A$.* $\diamond$

It may seem counterintuitive that $AR \not\subseteq \Gamma A$, but $\Gamma A^\prec \subsetneq AR^\prec$. Intuitively, the notion of safe positions from Definition 41 uses the rule dependency relation, which allows us to construct an example that is in $\Gamma A$ but not in $AR$. In $\Gamma A^\prec$, this extra condition is always applied to rules that occur in $\prec$ on a cycle; thus, the notion of safe positions collapses to a notion weaker than $AR$, which in turn allows $\Gamma A^\prec$ to be subsumed by $AR^\prec$.

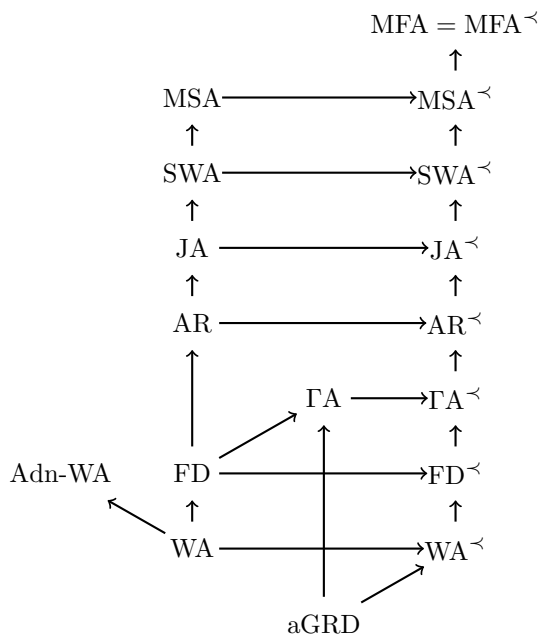

Figure 1: The Landscape of Acyclicity Notions

## 4.4 The Landscape of Acyclicity Notions

To obtain a complete picture of the relative expressiveness of the acyclicity notions considered in this paper, we make the following observations.

- The rule set from Example 15 is MFA but not MSA, and one can readily verify that $r_1 \prec r_2 \prec r_3 \prec r_4 \prec r_1$; but then, $\text{MSA}^\prec \subsetneq \text{MFA}^\prec = \text{MFA}$ by Proposition 27.

- The rule set from Example 20 is SWA but not JA, and one can readily verify that $r_1 \prec r_2 \prec r_3 \prec r_1$; but then, $\text{JA}^\prec \subsetneq \text{SWA}^\prec$ and $\text{SWA} \nsubseteq \text{JA}^\prec$ by Proposition 27.

- The rule set from Example 1 is MSA but not SWA, and one can readily verify that $r_1 \prec r_3 \prec r_4 \prec r_5 \prec r_2 \prec r_1$; but then, $\text{SWA}^\prec \subsetneq \text{MSA}^\prec$ and $\text{MSA} \nsubseteq \text{SWA}^\prec$ by Proposition 27.

- The rule set from Example 22 is aGRD: we have $r_1 \nprec r_1$, $r_1 \nprec r_2$, $r_2 \nprec r_1$, and $r_2 \nprec r_2$. Thus, $\text{aGRD} \nsubseteq \text{Adn-WA}$.

- The rule set from Example 22 is FD: we can assume all positions in the rule set to be finite domain without violating conditions of Definition 33. Thus, $\text{FD} \nsubseteq \text{Adn-WA}$.

The landscape of the acyclicity notions considered in this paper is shown in Figure 1. All inclusions between notions shown in the figure are strict: if a notion $X$ is reachable from a notion $Y$ via one or more (directed) arcs, then $X$ is strictly more general than $Y$. Furthermore, all inclusions are also complete: if a notion $X$ is not reachable from a notion $Y$ via one or more (directed) arcs, then $X$ does not contain $Y$.

## 5. Handling Equality via Singularisation

Most acyclicity notions presented so far provide no special provision for the equality predicate. If a set of rules $\Sigma$ contains the equality predicate, one can always axiomatise equality explicitly and then check acyclicity. More precisely, the acyclicity of $\Sigma \cup \Sigma_\approx$ (under any notion introduced thus far) guarantees termination of the skolem chase of $\Sigma$. Furthermore, note that MFA and MSA are defined as entailment checks in first-order logic with equality, which effectively incorporates the rules of equality into these checks even if rules (1)–(4) are not explicitly given; however, the effect of such a definition is the same.

While handling equality explicitly may be simple, such an approach does not ensure termination of the skolem chase in many practically relevant cases. In particular, the following example shows that the equalities between terms tend to proliferate during skolem chase, which can lead to non-termination.

**Example 47.** *Consider the set of rules $\Sigma$ containing rules* (95)–(96).

$$A(x) \wedge B(x) \to \exists y. [R(x, y) \wedge B(y)] \tag{95}$$

$$R(z, x_1) \wedge R(z, x_2) \to x_1 \approx x_2 \tag{96}$$

*The skolem chase of $I_\Sigma^*$ and $\Sigma$ derives the following infinite set of facts:*

$$
\begin{array}{llll}
R(*, f(*)) & B(f(*)) & * \approx f(*) & A(f(*)) \\
R(f(*), f(f*))) & B(f(f(*))) & \dots &
\end{array}
$$

*Thus, $\Sigma$ is not universally MFA by Proposition 5, and by Theorem 14 it is not universally MSA either.* ◇

It is worth noticing that in the presence of equality WA is no longer subsumed by MFA and hence both notions become incomparable. As explained in Section 2.4, WA can be applied to rules containing the equality predicate (and without an explicit axiomatisation of equality). Under such a treatment, the rules in Example 47 are WA. This, however, does not contradict the results from Section 4: WA does not require an explicit axiomatisation of equality because it ensures termination of nonoblivious chase—an optimised chase variant that expands existential quantifiers only if necessary and that handles equality by replacing equal terms with canonical representatives. In contrast, the results in Section 4 ensure termination of the skolem chase; since this chase variant uses an explicit axiomatisation of equality, all of our results hold only for equality-free rules (or, equivalently, for the rules containing an explicit axiomatisation of equality). The rules in Example 47 are not WA if equality is axiomatised explicitly, which explains the apparent mismatch with Section 4.

In order to use the skolem chase with rule sets such as the ones in Example 47, Marnette (2009) proposed the *singularisation* technique. Roughly speaking, singularisation replaces the equality predicate $\approx$ with a fresh binary predicate $\mathsf{Eq}$ to clarify that the two are to be treated differently; furthermore, it axiomatises $\mathsf{Eq}$ as reflexive, symmetric, and transitive, but it does not introduce replacement rules analogous to (4); finally, it modifies the rules in $\Sigma$ to take the lack of the replacement rules into account. The chase of the transformed rule set is not a model of $\Sigma$, but it can be used to answer queries over $\Sigma$ in a particular well-defined way. The modification of $\Sigma$, however, is nondeterministic: there are many ways to modify $\Sigma$ and, while some may ensure termination of the skolem chase, not all are required to do so. We next recapitulate the definition of singularisation by Marnette (2009).

**Definition 48.** *A* marking $M_r$ *of a rule $r$ of the form* (5) *is a mapping that assigns to each variable $w \in \vec{x} \cup \vec{z}$ a single occurrence of $w$ in $\varphi$; the marked occurrence of $w$ in a rule is written $w^\diamond$. All other occurrences of $w$ are* unmarked*, and all occurrences of constants are unmarked as well. For $\Sigma$ a set of rules, a* marking *$M$ of $\Sigma$ contains exactly one marking $M_r$ for each $r \in \Sigma$. Let $\mathsf{Eq}$ be a fresh binary predicate not occurring in $\Sigma$. The* singularisation *of $\Sigma$ under $M$ is the set $\mathsf{Sing}(\Sigma, M)$ that contain rules*

$$\rightarrow \mathsf{Eq}(x, x) \tag{97}$$

$$\mathsf{Eq}(x_1, x_2) \rightarrow \mathsf{Eq}(x_2, x_1) \tag{98}$$

$$\mathsf{Eq}(x_1, x_2) \wedge \mathsf{Eq}(x_2, x_3) \rightarrow \mathsf{Eq}(x_1, x_3) \tag{99}$$

*and, for each rule $r \in \Sigma$, the rule obtained from $r$ by replacing each atom $s \approx t$ with atom $\mathsf{Eq}(s, t)$, and by replacing each unmarked occurrence of a term $t$ in a body atom with a fresh variable $z'$ and then adding atom $\mathsf{Eq}(t, z')$ to the rule body.*

Note that $\mathsf{Sing}(\Sigma, M)$ is unique up to the renaming of the fresh variables. Furthermore, note that rule (97) can be transformed into a safe rule as explained in Section 2.2. Finally, note that $\mathsf{Sing}(\Sigma, M)$ is equality-free (since $\approx$ and $\mathsf{Eq}$ are different predicates); therefore, no specific treatment of equality is needed when computing its chase or checking its acyclicity.

**Example 49.** *Singularisation of the marked rule* (100) *produces rule* (101)*.*

$$A(x^\diamond) \wedge B(x) \wedge R(x, z^\diamond) \rightarrow C(x) \tag{100}$$

$$A(x) \wedge B(x_1) \wedge R(x_2, z) \wedge \mathsf{Eq}(x, x_1) \wedge \mathsf{Eq}(x, x_2) \rightarrow C(x) \tag{101}$$

*Note that singularisation should be applied 'globally' to all rules, including the ones that do not contain the equality predicate.* ◇

The properties of singularisation can be summarised as follows. Let $\Sigma$ be a set of rules, let $I$ be an instance, and let $M$ be a marking for $\Sigma$. Furthermore, let $\Sigma' = \mathsf{Sing}(\Sigma, M)$, and let $I' = I_{\Sigma'}^\infty$ be the chase of $I$ and $\Sigma'$. Finally, note that predicate $\mathsf{Eq}$ is interpreted in $I'$ as an equivalence relation, so let $\rho$ be a function that maps each term $t$ occurring in $I'$ to an arbitrarily chosen representative from the equivalence class of $t$. The first-order interpretation $\rho(I')$ is defined as follows, where $\mathsf{rng}(\rho)$ is the range of the mapping $\rho$, the set $\triangle^{\rho(I')}$ is the universe of $\rho(I')$, and $(P)^{\rho(I')}$ is the interpretation of a predicate $P$:

$$\begin{aligned}
\triangle^{\rho(I')} &= \mathsf{rng}(\rho) \\
(P)^{\rho(I')} &= \{\langle \rho(t_1), \ldots, \rho(t_n) \rangle \mid P(t_1, \ldots, t_n) \in I\} \text{ for each } P \text{ different from } \mathsf{Eq} \\
(\mathsf{Eq})^{\rho(I')} &= \{\langle x, x \rangle \mid x \in \triangle^{\rho(I')}\}
\end{aligned}$$

Note that $\rho(I')$ interprets $\approx$ as true equality—that is, each term $t$ is interpreted in $\rho(I')$ as a representative of the equivalence class that contains $t$; hence, $\rho(I')$ is not a Herbrand interpretation. Marnette (2010) showed that, for an arbitrary $\rho$, interpretation $\rho(I')$ is a *universal model* of $\Sigma$ and $I$—that is, $\rho(I')$ can be homomorphically embedded into an arbitrary model of $\Sigma$ and $I$. Thus, $\rho(I)$ can be used for query answering: for a Boolean conjunctive query $Q$, we have $I \cup \Sigma \models Q$ if and only if $\rho(I') \models Q$.

This result can be reformulated as follows. Let $\Sigma$, $I$, $M$, and $I'$ be as specified above, and let us assume that $Q$ is of the form $Q = \exists \vec{y}.\varphi(\vec{y})$. Furthermore, let $r$ be the following rule, and let $M'$ be an arbitrary marking of $r$:

$$r = \quad \varphi(\vec{y}) \rightarrow H \tag{102}$$

Then, the above characterisation of singularisation implies that

$$
\begin{aligned}
I \cup \Sigma &\models Q &&\text{if and only if} \\
I \cup \mathsf{Sing}(\Sigma \cup \{r\}, M \cup M') &\models H &&\text{if and only if} \\
I' \cup \mathsf{Sing}(\{r\}, M') &\models H.
\end{aligned}
$$

Hence, we can answer $Q$ w.r.t. $\Sigma$ and $I$ by evaluating $\mathsf{Sing}(\{r\}, M')$ in the chase of $I$ and $\mathsf{Sing}(\Sigma, M)$. It is straightforward to generalise this approach to non-Boolean queries.

The absence of replacement rules (4) often allows the skolem chase to terminate on $\mathsf{Sing}(\Sigma, M)$, but this may depend on the selected marking.

**Example 50.** *Rule* (95) *from Example 47 admits the following two markings:*

$$A(x^\diamond) \wedge B(x) \rightarrow \exists y.[R(x,y) \wedge B(y)] \tag{103}$$

$$A(x) \wedge B(x^\diamond) \rightarrow \exists y.[R(x,y) \wedge B(y)] \tag{104}$$

*The skolem chase does not universally terminate for the singularisation obtained from* (104) *and* (96)*. In contrast, the singularisation obtained from* (103) *and* (96) *is JA.* ◇

**Definition 51.** *For $X \in \{MFA, MSA, JA\}$, acyclicity notion $X^\exists$ (resp. $X^\forall$) contains each finite set of rules $\Sigma$ such that $\mathsf{Sing}(\Sigma, M) \in X$ for some (resp. each) marking $M$ of $\Sigma$.*

Clearly, $X^\forall \subseteq X^\exists$ for each $X \in \{\mathrm{MFA}, \mathrm{MSA}, \mathrm{JA}\}$, and Example 50 shows this inclusion to be proper. We next show that $\mathrm{JA}^\forall$ actually coincides with WA.

**Theorem 52.** *For $\Sigma$ an arbitrary finite set of rules, $\Sigma$ is $JA^\forall$ if and only if $\Sigma$ is WA.*

*Proof.* ($\mathrm{JA}^\forall \subseteq \mathrm{WA}$) We prove the contrapositive, so let $\Sigma$ be an arbitrary set of rules that is not WA; w.l.o.g. we assume that each variable in $\Sigma$ occurs in at most one rule. We consider each edge from $p$ to $q$ in the WA dependency graph $WA(\Sigma)$ to be a triple $e = \langle p, q, t \rangle$, where $t = \cdot$ if the edge is regular and $t = *$ if the edge is special. By the definition of WA, for each such $e$, a rule $r \in \Sigma$ and universally quantified variable $x$ occurring in the head and the body of $r$ exist such that $p \in \mathsf{Pos}_B(x)$, so let $x_e$ be one such arbitrarily chosen but fixed variable; furthermore, if edge $e$ is special, then an existentially quantified variable $y$ exists such that $q \in \mathsf{Pos}_H(y)$, so let $y_e$ be one such arbitrarily chosen but fixed variable.

A *cycle* in $WA(\Sigma)$ is a sequence of edges $e_1, \ldots, e_n$ of the form $e_i = \langle p_i, q_i, t_i \rangle$ such that $q_i = p_{i+1}$ for each $1 \leq i < n$ and and $q_n = p_1$. Such a cycle is *dangerous* if an edge $e_k$ exists that is special; and such a cycle is *simple* if $x_{e_i} \neq x_{e_j}$ for all $1 \leq i < j \leq n$.

Now let $\Pi' = e_1, \ldots, e_n$ be an arbitrary dangerous cycle in $WA(\Sigma)$. If $\Pi'$ is not simple, we show how to transform $\Pi'$ to a shorter dangerous cycle. Towards this goal, assume that $\Pi'$ contains edges $e_i = \langle p_i, q_i, t_i \rangle$ and $e_j = \langle p_j, q_j, t_j \rangle$ such that $1 \leq i < j \leq n$ and $x_{e_i} = x_{e_j}$; hence, some rule $r \in \Sigma$ contains body atoms in which $x_{e_i}$ occurs at positions $p_i$ and $p_j$. Furthermore, let $e_k$ be an arbitrarily chosen, but fixed special edge in $\Pi'$; such $e_k$ exists since $\Pi'$ is dangerous. We have the following possibilities.

---

3. Note that a cycle of length one is always simple.

- If $i \leq k < j$, let $\Pi'' = e, e_{i+1}, \ldots, e_{j-1}$ where $e = \langle p_j, q_i, t_i \rangle$. If $e_i$ is regular, then $x_{e_i}$ occurs in a head atom of $r$ at position $q_i$; furthermore, if $e_i$ is special, then some head atom of $r$ contains an existentially quantified variable at position $q_i$. Either way, $e$ is an edge of $WA(\Sigma)$, so $\Pi''$ is a cycle in $WA(\Sigma)$. Furthermore, $e$ is special if $k = i$, and $\Pi''$ contains $e_k$ otherwise; hence, $\Pi''$ is dangerous.

- Otherwise, let $\Pi'' = e_1, \ldots, e_{i-1}, e, e_{j+1}, \ldots, e_n$ where $e = \langle p_i, q_j, t_j \rangle$. Edges $e$ and $e_j$ are of the same type, so $e$ is an edge of $WA(\Sigma)$ and $\Pi''$ is a cycle in $WA(\Sigma)$. Furthermore, $e$ is special if $k = j$, and $\Pi''$ contains $e_k$ otherwise; hence, $\Pi''$ is dangerous.

In both cases, $\Pi''$ contains at least one edge less than $\Pi'$. Thus, we can iteratively transform an arbitrary dangerous cycle in $WA(\Sigma)$ to a simple dangerous cycle $\Pi$.

Now let $M$ be a marking for $\Sigma$ that marks each variable $w$ occurring in the body of a rule $r \in \Sigma$ as follows.

- If an edge $e = \langle p, q, t \rangle$ in $\Pi$ exists such that $w = x_e$, then $M$ marks an occurrence of $w$ in $r$ at position $p$ (if there are multiple such occurrences, one is chosen arbitrarily). Since $\Pi$ is simple, edge $e$ is unique, and so $M$ is correctly defined.

- Otherwise, $M$ marks an arbitrarily chosen occurrence of $w$ in $r$.

Let $\Sigma' = \mathsf{Sing}(\Sigma, M)$, and let $JA(\Sigma')$ be the JA dependency graph for $\Sigma'$. To show that $JA(\Sigma')$ contains a cycle, we first prove the following property $(\star)$.

For each subpath $e_1, \ldots, e_k$ of $\Pi$ where edge $e_1$ is special and each edge $e_i$ with $1 < i \leq k$ is regular, we have $\{q_i, \mathsf{Eq}|_1\} \subseteq \mathsf{Move}(y_{e_1})$ for each $1 \leq i \leq k$.[4]

Since $\Sigma'$ contains rule (97), we clearly have $\mathsf{Eq}|_1 \in \mathsf{Move}(y_{e_1})$. We next prove $(\star)$ by induction on $k$. For the base case $k = 1$, we have $q_1 \in \mathsf{Move}(y_{e_1})$ by the definition of JA. For the induction step, assume that the claim holds for all subpaths of length $k$, and consider a subpath $e_1, \ldots, e_k, e_{k+1}$. By the induction assumption and the fact that $q_k = p_{k+1}$, we have $p_{k+1} \in \mathsf{Move}(y_{e_1})$. Furthermore, variable $x_{e_{k+1}}$ occurs in the body and the head atom of some rule $r \in \Sigma'$ at positions $p_{k+1}$ and $q_{k+1}$, respectively. Finally, by the definition of $M$ and the properties of singularisation, we have that $\mathsf{Pos}_B(x_{e_{k+1}})$ contains $p_{k+1}$ and possibly $\mathsf{Eq}|_1$. But then, by the definition of JA, we have $q_{k+1} \in \mathsf{Move}(y_{e_1})$, as required.

To complete the proof, consider now an arbitrary subpath $e_1, \ldots, e_\ell$ of $\Pi$ where edges $e_1$ and $e_\ell$ are special and each edge $e_i$ with $1 < i < k$ is regular. By $(\star)$ and the fact that $q_{\ell-1} = p_\ell$, we have $\{p_\ell, \mathsf{Eq}|_1\} \subseteq \mathsf{Move}(y_{e_1})$. Furthermore, as in the previous paragraph, $\mathsf{Pos}_B(x_{e_\ell})$ contains $p_\ell$ and possibly $\mathsf{Eq}|_1$; but then, $JA(\Sigma')$ contains an edge from $y_{e_1}$ to $y_{e_\ell}$. Since $\Pi$ is a cycle, $JA(\Sigma')$ clearly contains a cycle, so $\Sigma'$ is not JA, as required.

$(\mathrm{JA}^\forall \supseteq \mathrm{WA})$ Assume that $\Sigma \notin \mathrm{JA}^\forall$, so there exists a marking $M$ for $\Sigma$ such that $\Sigma' = \mathsf{Sing}(\Sigma, M)$ is not JA. We assume that $\Sigma$ does not contain an existentially quantified variable that occurs in an equality atom; this is w.l.o.g. as we can always replace each equality atom $y \approx t$ with an atom $R(x, t)$ and add a rule $R(x_1, x_2) \rightarrow x_1 \approx x_2$ for $R$ a fresh binary predicate, and such a transformation clearly does not affect the membership of the rule set in $\mathrm{JA}^\forall$ and WA. Now consider an arbitrary existentially quantified variable $y$, and arbitrary positions $p \in \mathsf{Pos}_H(y)$ and $q \in \mathsf{Move}(y)$ that do not involve $\mathsf{Eq}$ (both sets are w.r.t. $\Sigma'$); by induction on the construction of $\mathsf{Move}(v)$, one can prove that $WA(\Sigma)$ then contains a sequence of regular edges from $p$ to $q$. The proof is straightforward, and we omit the details for the sake of brevity. Similarly, consider an arbitrary edge from $y_1$ to $y_2$ in $JA(\Sigma')$, and arbitrary positions $p \in \mathsf{Pos}_H(y_1)$ and $q \in \mathsf{Pos}_H(y_2)$ that do not involve $\mathsf{Eq}$; by the definition of JA, a variable $x$ occurring in the rule of $y_2$ and a position $s$ not involving $\mathsf{Eq}$ exist such that $s \in \mathsf{Move}(y_1)$ and $s \in \mathsf{Pos}_B(x)$. But then $WA(\Sigma)$ contains a path consisting of regular edges from $p$ to $s$, as well as a special edge from $s$ to $q$. Since $JA(\Sigma')$ is cyclic, $WA(\Sigma)$ clearly contains a cycle involving a special edge. $\qquad\square$

---

4. The notion of a subpath is defined in the obvious way; however, please note that, although $\Pi$ is defined as a sequence of edges, subpaths of $\Pi$ can 'wrap around' this sequence as $\Pi$ is a cycle.

Checking all possible markings may be infeasible: the number of candidates is exponential in the total number of variables that occur more than once in a rule body. Theorem 52 shows that $\mathrm{JA}^\forall$ can be decided using WA. For the other cases, the following simple observation shows how to reduce the number of markings.

**Definition 53.** *A variable $x$ is relevant for a rule $r \in \Sigma$ if $x$ occurs more than once in the body of $r$, and the head of $r$ contains an atom $P(\vec{t})$ such that $x \in \vec{t}$ and $P$ is not $\approx$.*

**Proposition 54.** *Let $M$ and $M'$ be markings for $\Sigma$ such that, for each rule $r \in \Sigma$, the markings for $r$ in $M$ and $M'$ coincide on each relevant variable in $r$. Then, for each instance $I$, the result of the skolem chase for $I$ and $\mathsf{Sing}(\Sigma, M)$ coincides with the result of the skolem chase for $I$ and $\mathsf{Sing}(\Sigma, M')$; furthermore, $\mathsf{Sing}(\Sigma, M)$ is JA/MSA/MFA if and only if $\mathsf{Sing}(\Sigma, M')$ is JA/MSA/MFA.*

*Proof.* Consider an arbitrary rule $r \in \Sigma$. If a variable $x$ occurs only in the body of $r$, then marking various occurrences of $x$ in $r$ clearly produces rules equivalent up to the renaming of variables. Furthermore, assume that a variable $x$ occurs in the head of $r$ only in an equality atom of the form $x \approx t$, and that the markings of $x$ differ. Then, the rules obtained from $r$ by singularisation will all have the same body (up to the renaming of variables); furthermore, the bodies contain atoms $\mathsf{Eq}(x_i, x)$, and the rule heads are of the form $\mathsf{Eq}(x, t)$. Since $\mathsf{Sing}(\Sigma, M)$ and $\mathsf{Sing}(\Sigma, M')$ contain rules (97)–(99), the skolem chase for $I$ and $\mathsf{Sing}(\Sigma, M)$ clearly derives the same ground atoms as the skolem chase for $I$ and $\mathsf{Sing}(\Sigma, M')$. $\qquad\square$

Despite this optimisation, the number of markings to check can still be exponential in the size of $\Sigma$, so we next describe a useful approximation. Let $\mathcal{M}$ be a maximal set of markings for $\Sigma$ such that, for all $M_1, M_2 \in \mathcal{M}$, each rule $r \in \Sigma$, and each variable $x$ that is not relevant in $r$, the markings of $x$ in $r$ under $M_1$ and $M_2$ coincide. Intuitively, such $\mathcal{M}$ contains all possible markings of the relevant variables, but the markings of all other variables coincide. By Proposition 54 it is clear that, given two such sets $\mathcal{M}_1$ and $\mathcal{M}_2$, the skolem chase of $\bigcup_{M \in \mathcal{M}_1} \mathsf{Sing}(\Sigma, M)$ and $\bigcup_{M \in \mathcal{M}_2} \mathsf{Sing}(\Sigma, M)$ coincides for an arbitrary instance $I$; thus, let $\mathcal{M}$ be one arbitrarily chosen such set of markings. Also, let $\mathsf{Sing}_\cup(\Sigma) = \bigcup_{M \in \mathcal{M}} \mathsf{Sing}(\Sigma, M)$, let $\mathrm{MFA}^\cup$ be the class containing each rule set $\Sigma$ such that $\mathsf{Sing}_\cup(\Sigma) \in \mathrm{MFA}$, and let $\mathrm{MSA}^\cup$ and $\mathrm{JA}^\cup$ be defined analogously. As the following proposition shows, $\mathsf{Sing}_\cup(\Sigma)$ provides a 'lower bound' on acyclicity that can be obtained via singularisation.

**Proposition 55.** *For each $X \in \{MFA, MSA, JA\}$, we have that $X^{\cup} \subseteq X^{\forall}$. Furthermore, the size of $\mathsf{Sing}_{\cup}(\Sigma)$ is exponential in the maximal number of relevant variables in a rule in $\Sigma$, and it is linear in the number of rules in $\Sigma$.*

*Proof.* The first claim follows from the fact that all considered notions of acyclicity are monotone in the sense that every subset of an acyclic rule set is also acyclic. The second claim follows from the fact that, if a rule $r$ exists that contains $k$ relevant variables and each variable occurs $m$ times in $r$, then $\mathcal{M}$ contains $m^k$ different markings for $r$. □

This result is interesting when dealing with rules that are obtained from DLs, where each rule has at most one relevant variable: on such rule sets, the size of $\mathsf{Sing}_{\cup}(\Sigma)$ is linear in the size of $\Sigma$. For the general case, the complexity of acyclicity checking does not increase despite the exponential increase in the number of rules.

**Theorem 56.** *Deciding whether $\Sigma$ is in $MFA^{\cup}$ ($MFA^{\exists}$, $MFA^{\forall}$) is 2ExpTime-complete. Deciding whether $\Sigma$ is in $MSA^{\cup}$ ($MSA^{\exists}$, $MSA^{\forall}$) is ExpTime-complete.*

*Proof.* If $\Sigma$ contains no equality, it is easy to see that $\Sigma$ is in $MFA^{\cup}$ ($MFA^{\exists}$, $MFA^{\forall}$) if and only if it is in MFA. The same can be observed for MSA. Thus, hardness follows from Theorems 8 and 13.

For membership, we first consider the cases of $MFA^{\exists}$, $MFA^{\forall}$, $MSA^{\exists}$, and $MSA^{\forall}$. Each of these properties can be decided by considering all of the at most exponentially many markings. Since $\mathsf{Sing}(\Sigma, M)$ is linear in the size of $\Sigma$, the property can be checked for each marking for MFA in 2ExpTime (cf. Theorem 8) and for MSA in ExpTime (cf. Theorem 13). This yields the required bound since an exponential factor is not significant for the considered complexity classes.

For $MFA^{\cup}$ and $MSA^{\cup}$, membership follows by observing that the membership of MFA and MSA in 2ExpTime and ExpTime, respectively, is obtained from the double/single exponential bound on the number of ground facts that potentially need to be derived in order to decide the required property. While $\mathsf{Sing}_{\cup}(\Sigma)$ is exponentially larger than $\Sigma$, the maximal number of relevant ground facts is still the same since no new predicates or constant symbols are introduced. The increased number of rules leads to an exponential increase of the time to check applicability of all rules in each of the doubly/singly exponentially many steps, but this exponential factor does not affect membership of the decision problem in 2ExpTime/ExpTime. □

We finish this section by examining the interaction between rule normalisation and singularisation. Note that normalisation reduces the number of variables in a rule, which at least at first sight suggests that normalisation could prevent one from finding a marking that ensures acyclicity of the singularised rules. We next show that this cannot happen if normalisation is used without structure sharing: if the original set of rules is MFA w.r.t. some set of markings, then the transformed set of rules is MFA w.r.t. a set of markings as well. Furthermore, we show that this does not hold if normalisation is used with structure sharing; hence, normalisation should be applied with care when used with singularisation.

**Theorem 57.** *Let $\Sigma$ be a set of existential rules, let $\Sigma'$ be obtained from $\Sigma$ by applying a single normalisation step without structure sharing, and let $I$ be an instance. Then each* *marking $M$ of $\Sigma$ for which $\mathsf{Sing}(\Sigma, M)$ is MFA w.r.t. $I$ can be extended to a marking $M'$ of $\Sigma'$ such that $\mathsf{Sing}(\Sigma', M')$ is MFA w.r.t. $I$.*

*Proof.* Let $M$ be a marking of $\Sigma$ such that $\mathsf{Sing}(\Sigma, M)$ is MFA, let $r \in \Sigma$ be the rule of the form (8) to which the normalisation step is applied, and let $\Sigma'$ be the set of rules obtained from $\Sigma$ after the application of a normalisation step to $r$. We next prove that the claim holds for both a head and a body normalisation step.

(*Head Normalisation*) Assume that the set of rules $\Sigma'$ is obtained by replacing a rule $r \in \Sigma$ with rules $r_1$ and $r_2$ of the following forms, where $\vec{x} = \vec{x}_3 \cup \vec{x}_4$:

$$r = \qquad \varphi(\vec{x}, \vec{z}) \to \exists \vec{y}_1, \vec{y}_2, \vec{y}_3.[\psi_1(\vec{x}_3, \vec{y}_1, \vec{y}_2) \wedge \psi_2(\vec{x}_4, \vec{y}_1, \vec{y}_3)]$$
$$r_1 = \qquad \varphi(\vec{x}, \vec{z}) \to \exists \vec{y}_1, \vec{y}_3.[Q(\vec{x}_3, \vec{y}_1) \wedge \psi_2(\vec{x}_4, \vec{y}_1, \vec{y}_3)]$$
$$r_2 = \qquad Q(\vec{x}_3, \vec{y}_1) \to \exists \vec{y}_2.\psi_1(\vec{x}_3, \vec{y}_1, \vec{y}_2)$$

Let $M'$ be a marking that coincides with $M$ on all rules different from $r$, that marks $r_1$ in the same way as $M$ marks $r$, and that marks $r_2$ in the only possible way (note that the body of this rule does not contain repeated occurrences of variables); furthermore, let $\Omega = \mathsf{Sing}(\Sigma, M)$ and $\Upsilon = \mathsf{Sing}(\Sigma', M')$. We assume that rule $r$ is skolemised by replacing each variable $y \in \vec{y}_1$ with $g_1^y(\vec{x})$, each variable $y \in \vec{y}_2$ with $g_2^y(\vec{x})$, and each variable $y \in \vec{y}_3$ with $g_3^y(\vec{x})$; rule $r_1$ is skolemised as $r$; and rule $r_2$ is skolemised by replacing each variable $y \in \vec{y}_2$ with $h^y(\vec{x}_3, \vec{y}_1)$. Thus, the skolemised and singularised rules have the following form; formula $\varphi'$ is a singularisation of $\varphi$, and all freshly introduced variables are contained in $\vec{z}_1$:

$$\varphi'(\vec{x}, \vec{z}_1) \to \psi_1(\vec{x}_3, \vec{g}_1(\vec{x}), \vec{g}_2(\vec{x})) \wedge \psi_2(\vec{x}_4, \vec{g}_1(\vec{x}), \vec{g}_3(\vec{x}))$$
$$\varphi'(\vec{x}, \vec{z}_1) \to Q(\vec{x}_3, \vec{g}_1(\vec{x})) \wedge \psi_2(\vec{x}_4, \vec{g}_1(\vec{x}), \vec{g}_3(\vec{x}))$$
$$Q(\vec{x}_3, \vec{y}_1) \to \psi_1(\vec{x}_3, \vec{y}_1, \vec{h}(\vec{x}_3, \vec{y}_1))$$

Finally, we inductively define a partial mapping $\mu$ from terms to terms as follows:

- $\mu(c) = c$ for each constant $c$,

- $\mu(f(\vec{t})) = f(\mu(\vec{t}))$ for each function symbol $f$ not of the form $h^y$ or $g_1^y$ and all terms $\vec{t}$ such that $\mu(\vec{t})$ is defined, and

- $\mu(h^y(\vec{s}, \vec{g}_1(\vec{s}, \vec{t}))) = g_2^y(\mu(\vec{s}), \mu(\vec{t}))$ for each function symbols of the form $h^y$, the corresponding symbol $g_2^y$, and all terms $\vec{s}$ and $\vec{t}$ such that $\mu(\vec{s})$ and $\mu(\vec{t})$ are defined.

We next show the following property ($\star$): for each $A(\vec{t}) \in I_\Upsilon^\infty$ where $A$ is a predicate occurring in $\Omega$ (i.e., $A$ was not introduced by the normalisation step), $\mu(t)$ is defined and $A(\mu(\vec{t})) \in I_\Omega^\infty$. The proof is by induction on the chase sequence $I_\Upsilon^0, I_\Upsilon^1, \ldots$ for $I$ and $\Upsilon$. The base case holds trivially. Furthermore, since $\Omega$ and $\Upsilon$ coincide on all rules apart from $r$, $r_1$, and $r_2$, the proof of the claim is trivial for each conclusion of a rule different from $r_1$ or $r_2$. For the remaining cases, we can assume w.l.o.g. that $I_\Upsilon^{i+1}$ is obtained from $I_\Upsilon^i$ by a single application of $r_1$ of substitution $\sigma$ and an application of $r_2$ to the result; thus, the rules together derive the following facts:

$$Q(\vec{x}_3\sigma, \vec{g}_1(\vec{x}\sigma))$$
$$\psi_1(\vec{x}_3\sigma, \vec{g}_1(\vec{x}\sigma), \vec{h}(\vec{x}_3\sigma, \vec{g}_1(\vec{x}\sigma)))$$
$$\psi_2(\vec{x}_4\sigma, \vec{g}_1(\vec{x}\sigma), \vec{g}_3(\vec{x}\sigma))$$

By the induction assumption, $\varphi'(\mu(\vec{x}\sigma), \mu(\vec{z}\sigma)) \subseteq I_\Omega^\infty$, and so $I_\Omega^\infty$ contains the following facts:

$$\psi_1(\mu(\vec{x}_3\sigma), \vec{g}_1(\mu(\vec{x}\sigma)), \vec{g}_2(\mu(\vec{x}\sigma)))$$
$$\psi_2(\mu(\vec{x}_4\sigma), \vec{g}_1(\mu(\vec{x}\sigma)), \vec{g}_3(\mu(\vec{x}\sigma)))$$

Clearly, each term $h^y(\vec{x}_3\sigma, \vec{g}_1(\vec{x}\sigma))$ is of the form $h^y(\vec{s}, \vec{g}_1(\vec{s}, \vec{t}))$, so the mapping $\mu$ is defined on the term. Furthermore, by the definition of $\mu$, it is clear that property $(\star)$ holds.

The proof of $(\star)$ also reveals that functions symbols $h^y$ occur in $I_\Upsilon^\infty$ always in (sub)terms of the form $h^y(\vec{s}, \vec{g}_1(\vec{s}, \vec{t}))$, and that $\mu(u)$ is defined for each term $u$ occurring in $I_\Upsilon^\infty$. This observation and the following property of $\mu$ clearly imply the claim of this theorem: if $u$ is a cyclic term and $\mu(u)$ is defined, then $\mu(u)$ is cyclic as well. To prove the latter, it suffices to consider the following two cases.

- Assume that $u$ is cyclic due to the repetition of a function symbol $f$ not of the form $h^y$. Thus, $u$ contains a subterm of the form $f(\vec{s})$, and some $s_i \in \vec{s}$ contains a subterm of the form $f(\vec{t})$. By the definition of $\mu$, then $\mu(u)$ contains a term of the form $f(\mu(\vec{s}))$, and some $s_i' \in \mu(\vec{s})$ contains a subterm of the form $f(\mu(\vec{t}))$. Clearly, $\mu(u)$ is cyclic.

- Assume that $u$ is cyclic due to the repetition of a function symbol of the form $h^y$. By the above observation, then $u$ contains a subterm of the form $h^y(\vec{s}, \vec{g}_1(\vec{s}, \vec{t}))$, and some $s_i \in \vec{s} \cup \vec{t}$ contains a subterm of the form $h^y(\vec{v}, \vec{g}_1(\vec{v}, \vec{w}))$. By the definition of $\mu$, then $\mu(u)$ contains a subterm of the form $g_2^y(\mu(\vec{s}), \mu(\vec{t}))$, and some $s_i' \in \mu(\vec{s}) \cup \mu(\vec{t})$ contains a subterm of the form $g_2^y(\mu(\vec{v}), \mu(\vec{w}))$. Clearly, $u$ is cyclic.

(*Body Normalisation*) Assume that the set of rules $\Sigma'$ is obtained by replacing a rule $r \in \Sigma$ with rules $r_1$ and $r_2$ of the following forms, where $\vec{x} = \vec{x}_1 \cup \vec{x}_2 \cup \vec{x}_3$, and $\vec{x}_1$, $\vec{x}_2$, $\vec{x}_3$, $\vec{z}_1$, $\vec{z}_2$, and $\vec{z}_3$ are all pairwise disjoint:

$$
\begin{aligned}
r = \quad & \varphi_1(\vec{x}_1, \vec{x}_2, \vec{z}_1, \vec{z}_2) \wedge \varphi_2(\vec{x}_1, \vec{x}_3, \vec{z}_1, \vec{z}_3) \to \exists \vec{y}.\psi(\vec{x}, \vec{y}) \\
r_1 = \quad & \varphi_1(\vec{x}_1, \vec{x}_2, \vec{z}_1, \vec{z}_2) \to Q(\vec{x}_1, \vec{x}_2, \vec{z}_1) \\
r_2 = \quad & Q(\vec{x}_1, \vec{x}_2, \vec{z}_1) \wedge \varphi_2(\vec{x}_1, \vec{x}_3, \vec{z}_1, \vec{z}_3) \to \exists \vec{y}.\psi(\vec{x}, \vec{y})
\end{aligned}
$$

For each marked variable $v$, let $\vec{u}_v$ be the variables used to replace $v$ in singularisation. Then, the singularised rule $r$ can be represented as follows, where for clarity we do not show the free variables of various formulae, $\varphi_1'$ and $\varphi_2'$ do not contain atoms with predicate $\mathsf{Eq}$, and $\Gamma_1$ and $\Gamma_2$ are the conjunctions of atoms with predicate $\mathsf{Eq}$ obtained by renaming unmarked occurrences of the variables in $\varphi_1(\vec{x}_1, \vec{x}_2, \vec{z}_1, \vec{z}_2)$ and $\varphi_2(\vec{x}_1, \vec{x}_3, \vec{z}_1, \vec{z}_3)$, respectively:

$$\varphi_1' \wedge \varphi_2' \wedge \Gamma_1 \wedge \Gamma_2 \to \exists \vec{y}.\psi(\vec{x}, \vec{y})$$

Now let $M'$ be a marking that coincides with $M$ on all rules different from $r$, and that, for each marked occurrence of a variable $w \in \vec{x}_1 \cup \vec{x}_2 \cup \vec{z}_1$ in $r$, marks $r_1$ and $r_2$ as follows.

- If the marked occurrence of $w$ appears in $\varphi_1(\vec{x}_1, \vec{x}_2, \vec{z}_1, \vec{z}_2)$, then the corresponding occurrence of $w$ is marked in $r_1$; in addition, if $w \in \vec{x}_1 \cup \vec{x}_2 \cup \vec{z}_1$, then the occurrence of $w$ in atom $Q(\vec{x}_1, \vec{x}_2, \vec{z}_1)$ is marked in $r_2$.

- If the marked occurrence of $w$ appears in $\varphi_2(\vec{x}_1, \vec{x}_3, \vec{z}_1, \vec{z}_3)$, then the corresponding occurrence of $w$ is marked in $r_2$; in addition, if $w \in \vec{x}_1 \cup \vec{x}_2 \cup \vec{z}_1$, then an arbitrary occurrence of $w$ is marked in $r_1$.

Since there is no structure sharing, $\Sigma$ does not contain $r_1$, so the above definition is well-formed. The singularisation of $r_1$ and $r_2$ under $M'$ can be represented as follows:

$$\varphi_1'' \wedge \Gamma_1'' \to Q(\vec{x}_1, \vec{x}_2, \vec{z}_1)$$
$$Q(\vec{x}_1', \vec{x}_2, \vec{z}_1') \wedge \varphi_2' \wedge \Gamma_2' \to \exists \vec{y}. \psi(\vec{x}, \vec{y})$$

By the definition of $M'$, it should be clear that $\varphi_1'' \wedge \Gamma_1''$ is isomorphic to a subset of $\varphi_1' \wedge \Gamma_1'$. Based on this observation, it is now routine to prove that, if $A(\vec{t}) \in I_{\mathsf{Sing}(\Sigma,M)}^{\infty}$ and $A$ is different from the newly introduced predicate $Q$, then $A(\vec{t}) \in I_{\mathsf{Sing}(\Sigma',M')}^{\infty}$, which clearly implies our claim. □

In contrast to Theorem 57, the following example shows that normalisation with structure sharing can prevent one from finding a marking that makes the normalised rules acyclic. This example shows that normalisation must be used with care in applications that use singularisation to deal with equality.

**Example 58.** *Let $\Sigma$ be the following set of rules marked by a marking $M$ shown below.*

$$A(x) \wedge T(x^\diamond, z) \wedge B(z^\diamond) \to \exists y.[R(x, y) \wedge A(y)] \tag{105}$$
$$A(x^\diamond) \wedge T(x, z) \wedge C(z^\diamond) \to \exists y_1 \exists y_2.[S(x, y_1) \wedge T(y_1, y_2)] \tag{106}$$
$$R(z^\diamond, x_1^\diamond) \wedge R(z, x_2^\diamond) \to x_1 \approx x_2 \tag{107}$$
$$S(z^\diamond, x_1^\diamond) \wedge S(z, x_2^\diamond) \to x_1 \approx x_2 \tag{108}$$
$$T(z^\diamond, x_1^\diamond) \wedge T(z, x_2^\diamond) \to x_1 \approx x_2 \tag{109}$$

*One can show that $\mathsf{Sing}(\Sigma, M)$ is MFA w.r.t. the instance $I$ given below.*

$$I = \{A(a),\, R(a, a),\, T(a, b),\, B(b),\, A(a'),\, S(a', a'),\, T(a', b'),\, C(b')\}$$

*Furthermore, let $M_1$ be a marking identical to $M$ but which marks $A(x^\diamond)$ in rule (105), and let $M_2$ be a marking identical to $M$ but which marks $T(x^\diamond, z)$ in rule (106). One can show that neither $\mathsf{Sing}(\Sigma, M_1)$ nor $\mathsf{Sing}(\Sigma, M_2)$ is MFA w.r.t. $I$.*

*Now let $\Sigma'$ be obtained from $\Sigma$ by applying normalisation with structure sharing to rules (105) and (106); thus, rules (105) and (106) are replaced with the following rules:*

$$Q(x, z) \wedge B(z) \to \exists y.[R(x, y) \wedge A(y)] \tag{110}$$
$$Q(x, z) \wedge C(z) \to \exists y_1 \exists y_2.[S(x, y_1) \wedge T(y_1, y_2)] \tag{111}$$
$$A(x) \wedge T(x, z) \to Q(x, z) \tag{112}$$

*Note that conjunction $A(x) \wedge T(x, z)$ occurs in $\Sigma'$ only in rule (112); therefore, variable $x$ in this conjunction can be marked in only one way. This, however, has the same effect as choosing $M_1$ or $M_2$ for $\Sigma$: no possible marking $M'$ will make $\mathsf{Sing}(\Sigma', M')$ MFA w.r.t. $I$. Intuitively, normalisation with structure sharing reduces the space of available markings, due to which it may be impossible to find a marking that makes the rules acyclic.* ◇

## 6. Applying Acyclicity to Horn Description Logics

In this section we apply various acyclicity notions to reasoning problems in *description logic* (DL) ontologies. Description logics are knowledge representation formalisms that underpin the Web Ontology Language (OWL). DL ontologies are constructed from *atomic concepts* (i.e., unary predicates), *atomic roles* (i.e., binary predicates), and *individuals* (i.e., constants). Special atomic concepts ⊤ and ⊥ denote universal truth and falsehood, respectively. For each atomic role $R$, expression $R^-$ is an *inverse role*; furthermore, a *role* is an atomic or an inverse role. DLs provide a rich set of *constructors* for building *concepts* (first-order formulae with one free variable) from atomic concepts and roles. A description logic *TBox* is a set of *axioms*, which correspond to first-order sentences. In this paper we consider only *Horn* description logics, in which TBoxes can be translated into existential rules. Furthermore, in this paper we will consider only normalised TBoxes, in which concepts do not occur nested in other concepts. The latter assumption is without loss of generality as each Horn description logic TBox can be normalised in linear time, and the normalised ontology is a model-conservative extension of the original one.

In this paper we consider several logics all of which are fragments of the description logic Horn-$\mathcal{SROIF}$, which provides the formal underpinning for a prominent subset of OWL. A normalised Horn-$\mathcal{SROIF}$ *TBox* $\mathcal{T}$ consists of axioms shown on the left-hand side of Table 1; in the table, $A$, $B$, and $C$ are atomic concepts (including possibly ⊤ and ⊥), $R$, $S$, and $T$ are (not necessarily atomic) roles, and $a$ is an individual. To guarantee decidability of reasoning, $\mathcal{T}$ must satisfy certain *global* conditions (Kutz, Horrocks, & Sattler, 2006), which we omit for the sake of brevity. Roughly speaking, only so-called *simple* roles are allowed to occur in axioms of Type 2, and axioms of Type 6 must be *regular* according to a particular condition that allows such axioms to be represented using a nondeterministic finite automaton. We also consider the following fragments of Horn-$\mathcal{SROIF}$.

- Horn-$\mathcal{SRI}$ TBoxes are not allowed to contain axioms of Type 2 or 7.

- Horn-$\mathcal{SHIF}$ TBoxes are not allowed to contain axioms of Type 7, and all axioms of Type 6 satisfy $R = S = T$. Note that all Horn-$\mathcal{SHIF}$ TBoxes are regular.

- Horn-$\mathcal{SHI}$ TBoxes inherit the restrictions from Horn-$\mathcal{SHIF}$ and are further not allowed to contain axioms of type 2.

To simplify the presentation, we do not consider general at-least number restrictions—that is, concepts of the form $\geq n\, R.A$ with $n > 1$. The translation of such concepts into rules would require an explicit inequality predicate. As explained in Section 2.2, the inequality predicate can be simulated using an ordinary predicate, and so the extension of our results to general at-least number restrictions is straightforward.

In the rest of this paper we allow inverse roles to occur in atoms, so we take an atom of the form $R^-(t_1, t_2)$ with $R$ an atomic role as an abbreviation for $R(t_2, t_1)$. Then, each Horn-$\mathcal{SROIF}$ axiom corresponds to an existential rule as shown in Table 1. As explained in Section 2.2, we treat ⊤ and ⊥ as ordinary unary predicates where ⊤ is explicitly axiomatised. Thus, we can take a substitution $\theta$ to be an answer to a CQ $Q(\vec{x})$ w.r.t. a $\mathcal{T}$ and $I$ if $\mathcal{T} \cup I \models Q(\vec{x})\theta$ or $I \cup \mathcal{T} \models \exists y.\bot(y)$; the latter condition takes into account that an unsatisfiable theory entails all possible formulae. Due to this close correspondence between description logic axioms and existential rules, in the rest of this paper we identify a TBox $\mathcal{T}$ with the corresponding set of rules.

| | | |
|---|---|---|
| 1. | $A \sqsubseteq \exists R.B$ | $A(x) \rightarrow \exists y.[R(x,y) \wedge B(y)]$ |
| 2. | $A \sqsubseteq \, \leq 1\, R.B$ | $A(z) \wedge R(z,x_1) \wedge B(x_1) \wedge R(z,x_2) \wedge B(x_2) \rightarrow x_1 \approx x_2$ |
| 3. | $A \sqcap B \sqsubseteq C$ | $A(x) \wedge B(x) \rightarrow C(x)$ |
| 4. | $A \sqsubseteq \forall R.B$ | $A(z) \wedge R(z,x) \rightarrow B(x)$ |
| 5. | $R \sqsubseteq S$ | $R(x_1,x_2) \rightarrow S(x_1,x_2)$ |
| 6. | $R \circ S \sqsubseteq T$ | $R(x_1,z) \wedge S(z,x_2) \rightarrow T(x_1,x_2)$ |
| 7. | $A \sqsubseteq \{a\}$ | $A(x) \rightarrow x \approx a$ |

Table 1: Axioms of normalised Horn-$\mathcal{SROIF}$ ontologies and corresponding rules

The complexity of answering Boolean conjunctive queries over general (i.e., not acyclic) DL TBoxes is 2EXPTIME- and EXPTIME-complete for Horn-$\mathcal{SROIF}$ (Ortiz et al., 2011) and Horn-$\mathcal{SHIF}$ (Eiter et al., 2008), respectively. In the rest of this section we investigate the complexity of this problem on acyclic ontologies, as well as the complexity of acyclicity checking. In particular, in Section 6.1 we consider the case when the TBox is expressed in Horn-$\mathcal{SROIF}$, for which we show that both BCQ answering and MFA checking are EXPTIME-complete. Then, in Section 6.2 we consider Horn-$\mathcal{SHIF}$ TBoxes, for which we show that the complexity of these problems drops to PSPACE.

## 6.1 Acyclic Horn-$\mathcal{SROIF}$ TBoxes

We start by showing that BCQ answering for WA Horn-$\mathcal{SRI}$ TBoxes is EXPTIME-hard. Intuitively, this is due to the axioms of Type 6, which can be used to axiomatise existence of non-tree-like structures. Although regularity ensures that axioms of Type 6 can be represented by a nondeterministic finite automaton, this automaton can be exponential; as a consequence, axioms of Type 6 can axiomatise exponential non-tree-like structures, which is the main source of complexity.

**Lemma 59.** *Let $\mathcal{T}$ be a WA Horn-$\mathcal{SRI}$ TBox, let $I$ be an instance, and let $F$ be a fact. Then, checking whether $I \cup \mathcal{T} \models F$ is* EXPTIME-*hard.*

*Proof.* Let $\mathcal{M} = (\mathcal{S}, \mathcal{Q}, \delta, Q_0, Q_a)$ be a deterministic Turing machine, where $\mathcal{S}$ is a finite set of symbols, $\mathcal{Q}$ is a finite set of states, $\delta : \mathcal{Q} \times \mathcal{S} \rightarrow \mathcal{Q} \times \mathcal{S} \times \{\leftarrow, \rightarrow\}$ is a transition function, $Q_0 \in \mathcal{Q}$ is the initial state, and $Q_a$ is the accepting state. Furthermore, assume that an integer $k$ exists such that $\mathcal{M}$ halts on each input of length $n$ in time $2^{n^k}$. Given an arbitrary input $S_{i_1}, \ldots, S_{i_n}$, we construct an MFA set of Horn-$\mathcal{SRI}$ rules $\mathcal{T}$ and an instance $I$ such that $I \cup \mathcal{T} \models Q_a(a)$ if and only if $\mathcal{M}$ accepts the input. To simplify the presentation, we will use a slightly more general rule syntax than what is allowed by Table 1; however, all such rules can be brought into the required form by renaming parts of the rules with fresh predicates.

Let $\ell = n^k$; since $k$ is a constant, $\ell$ is polynomial in $n$. Our construction uses a unary predicate for each symbol and state; for simplicity, we do not distinguish between the predicate and the symbol/state. In addition, the construction also uses binary predicates $L_i$, $R_i$, $T_i$, $U_i$, $D_i$, $H_i$, and $V_i$ for $1 \leq i \leq \ell$, unary predicates $A_i$ and $B_i$ for $0 \leq i \leq \ell$, and unary predicates $O_1, \ldots, O_{n+1}$, $N_1$, and $N_2$. Instance $I$ contains only the fact $A_0(a)$. We next present the rules of $\mathcal{T}$. The set $\mathcal{T}$ will contain only Horn rules without empty heads, so it will be satisfiable in a minimal Herbrand model. For readability, we divide $\mathcal{T}$ into groups of rules and prove for each group various facts about this minimal Herbrand model of $\mathcal{T} \cup I$, which is shown schematically in Figure 2. The construction of $\mathcal{T}$ proceeds along the following lines.

- The first, the second, and the third group of rules construct the exponential grid shown at the bottom of Figure 2, whose edges are labelled with $H_\ell$ and $V_\ell$. Each sequence of $V_\ell$-edges will be used to encode the contents of the tape of the Turing machine at some point in time; furthermore, precisely one vertex in each such sequence will be labelled with a state, thus representing the position of the head. In contrast, $H_\ell$-edges will connect different points in time and will be used to encode the transitions of the Turing machine.

- The fourth group labels the right-most $V_\ell$-chain with $S_{i_1}, \ldots, S_{i_n}, S_\sqcup, S_\sqcup, \ldots, S_\sqcup, S_\sqcup$, where $S_\sqcup$ represents the empty tape symbol.

- The fifth and the sixth groups ensure that the symbols on the tape that are not modified by a move of a Turing machine are propagated between time points.

- The seventh and the eighth group encode the transitions of the Turing machine.

- The ninth group propagates the acceptance condition to the top of the figure by labelling the individual $a$ with the accepting state $Q_a$.

We next present the rules of $\mathcal{T}$ in detail.

The first group of rules in $\mathcal{T}$ contains rules (113)–(115) for each $0 < i \leq \ell$, and rule (116) for each $1 < i \leq \ell$.

$$A_{i-1}(x) \rightarrow \exists y.[L_i(x, y) \wedge A_i(y)] \tag{113}$$

$$A_{i-1}(x) \rightarrow \exists y.[R_i(x, y) \wedge A_i(y)] \tag{114}$$

$$R_i(z, x) \wedge L_i(z, x') \rightarrow T_i(x, x') \tag{115}$$

$$L_i(z, x) \wedge T_{i-1}(z, z') \wedge R_i(z', x') \rightarrow T_i(x, x') \tag{116}$$

On $I$, these rules axiomatise existence of a triangular structure in the top part of Figure 2 containing $T_i$ links.

The second group of rules in $\mathcal{T}$ contains rule (117), rules (118)–(120) for each $0 < i \leq \ell$, and rule (121) for each $1 < i \leq \ell$.

$$A_\ell(x) \rightarrow B_0(x) \tag{117}$$

$$B_{i-1}(x) \rightarrow \exists y.[U_i(x, y) \wedge B_i(y)] \tag{118}$$

$$B_{i-1}(x) \rightarrow \exists y.[D_i(x, y) \wedge B_i(y)] \tag{119}$$

$$U_i(z, x) \wedge D_i(z, x') \rightarrow V_i(x, x') \tag{120}$$

$$D_i(z, x) \wedge V_{i-1}(z, z') \wedge U_i(z', x') \rightarrow V_i(x, x') \tag{121}$$

These rules axiomatise existence of triangular structures in the bottom part of Figure 2 containing $V_i$ links.

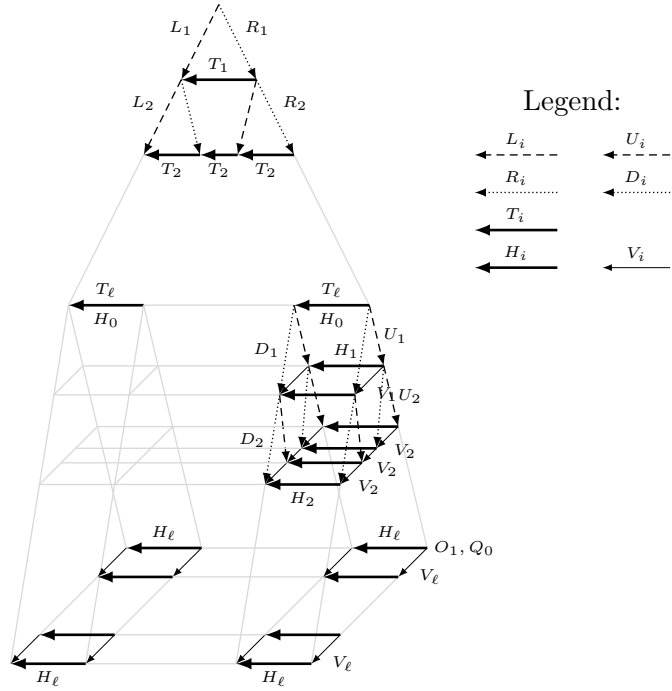

Figure 2: Grid Model of $\mathcal{T}$

The third group of rules in $\mathcal{T}$ contains rule (122), and rules (123) and (124) for each $0 < i \leq \ell$.

$$T_\ell(x, x') \to H_0(x, x') \tag{122}$$

$$U_i(z, x) \wedge H_{i-1}(z, z') \wedge U_i(z', x') \to H_i(x, x') \tag{123}$$

$$D_i(z, x) \wedge H_{i-1}(z, z') \wedge D_i(z', x') \to H_i(x, x') \tag{124}$$

These rules axiomatise existence of $H_i$ links, which with $V_i$ links form a grid of size $2^i \times 2^i$ shown in Figure 2.

In the rest of this proof, for variables $w_0$ and $w_\ell$, we use $R^\ell(w_0, w_\ell)$ as an abbreviation for $R_1(w_0, w_1) \wedge \ldots \wedge R_\ell(w_{\ell-1}, w_\ell)$, where each $w_i$ with $0 < i < m$ is a variable not occurring outside the conjunction. Furthermore, we analogously use $U^\ell(w_0, w_\ell)$ as an abbreviation for $U_1(w_0, w_1) \wedge \ldots \wedge U_\ell(w_{\ell-1}, w_\ell)$.

The fourth group of rules in $\mathcal{T}$ contains rule (125), rules (126) and (127) for each $1 \leq j \leq n$, and rules (128)–(129), where $S_\sqcup$ is the empty tape symbol. Remember that $S_{i_1}, \ldots, S_{i_n}$ encodes the input to $\mathcal{M}$.

$$A_0(z) \wedge R^\ell(z, z') \wedge U^\ell(z', x) \to O_1(x) \wedge Q_0(x) \tag{125}$$

$$O_j(z) \wedge V_\ell(z, x) \to O_{j+1}(x) \tag{126}$$

$$O_j(x) \to S_{i_j} \tag{127}$$

$$O_{n+1}(z) \wedge V_\ell(z, x) \to O_{n+1}(x) \tag{128}$$

$$O_{n+1}(x) \to S_\sqcup(x) \tag{129}$$

Rule (125) labels the grid origin and sets the initial state as shown in Figure 2. Rules (126) ensure that the $n$ subsequent nodes are labelled with $O_2, \ldots, O_{n+1}$, and rule (128) propagates $O_{n+1}$ to the rest of the $V_\ell$-chain. Finally, rules (127) and (129) ensure that nodes labelled with $O_j$ are also labelled with $S_{i_j}$, and that nodes labeled with $O_{n+1}$ are labeled with $S_\sqcup$. Thus, this group of rules in $\mathcal{T}$ ensures that the right-most $V_\ell$-chain in the grid contains the initial state of the tape of $\mathcal{M}$.

The fifth group of rules in $\mathcal{T}$ contains rules (130)–(131) for each state $Q_k \in \mathcal{Q}$, and rules (132)–(133). These rules essentially ensure that all nodes before and after a node labelled with some state $Q_k \in \mathcal{Q}$ are labeled with $N_1$ and $N_2$, respectively, thus indicating that the head is not above the node.

$$Q_k(z) \wedge V_\ell(x, z) \rightarrow N_1(x) \tag{130}$$

$$Q_k(z) \wedge V_\ell(z, x) \rightarrow N_2(x) \tag{131}$$

$$N_1(z) \wedge V_\ell(x, z) \rightarrow N_1(x) \tag{132}$$

$$N_2(z) \wedge V_\ell(z, x) \rightarrow N_2(x) \tag{133}$$

The sixth group of rules in $\mathcal{T}$ contains rules (134)–(135) instantiated for each symbol $S_k \in \mathcal{S}$; these rules ensure that the contents of the tape is copied between successive time points for all points in the grid not containing the head.

$$N_1(z) \wedge S_k(z) \wedge H_\ell(z, x) \rightarrow S_k(x) \tag{134}$$

$$N_2(z) \wedge S_k(z) \wedge H_\ell(z, x) \rightarrow S_k(x) \tag{135}$$

The seventh group of rules in $\mathcal{T}$ contains rules (136)–(137) instantiated for each symbol $S_k \in \mathcal{S}$ and each state $Q_k \in \mathcal{Q}$ such that $\delta(Q_k, S_k) = (Q_{k'}, S_{k'}, \leftarrow)$. These rules encode moves of $\mathcal{M}$ where the head moves left.

$$Q_k(z) \wedge S_k(z) \wedge H_\ell(z, x) \rightarrow S_{k'}(x) \tag{136}$$

$$Q_k(z) \wedge S_k(z) \wedge H_\ell(z, z') \wedge V_\ell(x, z') \rightarrow Q_{k'}(x) \tag{137}$$

The eighth group of rules in $\mathcal{T}$ contains rules (138)–(139) instantiated for each symbol $S_k \in \mathcal{S}$ and each state $Q_k \in \mathcal{Q}$ such that $\delta(Q_k, S_k) = (Q_{k'}, S_{k'}, \rightarrow)$. These rules encode moves of $\mathcal{M}$ where the head moves right.

$$Q_k(z) \wedge S_k(z) \wedge H_\ell(z, x) \rightarrow S_{k'}(x) \tag{138}$$

$$Q_k(z) \wedge S_k(z) \wedge H_\ell(z, z') \wedge V_\ell(z', x) \rightarrow Q_{k'}(x) \tag{139}$$

The ninth group of rules in $\mathcal{T}$ contains rules (140)–(143) for each $1 \leq i \leq \ell$; these rules simply ensure that acceptance is propagated back to the root of the upper tree.

$$Q_a(z) \wedge U_i(x, z) \rightarrow Q_a(x) \tag{140}$$

$$Q_a(z) \wedge D_i(x, z) \rightarrow Q_a(x) \tag{141}$$

$$Q_a(z) \wedge L_i(x, z) \rightarrow Q_a(x) \tag{142}$$

$$Q_a(z) \wedge R_i(x, z) \rightarrow Q_a(x) \tag{143}$$

The above discussion shows that labelling of the nodes in the grid shown in Figure 2 simulates the execution of $\mathcal{M}$ on input $S_{i_1}, \ldots, S_{i_n}$, where the contents of the tape at some time instant is represented by a $V_\ell$-chain, and $H_\ell$-links connect tape cells at successive time instants. Thus, $I \cup \mathcal{T} \models Q_a(a)$ if and only if $\mathcal{M}$ accepts $S_{i_1}, \ldots, S_{i_n}$ in time $2^\ell$. It is straightforward to see that $\mathcal{T}$ is WA, so the claim of this theorem holds.  □

The proof of Lemma 59 can be adapted to obtain the lower bound for checking MFA of Horn-$\mathcal{SRI}$ rules.

**Lemma 60.** *Checking whether a Horn-$\mathcal{SRI}$ TBox is universally MFA is* ExpTime-*hard.*

*Proof.* Let $\mathcal{M}$ be an arbitrary deterministic Turing machine and let $S_{i_1}, \ldots, S_{i_n}$ be an input string on which $\mathcal{M}$ terminates in time $2^{n^k}$. For such $\mathcal{M}$ and $S_{i_1}, \ldots, S_{i_n}$, let $\mathcal{T}$ be as in the proof of Lemma 59. TBox $\mathcal{T}$ is WA, it contains only constant-free, equality-free, and connected rules, and no predicate in $\mathcal{T}$ is of zero arity; hence, by Lemma 7, a Horn-$\mathcal{SRI}$ TBox $\mathcal{T}'$ exists such that $\mathcal{M}$ accepts $S_{i_1}, \ldots, S_{i_n}$ if and only if $\mathcal{T}'$ is not universally MFA.  □

Note that Lemmas 59 and 60 apply to Horn-$\mathcal{SRI}$ and thus do not rely on a particular treatment of equality. We can deal with the equality predicate in Horn-$\mathcal{SROIF}$ TBoxes using singularisation as described in Section 5, which leads us to the following result.

**Theorem 61.** *Let $\mathcal{T}$ be a Horn-$\mathcal{SROIF}$ TBox, let $M$ be a marking of $\mathcal{T}$, let $I$ be an instance, and let $Q$ be a BCQ. Then, checking whether $\mathsf{Sing}(\mathcal{T}, M)$ is MFA w.r.t. $I$ is* ExpTime-*complete. Furthermore, if $\mathsf{Sing}(\mathcal{T}, M)$ is MFA w.r.t. $I$, then checking whether $I \cup \mathcal{T} \models Q$ holds is* ExpTime-*complete as well.*

*Proof.* Note that all rules in Table 1 are $\exists$-1 rules. Since all rules in $\mathsf{Sing}(\mathcal{T}, M)$ are $\exists$-1 rules as well, Theorem 10 gives us an ExpTime upper bound for both of our problems. The matching lower bounds follow from Lemmas 59 and 60 (note that every Horn-$\mathcal{SRI}$ TBox is also a Horn-$\mathcal{SROIF}$ TBox) and the fact that their proofs do not use predicate $\approx$.  □

In fact, Theorem 10 provides us with even stronger complexity bounds. In particular, even if $\mathcal{T}$ does not satisfy all the required global conditions, and even if $\mathcal{T}$ is extended with SWRL rules (Horrocks, Patel-Schneider, Bechhofer, & Tsarkov, 2005), the rules in $\mathcal{T}$ are all still $\exists$-1 rules. Thus, one can decide whether such $\mathcal{T}$ is MFA (universally or w.r.t. an instance) in ExpTime, and if that is the case, one can answer BCQs in ExpTime as well. Consequently, ontology-based applications can freely use the expressivity beyond what is currently available in OWL without an increase in the complexity of reasoning, assuming that the resulting TBox is acyclic.

We conclude this section by observing that MSA provides us with a tractable notion for Horn-$\mathcal{SROIF}$ rules. Intuitively, all rules in $\mathsf{MSA}(\mathcal{T})$ have a bounded number of variables and all predicates in $\mathsf{MSA}(\mathcal{T})$ are of bounded arity, which eliminates all sources of intractability in datalog reasoning. We prove the matching lower bound in Section 6.2 for the more specific case of Horn-$\mathcal{SHIF}$ ontologies.

**Theorem 62.** *Let $\mathcal{T}$ be Horn-$\mathcal{SROIF}$ TBox, let $M$ be a marking, and let $I$ be an instance. Then, checking whether $\mathsf{Sing}(\mathcal{T}, M)$ is MSA w.r.t. $I$ is in* PTime.

*Proof.* As one can see in Table 1, the rules in $\mathcal{T}$ all contain a bounded number of variables and atoms in the body, and so the number of variables in the body of each rule in $\mathsf{MSA}(\mathsf{Sing}(\mathcal{T}, M))$ is bounded as well. Furthermore, the datalog program $\mathsf{MSA}(\mathsf{Sing}(\mathcal{T}, M))$ contains predicates of bounded arity, so its chase w.r.t. $I$ is polynomial in size. Thus, the chase of $I$ and $\mathsf{MSA}(\mathcal{T})$ can be computed in polynomial time, which implies our claim. □

## 6.2 Acyclic Horn-$\mathcal{SHIF}$ TBoxes

The exponential lower bound of Lemmas 59 and 60 critically depend on axioms of Type 6, which can be used to encode exponential structures; furthermore, a combination of inverse roles and axioms of Types 2 and 7 (i.e., of inverse roles, number restrictions, and nominals) is also well known to be problematical (Horrocks & Sattler, 2007). In practice, however, TBoxes are often expressed in Horn-$\mathcal{SHIF}$, which disallows such axioms in TBoxes. We next show that, for such TBoxes, the complexity of both problems drops to PSPACE.

We first prove PSPACE-hardness for both problems. Note that the PSPACE-hardness proof of concept satisfiability checking by Baader, Calvanese, McGuinness, Nardi, and Patel-Schneider (2007) is not applicable to Horn ontologies since it uses disjunctive concepts. Nonetheless, PSPACE-hardness can be proved by a reduction from checking QBF validity.

**Lemma 63.** *Let $\mathcal{T}$ be a WA Horn-$\mathcal{SHI}$ TBox, let $I$ be an instance, and let $F$ be a fact. Then, checking whether $I \cup \mathcal{T} \models F$ is PSPACE-hard.*

*Proof.* Let $\varphi = Q_1 x_1 \ldots Q_n x_n.C_1 \wedge \ldots \wedge C_k$ be an arbitrary quantified Boolean formula defined over variables $x_1, \ldots, x_n$, where each $Q_i \in \{\exists, \forall\}$, $1 \leq i \leq n$ is a quantifier, and each $C_j$, $1 \leq j \leq k$ is a clause of the form $C_j = L_{j,1} \vee L_{j,2} \vee L_{j,3}$. Checking validity of $\varphi$ is the canonical PSPACE-hard problem.

In the rest of this proof, for a binary predicate $P$ and variables $w_0$ and $w_m$, we use $P^m(w_0, w_m)$ as an abbreviation for $P(w_0, w_1) \wedge \ldots \wedge P(w_{m-1}, w_m)$, where each $w_i$ with $0 < i < m$ is a variable not occurring outside the conjunction. Let $\mathcal{T}$ be the Horn-$\mathcal{SHI}$ TBox containing rules (144)–(147) for each $1 \leq i \leq n$, rule (148) for each clause $C_j$ and each literal $L_{j,m} = x_\ell$ occurring in $C_j$, rule (149) for each clause $C_j$ and each literal $L_{j,m} = \neg x_\ell$ occurring in $C_j$, rule (150), rule (151) for each $1 \leq i \leq n$ such that $Q_i = \exists$, and rule (152) for each $1 \leq i \leq n$ such that $Q_i = \forall$.

$$A_{i-1}(x) \rightarrow \exists y.[X_i^+(x, y) \wedge A_i(y)] \tag{144}$$

$$A_{i-1}(x) \rightarrow \exists y.[X_i^-(x, y) \wedge A_i(y)] \tag{145}$$

$$X_i^+(x, x') \rightarrow P(x, x') \tag{146}$$

$$X_i^-(x, x') \rightarrow P(x, x') \tag{147}$$

$$X_\ell^+(z', z) \wedge P^{n-\ell}(z, x) \wedge A_n(x) \rightarrow C_j(x) \tag{148}$$

$$X_\ell^-(z', z) \wedge P^{n-\ell}(z, x) \wedge A_n(x) \rightarrow C_j(x) \tag{149}$$

$$C_1(x) \wedge \ldots \wedge C_k(x) \rightarrow F_n(x) \tag{150}$$

$$P(x, z) \wedge F_i(z) \rightarrow F_{i-1}(x) \tag{151}$$

$$X_i^+(x, z) \wedge F_i(z) \wedge X_i^-(x, z') \wedge F_i(z') \rightarrow F_{i-1}(x) \tag{152}$$

Strictly speaking, rules (148), (149), (150), and (152) are not Horn-$\mathcal{SHI}$ rules, but they can be transformed into Horn-$\mathcal{SHI}$ rules by replacing parts of their bodies with fresh concepts. It is straightforward to see that $\mathcal{T}$ is WA.

Let $I = \{A_0(a)\}$, and let $I_{\mathcal{T}}^{\infty}$ be the chase of $I$ and $\mathcal{T}$. Due to rules (144)–(145), $I_{\mathcal{T}}^{\infty}$ contains a binary tree of depth $n$ in which each leaf node is reachable from $a$ via a path that, for each $1 \leq i \leq n$, contains either $X_i^+$ or $X_i^-$. If we interpret the presence of $X_i^+$ and $X_i^-$ as assigning variable $x_i$ to $\mathsf{t}$ and $\mathsf{f}$, respectively, then each leaf node corresponds to one possible assignment of $x_1, \ldots, x_n$. Rules (148) and (149) then clearly label each leaf node with the clauses that are true in the node, and rule (150) labels each leaf node with $F_n$ for which all clauses are true. Finally, rules (151) and (152) label each interior node of the tree with $F_{i-1}$ according to the semantics of the appropriate quantifier of $\varphi$. Clearly, $\varphi$ is valid if and only if $I \cup \mathcal{T} \models F_0(a)$, which implies our claim. $\square$

We next turn our attention to the upper bounds on the complexity of answering a BCQ over an MFA TBox, and checking whether a TBox is MFA. While in Section 6.1 we considered a TBox $\mathcal{T}$ singularised according to some marking $M$, in this section we assume that equality in $\mathcal{T}$ is handled by means of an explicit axiomatisation $\mathcal{T}_{\approx}$. As we explain next, this is because singularised rules are not 'local', which makes a PSPACE membership proof quite difficult. For example, consider the following singularised rule:

$$A(x) \wedge x \approx x' \wedge B(x') \rightarrow C(x) \tag{153}$$

Atoms $A(x)$ and $B(x')$ in the rule do not share variables and therefore need not be matched 'locally' in the chase of $\mathsf{Sing}(\mathcal{T}, M)$ and $I$; furthermore, the chase can be exponential in size, so it is not trivial to see how it can be explored using polynomial space. Nevertheless, we conjecture that it is possible to extend our proof to singularised rules as well; however, the details involved seem quite technical, without explaining much about the nature of BCQ answering under equality. Therefore, we leave this problem open and restrict ourselves to the technically simpler case when equality in $\mathcal{T}$ is encoded explicitly using $\mathcal{T}_{\approx}$.

We next show that answering a BCQ $Q$ over an MFA Horn-$\mathcal{SHIF}$ TBox $\mathcal{T}$ and an instance $I$ can be performed in polynomial space. The proof uses the well-known tracing technique of inspecting a model of $\mathcal{T} \cup I$ using polynomial space. The key aspect of this result, however, is dealing with the transitive roles in the query, which allow the query to be embedded non-locally into the chase of $\mathcal{T}$ and $I$. Note, however, that we can guess an embedding of $Q$ into the result of $I_{\mathcal{T}}^{\infty}$ using nondeterministic polynomial time; furthermore, since $I_{\mathcal{T}}^{\infty}$ is a minimal Herbrand model of $\mathcal{T}$ (i.e., since $\mathcal{T}$ is Horn), we can check the entailment of each mapped atom of $Q$ separately, and in doing so we can use the well-known encoding by Demri and de Nivelle (2005) to handle transitive roles.

**Theorem 64.** *Let $\mathcal{T}$ be Horn-$\mathcal{SHIF}$ TBox, let $I$ be an instance such that $\mathcal{T}$ is MFA w.r.t. $I$, and let $Q$ be a BCQ. Then, checking whether $I \cup \mathcal{T} \models Q$ is* PSPACE*-complete.*

*Proof.* Hardness follows from Lemma 63. We next present a nondeterministic polynomial space algorithm that decides $I \cup \mathcal{T} \models Q$; by Savitch's Theorem, this algorithm can be transformed into a deterministic polynomial space algorithm, which proves our claim.

Assume that BCQ $Q$ is of the form $Q = \exists \vec{y}.B_1 \wedge \ldots \wedge B_n$. Furthermore, let $\Upsilon = \mathsf{sk}(\mathcal{T})$. Since $\bot$ is just a regular atomic concept, $I \cup \mathcal{T}$ is always satisfiable in the chase $I_{\mathcal{T}}^{\infty}$ of $I$ and $\mathcal{T}$. Furthermore, $I \cup \mathcal{T} \models Q$ if and only if a substitution $\theta$ from the variables in $\vec{y}$ to the terms in $I_{\mathcal{T}}^{\infty}$ exists such that $B_i \theta \in I_{\mathcal{T}}^{\infty}$ for each $1 \le i \le n$; the latter clearly holds if and only if $I \cup \Upsilon \models B_i \theta$. As shown in the proof of Theorem 10, each term in $I_{\mathcal{T}}^{\infty}$ is of the form $g_1(\ldots g_\ell(a) \ldots)$, where $\ell$ is less than or equal to the number of function symbols in $\Upsilon$. Thus, the first step in deciding $I \cup \mathcal{T} \models Q$ is to examine all possible $\theta$ and then check $I \cup \Upsilon \models B_i \theta$ for each $1 \le i \le n$; this can clearly be done using a deterministic Turing machine that uses polynomial space to store $\theta$, provided that each individual check $I \cup \Upsilon \models B_i \theta$ can also be decided in polynomial space.

If $B_i \theta$ is of the form $C(t)$, then let $\Upsilon' = \Upsilon$, and let $D = C$. Alternatively, if $B_i \theta$ is of the form $R(t', t)$, then let $\Upsilon'$ be $\Upsilon$ extended with the following rules, where $D$ and $E$ are fresh concepts not occurring in $\Upsilon$ and $I$:

$$\to E(t') \tag{154}$$
$$E(z) \wedge R(z, x) \to D(x) \tag{155}$$

It is straightforward to see that $I \cup \Upsilon \models R(t', t)$ if and only if $I \cup \Upsilon' \models D(t)$. Let $\Upsilon''$ be obtained from $\Upsilon'$ by deleting each rule in $\Upsilon'$ of the form

$$R(x_1, z) \wedge R(z, x_2) \to R(x_1, x_2) \tag{156}$$

and, for each role $R$ occurring in such a rule, replacing each rule of the form

$$A(z) \wedge R(z, x) \to B(x) \tag{157}$$

with the following rules, where $Q_{A,R,B}$ is a fresh concept unique for $A$, $R$, and $B$:

$$A(z) \wedge R(z, x) \to Q_{A,R,B}(x) \tag{158}$$
$$Q_{A,R,B}(z) \wedge R(z, x) \to Q_{A,R,B}(x) \tag{159}$$
$$Q_{A,R,B}(x) \to B(x) \tag{160}$$

This transformation corresponds to the well-known elimination of transitivity by Demri and de Nivelle (2005), so $I \cup \Upsilon' \models D(t)$ if and only if $I \cup \Upsilon'' \models D(t)$; the proof of this claim is straightforward and we omit it for the sake of brevity.

Let $\Xi$ be $\Upsilon''$ extended with the equality axioms (2) and (4). Since $\approx$ does not occur in the body of the rules in $\Upsilon''$, we have that $I \cup \Upsilon'' \not\models D(t)$ if and only if $I \cup \Xi \not\models_{\approx} D(t)$. Let $I_{\Xi}^{\infty}$ be the chase for $I$ and $\Xi$; then $I \cup \Xi \not\models_{\approx} D(t)$ if and only if $D(t) \notin I_{\Xi}^{\infty}$. Note that $\Xi$ contains rules of Types 1–5 from Table 1, rules (2) and (4), and possibly rules of the form $\to E(t_1)$. These facts can be used to show that each assertion in $I_{\Xi}^{\infty}$ is of one of the following forms, where $a$ and $b$ are constants, $t$ is a constant or a term that contains only unary function symbols, $f$ and $g$ are unary function symbols, $C$ is an atomic concept, and $R$ is an atomic role:

- $C(t)$,

- $R(a, b)$, $R(a, f(b))$, $R(f(b), a)$, $R(t, f(t))$, $R(f(t), t)$, or

- $t \approx f(g(t))$, $f(t) \approx g(t)$, $a \approx b$, $a \approx f(b)$, or an equality symmetric to these ones.

The proof is by induction on the length of the chase sequence for $I$ and $\Xi$, and the claim follows straightforwardly from the $I_\Xi^{\approx}$ form of rules of Types 1–5. Motik et al. (2009b) prove an analogous claim for a more general description logic, and their proof carries over to the above setting with only syntactic changes.

We say that $x$ is the *central variable* in a rule of Type 1 or 3, and that $z$ is the *central variable* in a rule of Type 2 or 4. W.l.o.g. we assume that the body of a rule of Type 5 does not contain inverse roles; then, $x_1$ is the *central variable* of a rule of Type 5. Finally, in the equality replacement rules (4), the *central variable* is the variable being replaced.

Clearly, $D(t) \notin I_\Xi^{\approx}$ if and only if a Herbrand interpretation $J$ exists in which all assertions are of the form mentioned above, such that $I_\Xi^{\approx} \subseteq J$, $J \models_{\approx} \Xi$, and $D(t) \notin J$. We next show how to check the existence of such $J$ using a nondeterministic Turing machine that runs in polynomial space.

Let $f_1, \ldots, f_m$ be all function symbols occurring in $\Xi$. We first guess a Herbrand interpretation $J^0$ over the constants of $I$ satisfying $I \subseteq J^0$, and we check whether all rules in $\Xi$ not of Type 1 are satisfied in $J^0$. If that is the case, we consider each constant $c$ in $J_0$ and call the following procedure for $s = c$ and $i = 1$:

1. If $i = m + 1$ return *true*.

2. Guess a Herbrand interpretation $J^i$ such that each assertion in $J^i$ is of a form as specified earlier and involves at least one term among $f_1(s), \ldots, f_m(s)$.

3. If $D(t) \in J^i$, return *false*.

4. Check whether the equality symmetry rule (4) is satisfied in $J^i$; if not, return *false*.

5. Check whether $J^i \cup J^{i-1} \cup \ldots \cup J^0$ satisfies each rule in $\Xi$ if the central variable of the rule is mapped to $s$; if this is not the case for each rule, return *false*.

6. For each $1 \leq k \leq m$, recursively call this procedure for $f_k(s)$ and $i + 1$; if one of this calls returns *false*, return *false* as well.

7. Return *true*.

Assume that this procedure returns *true* for each constant $c$, and let $J$ be the union of all $J^i$ considered in the process. It is straightforward to see that $I \subseteq J$ and $D(t) \notin J$; furthermore, $J \models_{\approx} \Xi$ holds since the satisfaction of each rule $r \in \Xi$ in $J$ can be ascertained 'locally', by inspecting the vicinity of the ground term that is mapped to the central variable of $r$. Furthermore, the recursion depth of our algorithm is $m$ and at each recursion level we need to keep a polynomially sized interpretation $J^i$, so our algorithm can be implemented using a nondeterministic Turing machine that uses polynomial space. □

**Theorem 65.** *Let $\mathcal{T}$ be Horn-$\mathcal{SHIF}$ TBox, and let $I$ be an instance. Then, deciding whether $\mathcal{T}$ is MFA w.r.t. $I$ is in* PSPACE, *and deciding whether $\mathcal{T}$ is universally MFA is* PSPACE-*hard.*

*Proof.* (Membership) Rules in MFA($\mathcal{T}$) are 'almost' Horn-$\mathcal{SHIF}$ rules: rule (19) can be made a Horn-$\mathcal{SHIF}$ rule by replacing S in the body with D (which clearly does not affect the consequences of the rule), and the fact that rule (20) contains a nullary atom in the head is immaterial. Thus, the claim can be proved by a straightforward adaptation of the membership proof of Theorem 64. The main difference in the algorithm is that, with $n$ function symbols, we need to examine the models to depth $n+1$; however, such an algorithm still uses polynomial space.

(Hardness) Let $\varphi$ be an arbitrary QBF, and let $\mathcal{T}$ be as in the hardness proof of Lemma 63. TBox $\mathcal{T}$ is WA; it contains only constant-free, equality-free, and connected rules; and it does not contain a predicate of zero arity. Hence, by Lemma 7, a Horn-$\mathcal{SHI}$ TBox $\mathcal{T}'$ exists such that $\varphi$ is valid if and only if $\mathcal{T}'$ is not universally MFA. $\qquad\square$

We finish this section by proving that checking whether a set of Horn-$\mathcal{SHI}$ rules is universally MSA is PTime-hard; in this way, we also obtain a matching lower bound for theorem Theorem 62 from Section 6.1.

**Theorem 66.** *Checking whether a Horn-$\mathcal{SHI}$ TBox $\mathcal{T}$ is universally MSA is* PTime-*hard.*

*Proof.* Let $\mathcal{N}$ be a set of Horn propositional clauses of the form $\neg v_1 \vee \ldots \vee \neg v_n \vee v_{n+1}$ and let $v$ be a propositional variable; deciding $\mathcal{N} \models v$ is well known to be PTime-hard. Let $V_i$ be a concept uniquely associated with each propositional variable $v_i$; let $A$ be a fresh concept; and let $\mathcal{T}$ be the TBox obtained by transforming each propositional clause in $\mathcal{N}$ of the above form into rule (161).

$$A(x) \wedge V_1(x) \wedge \ldots \wedge V_n(x) \rightarrow V_{n+1}(x) \tag{161}$$

Finally, let $I = \{A(a)\}$. Clearly, $\mathcal{N} \models v$ holds if and only if $I \cup \mathcal{T} \models V(a)$ holds. TBox $\mathcal{T}$ is WA, it contains only constant-free, equality-free, and connected rules, and no predicate in $\mathcal{T}$ is of zero arity; hence, by Lemma 7, a Horn-$\mathcal{SHI}$ TBox $\mathcal{T}'$ exists such that $\mathcal{N} \not\models v$ holds if and only if $\mathcal{T}'$ is universally MFA. Finally, the only existential variable in $\mathcal{T}'$ occurs in a rule of the form (23), so it is straightforward to see that $\mathcal{T}'$ is universally MFA if and only if $\mathcal{T}'$ is universally MSA. $\qquad\square$

## 7. Experiments

To estimate the extent to which various acyclicity notions can be used in practice, we conducted two sets of experiments. First, we implemented MFA, MSA, JA, and WA checkers, and we used them to check acyclicity of a large corpus of Horn ontologies. Our goal was to see how many ontologies are acyclic and could thus be used with (suitably extended) materialisation-based OWL reasoners. Second, we computed the materialisation of the acyclic Horn ontologies and compared the number of facts before and after materialisation. The goal of these tests was to see whether materialisation-based reasoning with acyclic ontologies is practically feasible.

Tests were performed on a Windows R2 Server with two Intel Xeon 3.06GHz processors. We used a repository of 336 OWL ontologies whose TBox axioms can be transformed into existential rules where at least one rule contains an existential quantifier in the head. These ontologies include a large subset of the Gardiner ontology corpus (Gardiner, Tsarkov, & Horrocks, 2006), the LUBM ontology, several Phenoscape ontologies, and a number of ontologies from two versions of the Open Biomedical Ontology (OBO) corpus. Please note that no test ontology has been obtained from conceptual models (e.g., the ER models or UML diagrams): due to the specific modelling patterns used in conceptual modelling, such ontologies are less likely to be acyclic. Each test ontology can be accessed online from our ontology repository by means of a unique ID.[5] Each ID identifies one self-contained OWL ontology 'frozen in time' with all of its imports resolved at the time the ontology was added to the repository; furthermore, any possible future version of the ontology will be assigned a fresh ID. These measures should ensure that our experiments can be independently repeated at any point in the future.

## 7.1 Acyclicity Tests

We implemented all acyclicity checks by adapting the HermiT reasoner.[6] HermiT was used to transform an ontology into DL-clauses—formulae quite close to existential rules. In the result, at-least number restrictions in head atoms were replaced with existential quantification, atoms involving datatypes were eliminated, and the DL-clauses with no head atoms were removed: datatypes and empty heads can cause inconsistencies, but they cannot prevent the skolem chase from terminating.

Each set of rules $\Sigma$ obtained by the above preprocessing steps was considered in combination with each acyclicity notion $X \in \{\text{WA}, \text{JA}, \text{MSA}, \text{MFA}\}$ as follows. If $\Sigma$ did not contain the equality predicate, we simply checked whether $\Sigma \in X$. If $\Sigma$ contained the equality predicate, we checked whether $\Sigma \in X^{\cup}$, and we also checked whether $\Sigma' \in X$ for $\Sigma' \subseteq \Sigma$ the set of all rules of $\Sigma$ that do not contain the equality predicate; these tests provided us with a 'lower' and an 'upper' bound for acyclicity, respectively. Each acyclicity test was performed by modifying $\Sigma$ (or $\Sigma'$) as required by $X$ and then running HermiT to check for a particular logical entailment on the critical instance.

Our tests revealed MFA and MSA to be indistinguishable for all 336 test ontologies; that is, all MFA ontologies were found to be MSA as well (the converse holds per Theorem 14). A total of 213 (63.4%) ontologies were found to be MSA, including 43 of the 49 (87.8%) ontologies from the Gardiner corpus, 164 of the 208 (78.8%) OBO ontologies, and the LUBM ontology. In contrast, the GALEN ontology and its variants, the GO ontology and its extensions, and the 55 Phenoscape ontologies were found not to be MFA. These results are summarised in Table 2. Given the large number of ontologies tested, it would be impractical to present the results for each ontology individually. Instead, the ontologies are grouped by number of generating rules (G-rules), which are the rules containing an existential quantifier; for each group, Table 2 shows the total number of ontologies, as well as the numbers of ontologies found to be MSA, JA, and WA. Of the 123 ontologies that are not MFA, seven ontologies are in $\mathcal{ELH}^r$, so CQ answering over these ontologies can be realised using the combined approaches by Lutz et al. (2009) and Kontchakov et al. (2011).

The five older versions of OBO ontologies (IDs 00359, 00374, 00376, 00382, and 00486) are MSA, whereas their newer versions (IDs 00360, 00375, 00377, 00383, and 00487) are not MFA. In contrast, two older versions of OBO ontologies are not MFA (IDs 00432 and 00574), but their newer versions (IDs 00433 and 00575) are MSA.

---

| | ontologies without equality | | | | ontologies with equality | | | |
|---|---|---|---|---|---|---|---|---|
| G-rules | Total | MSA | JA | WA | Total | MSA | JA | WA |
| < 100 | 44 | 42 | 42 | 41 | 48 | 39 | 39 | 39 |
| 100–1K | 69 | 62 | 62 | 41 | 41 | 1 | 0 | 0 |
| 1K–5K | 38 | 31 | 30 | 24 | 17 | 2 | 1 | 1 |
| 5K–12K | 28 | 23 | 16 | 14 | 10 | 5 | 0 | 0 |
| 12K–160K | 18 | 7 | 6 | 5 | 23 | 1 | 0 | 0 |

Table 2: Results of acyclicity tests

Finally, we found 15 large OBO ontologies (including different versions of the same ontologies) that are MSA but not JA. Thus, MSA seems to be particularly useful on complex ontologies since it analyses implications between existentially quantified variables more precisely than the previously known notions. Table 3 shows for each of these ontologies the number of generating rules (G-rules), whether the ontology uses the equality predicate (Eq), the ontology expressivity in the description logic family of languages (DL), and the number of classes (C), properties (P), and axioms (A) that the ontology contains. Different versions of the same ontology are distinguished in the table as 'old' and 'new'. Two further ontologies (IDs 00762 and 00766) containing the equality predicate are MSA$^\cup$, but their status regarding joint acyclicity is unknown: they are JA when the rules involving the equality predicate are deleted, but are not JA$^\cup$.

## 7.2 Materialisation Tests

To estimate the practicability of materialisation in acyclic ontologies, we measured the maximal depth of function symbol nesting in terms generated by materialisation on critical instances. This measure, which we call *ontology depth*, is of interest as it provides us with a bound on the size of the materialisation. Out of the 213 MSA ontologies, our test succeeded on 207 of them (tests were aborted if they did not finish in three hours). On the latter ontologies, depth was distributed as follows:

- 123 (59.4%) ontologies have depth less than 5;

- 30 (14.5%) ontologies have depth between 5 and 9;

- 47 (22.7%) ontologies have depth between 10 to 19;

- 5 (2.4%) ontologies have depth between 20 and 49; and

- 2 (1.0%) ontologies have depth between 50 to 70.

These results leads us to believe that many (but clearly not all) ontologies have manageable depths, which should allow for successful materialisation-based query answering.

We also computed the materialisation for several acyclic ontologies. As our implementation is prototypical, our primary goal was not to evaluate the performance of computing the materialisation, but rather to estimate the blowup in the number of facts. We clearly do not expect this blowup to depend linearly on size of the input number of facts; however, our results should provide us with a rough estimate of the performance of materialisation-based reasoning in practice. Most of our test ontologies, however, do not contain many facts: ontologies are often constructed as general vocabularies, while facts are often application-specific and are thus not publicly available. To overcome this problem, we conducted two kinds of experiments.

| Ontology ID | G-rules | Eq | DL | C | P | A |
|---|---|---|---|---|---|---|
| biological_process_xp_cell.imports-local.owl | | | | | | |
| 00371 | 7464 | yes | $\mathcal{SHIF}$ | 17296 | 178 | 117925 |
| biological_process_xp_cellular_component.imports-local.owl (old) | | | | | | |
| 00374 | 8270 | yes | $\mathcal{SHIF}$ | 18673 | 186 | 126796 |
| biological_process_xp_multi_organism_process.imports-local.owl (old) | | | | | | |
| 00382 | 8378 | no | $\mathcal{EL}^{++}$ | 27900 | 18 | 295396 |
| biological_process_xp_plant_anatomy.imports-local.owl (old) | | | | | | |
| 00386 | 7559 | yes | $\mathcal{SHIF}$ | 19146 | 193 | 122062 |
| biological_process_xp_plant_anatomy.imports-local.owl (new) | | | | | | |
| 00387 | 12025 | yes | $\mathcal{SRIF}$ | 27412 | 215 | 213956 |
| bp_xp_cell.imports-local.owl | | | | | | |
| 00398 | 7419 | yes | $\mathcal{SHIF}$ | 17296 | 177 | 117881 |
| bp_xp_cellular_component.imports-local.owl | | | | | | |
| 00400 | 7999 | yes | $\mathcal{SHIF}$ | 18676 | 175 | 126540 |
| cellular_component_xp_go.imports-local.owl (old) | | | | | | |
| 00415 | 7752 | no | $\mathcal{EL}^{++}$ | 27890 | 8 | 210765 |
| cellular_component_xp_go.imports-local.owl (new) | | | | | | |
| 00416 | 12269 | no | $\mathcal{EL}^{++}$ | 37254 | 9 | 334762 |
| fypo.owl | | | | | | |
| 00476 | 1834 | no | $\mathcal{EL}^{++}$ | 1677 | 22 | 8027 |
| go_xp_regulation.imports-local.owl (old) | | | | | | |
| 00486 | 7777 | no | $\mathcal{EL}^{++}$ | 27891 | 5 | 295138 |
| go_xp_regulation.owl (old) | | | | | | |
| 00488 | 7777 | no | $\mathcal{EL}^{++}$ | 27883 | 5 | 214080 |
| go_xp_regulation.owl (new) | | | | | | |
| 00489 | 9507 | no | $\mathcal{EL}^{++}$ | 30170 | 6 | 238200 |
| molecular_function_xp_regulators.imports-local.owl (old) | | | | | | |
| 00536 | 6762 | no | $\mathcal{EL}^{++}$ | 25521 | 5 | 198170 |
| molecular_function_xp_regulators.imports-local.owl (new) | | | | | | |
| 00537 | 11089 | no | $\mathcal{EL}^{++}$ | 34135 | 8 | 316057 |

Table 3: MSA but not JA ontologies

First, we computed the materialisation of two ontologies that contain facts: LUBM with one university (ID 00347), and the 'kmi-basic-portal' ontology (ID 00078). The TBox of LUBM has eight generating rules and depth one, and there are $100,543$ facts before ma-terialisation. Materialisation took only 2 seconds, and it produced $231,200$ new facts, of which $97,860$ were added by the generating rules. The 'kmi-basic-portal' ontology has ten generating rules and depth two, and there were 198 facts before materialisation. Materialisation took only 0.03 seconds, and it produced 744 new facts, of which 145 were added by the generating rules.

| Depth | # | time | | gen. size | | mat. size | |
|---|---|---|---|---|---|---|---|
| | | max | avg | max | avg | max | avg |
| $< 5$ | 123 | 38 | 0.7 | 65 | 2 | 89 | 8 |
| 5–9 | 30 | 71 | 7 | 122 | 30 | 132 | 38 |
| 10–70 | 54 | 9396 | 1807 | 1286 | 175 | 1297 | 189 |

Table 4: Materialisation times (in seconds) and sizes

Second, for each of the ontologies identified as MSA, we instantiated each class and each property with fresh individuals. We then computed the materialisation and measured the *generated size* (the number of facts introduced by the generating rules divided by the number of facts before materialisation), the *materialisation size* (the number of facts after materialisation divided by the number of facts before materialisation), and the time needed to compute the materialisation. Since most generating rules in these ontologies have singleton body atoms (i.e., they are of the form $A(x) \rightarrow \exists R.C(x)$), these measures should provide a reasonable estimate of the increase in the number of facts during materialisation. Table 4 summarises the results of our tests for the 207 ontologies on which the test succeeded. Ontologies are grouped by their depth, and each group shows the number of ontologies (#), and the maximal and average materialisation time, generated size, and materialisation size.

Thus, materialisation seems practically feasible for many ontologies: for 123 ontologies with depth less than 5, materialisation increases the ontology size by a factor of 8. This suggests that principled, materialisation-based reasoning for ontologies beyond the OWL 2 RL profile may be feasible, especially for ontologies with relatively small depths.

## 8. Conclusions

In this paper, we investigated acyclicity notions—sufficient conditions that ensure termination for skolem chase on existential rules. We proposed two novel notions, called MFA and MSA, for which we determined tight complexity bounds for membership checking, as well as for conjunctive query answering over acyclic existential rules.

We also conducted a thorough investigation of the acyclicity notions known in the literature, and we produced a complete taxonomy of their relative expressiveness. Our results show that MFA and MSA generalise most of the previously considered notions.

We next investigated ways to ensure acyclicity of existential rules that contain the equality predicate. To this end, we presented several optimisations of the singularisation technique by Marnette (2009). Our optimisations can often reduce the number of acyclicity checks needed, thus making the singularisation technique more suitable for practical use.

Finally, we studied the problem of answering conjunctive queries over acyclic DL ontologies. On the theoretical side, we showed that acyclicity can make this problem computation-ally easier; furthermore, provided that the result is acyclic, one can extend Horn ontologies with arbitrary SWRL rules without affecting decidability and the worst-case complexity of query answering. On the practical side, we investigated the extent to which acyclicity notions enable principled extensions of materialisation-based ontology reasoners with support for existential quantification. Our tests show that many ontologies commonly used in practice are acyclic, and that the blowup in the number of facts due to materialisation is manageable. This suggests that principled extensions of materialisation-based ontology reasoners are practically feasible and useful.

An interesting topic for future work is to see whether our acyclicity notions can be used in a more general logic programming setting. We see several main sources of technical difficulties towards this goal. First, general logic programs can contain functional terms in body atoms. Such terms can 'cancel out' function symbols introduced by head atoms, and it is not clear how to take this into account in an acyclicity test. Second, logic programs can contain atoms under nonmonotonic negation, which are likely to need special treatment; Magka, Krötzsch, and Horrocks (2013) recently made a first step in that direction. Third, it might be desirable to modularise the ways in which these different concerns are handled and thus arbitrarily combine the approaches for handling function symbols in the body and/or the head with the approaches for dealing with nonmonotonic negation.

## Acknowledgments

This work was supported by the Royal Society, the Seventh Framework Program (FP7) of the European Commission under Grant Agreement 318338, 'Optique', and the EPSRC projects ExODA, Score!, and MaSI[3].